\documentclass[a4paper,11pt]{article}
\usepackage{jheppub}
\pdfoutput=1

\usepackage[T1]{fontenc} 

\usepackage{comment}
\usepackage{amsmath}    
\usepackage{amssymb}    %
\usepackage{graphicx}   
\usepackage{verbatim}   
\usepackage{color}      
\usepackage{subfigure}  
\usepackage{hyperref}
\usepackage{multirow}
\usepackage{comment}
\usepackage{pdflscape}
\usepackage{rotating}
\usepackage{booktabs,tabularx}
\usepackage{float}	
\usepackage{color}
\definecolor{rosso}{RGB}{210,0,0}

 \definecolor{X575}{rgb}{0.05, 0.7, 0.05}

 \newcommand\sss{\scriptscriptstyle}

 \newcommand{\nn}{\nonumber}
 \newcommand{\lt}{\lambda_3}
 \newcommand{\tril}{\lt}
 \newcommand{\lf}{\lambda_4}
 \newcommand{\qual}{\lf}
 \newcommand{\kt}{\kappa_3}
 \newcommand{\kf}{\kappa_4}
 
 \newcommand{\ktre}{\kt}
 \newcommand{\kqual}{\kf}
 \newcommand{\trilsm}{\tril^{\rm SM}}
\newcommand{\qualsm}{\qual^{\rm SM}}
\newcommand{\cbs}{\bar{c}_6}
\newcommand{\cbe}{\bar{c}_8}
\newcommand{\cbstr}{\bar{c}_6^{\rm true}}
\newcommand{\cbetr}{\bar{c}_8^{\rm true}}

\newcommand{\mh}{m_{ \sss H}}

\newcommand{\mt}{m_{t}}

\newcommand{\inmath}[1]{\relax\ifmmode#1\else$#1$\fi}
\def\ee/{\inmath{e^{+}e^{-}}}
\def\zh/{\inmath{ZH}}
\def\zhh/{\inmath{ZHH}}
\def\zhhh/{\inmath{ZHHH}}
\def\vvh/{\inmath{\nu_e\bar{\nu}_eH}}
\def\vvhh/{\inmath{\nu_e\bar{\nu}_eHH}}
\def\vvhhh/{\inmath{\nu_e\bar{\nu}_eHHH}}
\def\eezh/{\ee/\inmath{\to}\zh/}
\def\eezhh/{\ee/\inmath{\to}\zhh/}
\def\eezhhh/{\ee/\inmath{\to}\zhhh/}
\def\eevvh/{\ee/\inmath{\to}\vvh/}
\def\eevvhh/{\ee/\inmath{\to}\vvhh/}
\def\eevvhhh/{\ee/\inmath{\to}\vvhhh/}

\def\beq{\begin{equation}}
\def\beqn{\begin{eqnarray}}
\def\eeq{\end{equation}}
\def\eeqn{\end{eqnarray}}
\def\beal{\begin{align}}
\def\endal{\end{align}}

\newcommand{\MSbar}{{\rm \overline{MS}}}
\newcommand{\M}{\mathcal{M}}

\title{Probing the scalar potential via double Higgs boson production at hadron colliders}

 \author[a]{Sophia Borowka,}
 \author[a,b]{Claude Duhr,}
 \author[b,c]{Fabio Maltoni,}
 \author[d]{Davide Pagani,}
 \author[b]{Ambresh Shivaji,}
 \author[b]{Xiaoran Zhao.}

\affiliation[a] {\small Theoretical Physics Department, CERN, CH-1211 Geneva 23, Switzerland}
\affiliation[b]{\small Centre for Cosmology, Particle Physics and Phenomenology (CP3), Universit\'{e} Catholique de Louvain, B-1348 Louvain-la-Neuve, Belgium}
\affiliation[c]{\small Dipartimento di Fisica e Astronomia, Universit\`a di Bologna and INFN, Sezione di Bologna, via Irnerio 46, 40126 Bologna, Italy}
\affiliation[d]{\small Technische Universit\"{a}t M\"{u}nchen, James-Franck-Str.~1, D-85748 Garching, Germany}
\emailAdd{sophia.borowka@cern.ch}
\emailAdd{claude.duhr@cern.ch}
\emailAdd{fabio.maltoni @uclouvain.be \& @unibo.it}
\emailAdd{davide.pagani@tum.de}
\emailAdd{ambresh.shivaji@uclouvain.be}
\emailAdd{xiaoran.zhao@uclouvain.be}

 \note{Preprint:  TUM-HEP-1176/18, CP3-18-69, CERN-TH-2018-258, MCnet-18-31}

\abstract{
We present a sensitivity study on the cubic and quartic self couplings in double Higgs production via gluon fusion at hadron colliders. Considering the relevant operators in the Standard Model Effective Field Theory up to dimension eight, we calculate the dominant contributions up to two-loop level, where the first dependence on the quartic interaction appears.  Our approach allows to study the independent variations of the two self couplings and to clearly identify the terms necessary to satisfy gauge invariance and to obtain UV-finite results order by order in perturbation theory. We focus on the $b \bar b \gamma \gamma$ signature for simplicity and provide the expected bounds for the cubic and quartic self couplings at the 14 TeV LHC with 3000 fb$^{-1}$ (HL-LHC) and for a future 100 TeV collider (FCC-100) with 30 ab$^{-1}$. We find that while the HL-LHC will provide very limited sensitivity on the quartic self coupling, precision measurements of double Higgs production at a FCC-100 will offer the opportunity to set competitive  bounds. We show that combining information from double and triple Higgs production leads to significantly improved prospects for the determination of the quartic self coupling.}

\begin{document} 
\maketitle
\flushbottom

 \section{Introduction}\label{sec:intro}
 
Since the discovery in 2012 by the ATLAS and CMS collaboration \cite{Aad:2012tfa,Chatrchyan:2012xdj}, the Large Hadron Collider (LHC) has already disclosed an impressive amount of information on the properties of the resonance at 125 GeV, confirming so far the expectations of the Standard Model (SM). The new  particle is a narrow scalar \cite{Aad:2013xqa, Chatrchyan:2013mxa}, interacting with (third generation) fermions and vector bosons with a strength proportional to the mass of the particle \cite{Khachatryan:2016vau, Sirunyan:2018koj}. All the expected main production and decay modes have been observed \cite{Aad:2012tfa,Chatrchyan:2012xdj, Sirunyan:2017khh, ATLAS:2018lur, Sirunyan:2018kst, Aaboud:2018zhk, Aaboud:2018urx, Sirunyan:2018hoz}. Future runs at the LHC and future colliders will provide new information (such as the coupling to second generation fermions) and higher accuracy on the known couplings. The current measurements  already indicate that New Physics (NP) effects cannot substantially affect the couplings of the Higgs boson with vector bosons and third generation fermions, placing the scale of NP well above the electroweak symmetry breaking scale.  

The situation, however, is very different for the scalar potential on which we have not gained any relevant information so far and which is therefore largely unexplored. The reason is simply that the scalar potential, whose shape  is ultimately responsible for Electroweak Symmetry Breaking (EWSB), can be probed only by measuring the Higgs self couplings. At hadron and lepton colliders, a direct sensitivity on the cubic or quartic Higgs self couplings can be achieved only via the simultaneous production of two or three Higgs bosons, respectively. Due to the smallness of the corresponding cross sections (around 32 fb in the case of $pp \to HH$ at 13 TeV~\cite{Maltoni:2014eza,deFlorian:2015moa,Borowka:2016ehy,Grazzini:2018bsd} and 0.05 fb for $pp \to HHH$ ~\cite{Maltoni:2014eza,deFlorian:2016sit}) these processes have not yet been observed at the LHC. Therefore the study of the Higgs self couplings is currently not only far from the precision level but also very challenging for the future.

In the case of double Higgs production, only exclusion limits are currently available and the most stringent result has been obtained by the ATLAS collaboration. Combining  three different analyses ($4b, b \bar b \tau \tau, $ and $ b \bar b \gamma \gamma$ signatures) based on 27.5-36.1 fb$^{-1}$ of data accumulated at 13 TeV \cite{Aaboud:2018knk, Aaboud:2018ftw, ATLAS:2018otd, Aaboud:2018sfw}, cross sections larger than 6.7 times the SM one can be excluded. This limit translates into the bound $-5.0 ~\lt^{\rm SM} < \lt < 12.1 ~\lt^{\rm SM}    $, where $\lt$ is the cubic coupling and $\lt^{\rm SM}$ is its SM prediction. 
With a collected luminosity of 300 fb$^{-1}$, or even with 3000 fb$^{-1}$ in the case of the High-Luminosity (HL) option, it is not still clear if the observation of SM production can be achieved. Although many phenomenological studies have been performed \cite{Baur:2003gp, Baglio:2012np, Yao:2013ika, Barger:2013jfa, Azatov:2015oxa, Lu:2015jza, Dolan:2012rv, Papaefstathiou:2012qe, deLima:2014dta, Wardrope:2014kya, Behr:2015oqq, Englert:2014uqa, Liu:2014rva, Cao:2015oxx, Englert:2015hrx, Bishara:2016kjn, Cao:2016zob, Huang:2017jws, Adhikary:2017jtu, Goncalves:2018yva, Chang:2018uwu, Arganda:2018ftn, Homiller:2018dgu}, the best experimental predictions for HL-LHC only provide upper limits on the SM cross sections. 

In the case of $\lf$, the prospects are very  uncertain. At the LHC, inferring information from triple Higgs production will be extremely challenging
\cite{Plehn:2005nk, Binoth:2006ym}. Its cross section is very small~\cite{Maltoni:2014eza,deFlorian:2016sit} and depends on the quartic interaction very weakly. 
Even a future 100 TeV proton--proton collider will need a considerable amount of integrated luminosity in order to obtain rather loose bounds~\cite{Chen:2015gva,Kilian:2017nio,Fuks:2017zkg}.

Given the current and expected future results, new complementary strategies for the determination of the Higgs self couplings would be desirable. Recently, the possibility of probing the cubic Higgs self coupling $\lt$ via precision measurements of single Higgs production channels at future lepton colliders \cite{McCullough:2013rea} and at the LHC and future hadron colliders \cite{Gorbahn:2016uoy,Degrassi:2016wml} has been suggested, exploiting the fact that next-to-leading order (NLO) EW corrections to the single Higgs production and decay modes involve $\lt$. The turning point for the possibility of determining the cubic interaction from single Higgs production measurements at the LHC has been the understanding that the different production channels depend on $\lt$ in a very different way and that the effects are differential, the sensitivity being enhanced at threshold~\cite{Degrassi:2016wml}. Even though the expected effects are small, a competitive sensitivity can be obtained by combining globally information from single Higgs measurements, total cross sections as well as distributions~\cite{Degrassi:2016wml}.  Since then, considerable effort has been invested in studying the feasibility of this strategy: predictions for the differential distributions for all the Higgs production channels have become available~\cite{Bizon:2016wgr, Maltoni:2017ims}, and studies with more general (and realistic) scenarios for the existence of anomalous Higgs interactions \cite{DiVita:2017eyz, Barklow:2017awn, Maltoni:2017ims,DiVita:2017vrr} have appeared, also in combination with the direct double Higgs information \cite{DiVita:2017eyz, Barklow:2017awn, DiVita:2017vrr,Maltoni:2018ttu}. Following the same logic, the $\lt$ bounds have been extracted also from EW precision observables~\cite{Degrassi:2017ucl,Kribs:2017znd}. It is now clear that the indirect determination of $\lt$ via precision measurements is expected to provide comparable bounds to those that are currently obtained via the direct searches for double Higgs production. Very recently it has been proposed that double Higgs production could be exploited for probing the quartic Higgs self coupling $\lf$ via precise measurements~\cite{Maltoni:2018ttu, Liu:2018peg}. The first studies at lepton colliders show that coarse bounds on $\lf$ could be obtained and would complement the information from triple Higgs production, improving the ultimate results via a combination.  At variance with the case of $\lt$, $\lf$-dependent loop corrections are ultraviolet (UV) divergent and in order to be renormalised they have to be performed in a Effective-Field-Theory (EFT) framework. The renormalisation procedure and the relevant counterterms have been provided in Ref.~\cite{Maltoni:2018ttu}. This framework has to be used also when the interest is focused only on independent variations of $\lt$ and $\lf$, so that UV-finite results can be obtained.

The similar calculation for the case of hadronic collisions is computationally more involved, since the process $pp\to HH$ involves the loop-induced $gg\to HH$ partonic process at Born level and therefore the sensitivity on $\lf$ originates from two-loop amplitudes. The first incomplete estimation of these effects has been presented in Ref.~\cite{Bizon:2018syu}, showing the possibility of following this strategy also at future hadron colliders.

\medskip

In this paper we analyse this strategy in detail and provide the first complete and consistent computation of the relevant contributions to $g g \to HH$ at two loops.  All the two-loop diagrams involving $\lf$  are taken into account and  numerically evaluated without any further approximation via {\sc\small pySecDec} \cite{Borowka:2017idc,Borowka:2017esm}. Moreover, following the approach of Ref.~\cite{Maltoni:2018ttu},  we take into account also corrections induced by additional $\lt$ effects at two loops, which are non negligible for large values of $\lt$, and we renormalise the ensuing UV divergences. We perform this calculation at the differential level and we consider the $b \bar b \gamma \gamma$ signature emerging from the decays of the Higgs bosons as a first application.  This channel has been identified as  the most promising one~\cite{Contino:2016spe,Goncalves:2018qas,Azatov:2015oxa,Barr:2014sga,He:2015spf,Mangano:2016jyj,Chang:2018uwu} and it allows for the reconstruction of the di-Higgs invariance mass $m(HH)$. Following the analyses in Ref.~\cite{Azatov:2015oxa}, we study the constraints that can be set on $\lt$ and $\lf$ via the measurement of the $m(HH)$ distribution from $b \bar b \gamma \gamma$ events for two different experimental setups:  the LHC with 3000 fb$^{-1}$ integrated luminosity (HL-LHC) and at a 100 TeV collider with  30 ab$^{-1}$ integrated luminosity. The EFT parametrisation allows us to consider both the generic case, where $\lt$ and $\lf$ can vary independently, and a ``well-behaved'' EFT approach, where higher dimension operators induce smaller effects and $\lf$ depends on $\lt$. In both cases we assume that the dominant BSM effects originate from the distortion of the Higgs potential, namely, anomalous interactions of the Higgs boson with other SM particles lead to subdominant effects. This approach is adequate to establish the sensitivity. 

The paper is organised as follows. We first provide details on the computational framework, clarifying the theoretical assumptions, identifying the most relevant terms and describing the most important elements and features of the two-loop computation in sec.~\ref{sec:calculation}. Section~\ref{sec:num} presents the results of the computation at the total as well as differential level, while in sec.~\ref{sec:constraints} constraints that can be derived from future measurements at the LHC and at 100 TeV FCC are discussed in two different scenarios. We summarise our findings in sec.~\ref{sec:conclusion}. Three appendices contain complementary and technical information.

\section{Calculation}\label{sec:calculation}
\subsection{Parametrisation of $\lt$ and $\lf$ effects}

As already mentioned in the introduction, in order to vary the cubic and quartic Higgs self couplings $\lt$ and $\lf$ independently at all orders in perturbation theory in a consistent way, an EFT approach where operators are defined above the EWSB scale and respect all symmetries, hidden or not,  has to be employed. This allows one to systematically identify gauge invariant and UV finite subsets of diagrams. For this reason, we will use the computational framework introduced and described in detail in Ref.~\cite{Maltoni:2018ttu}. In this section we summarise the most important aspects and we highlight some differences w.r.t.~Ref.~\cite{Maltoni:2018ttu}.

Starting from the SM Higgs potential
\begin{equation}
V^{\rm SM}(\Phi)=-\mu^2 (\Phi^\dagger \Phi)+\lambda (\Phi^\dagger \Phi)^2\, ,
\label{VSM}
\end{equation}
 we denote NP effects as $V^{\rm NP}$ so that the general form of the potential can be written as
\begin{equation}
V(\Phi)=V^{\rm SM}(\Phi)+ V^{\rm NP}(\Phi)\, ,\qquad \Phi =  \begin{pmatrix}G^+ \cr  \frac{1}{\sqrt{2}}(  v+H+i G^0)\, 
 \end{pmatrix}\  \, ,
\label{V}
\end{equation}
where the symbol $\Phi$ refers to the Higgs doublet.
Using the conventions of  Ref.~\cite{Boudjema:1995cb}, the most general form of an $SU(2)$-invariant  $V^{\rm NP}$ potential reads
\begin{equation}
V^{\rm NP}(\Phi)\equiv  \sum_{n=3}^{\infty}\frac{c_{2n}}{\Lambda^{2n-4}}\left(\Phi^\dag\Phi -\frac{1}{2}v^2 \right)^n \, .
\label{VNP}
\end{equation}
One of the advantages of this parameterisation is that at tree-level $\lt$ only depends on $c_6$ and $\lf$ only on $c_6$ and $c_8$.
Indeed, after EWSB, we can rewrite $V(\Phi)$ as
\begin{equation}
 V(H) = \frac{1}{2} \mh^2 H^2 + \tril v H^3 + \frac{1}{4}\qual H^4 +  {\lambda_5} \frac{H^5}{v} + O(H^6)\,,
\end{equation}
and thus define the self couplings $\lt$ and $\lf$ via
\begin{eqnarray}
\label{k3}
\ktre \equiv\frac{\tril}{\trilsm} = 1+ \frac{c_6 v^2}{\lambda \Lambda^2} &\equiv& 1+\bar c_6,\\
\label{k4}
\kqual  \equiv\frac{\qual}{\qualsm}  = 
1+ \frac{ 6 c_6 v^2}{\lambda \Lambda^2} +
\frac{ 4 c_8 v^4}{\lambda \Lambda^4}
&\equiv& 1+ 6 \bar c_6 + \bar c_8 
  \, . 
\end{eqnarray}   
The quantities $\trilsm$ and $\qualsm$ are the values of $\tril$ and $\qual$ in the SM, respectively, and read
\begin{equation}\label{l34}
\trilsm=\qualsm=\lambda=\frac{\mh^2}{2 v^2}\,.
\end{equation}
In other words, the barred quantities $\bar c_{6}$ and $\bar c_{8}$  are simply  $c_6$ and $c_8$  normalised in such a way that relations to $\ktre$ and  $\kqual$ are simple. In particular
\begin{eqnarray}
    \bar c_6&\equiv &\frac{c_6 v^2}{\lambda \Lambda^2}=\ktre-1\,, \label{k3inv} \\
    \quad\bar c_8&\equiv& \frac{4c_8v^4}{\lambda\Lambda^4}=\kqual-1-6(\ktre-1)\,. \label{k4inv}
\end{eqnarray}

Using  the parameterisation in eq.~\eqref{VNP} and eqs.~\eqref{k3} and \eqref{k4}, or equivalently eqs.~\eqref{k3inv} and \eqref{k4inv},  we can trade $\ktre$ and $\kqual$ with only two other parameters, $\cbs$ and $\cbe$. In so doing, we can always think of using the EFT approach as a way to obtain gauge invariant and UV-finite results in the anomalous coupling approach.\footnote{Note that using the alternative parameterisation $V^{\rm NP}(\Phi)\equiv  \sum_{n=3}^{\infty}\frac{c'_{2n}}{\Lambda^{2n-4}}(\Phi^\dag\Phi)^n$ both $\lt$ and $\lf$ would depend on all the $c'_{i}$ coefficients already at the tree level.} 
We note that, \emph{a priori}, in a well-behaved EFT  higher dimensional effects are expected to suppressed by a large scale $\Lambda$. Thus, in the first approximation, deviations in $\ktre$ and $\kqual$ are strongly correlated, {\it i.e.}, $(\kqual -1) \simeq 6  (\ktre -1)$, see also eq.~\eqref{k4inv}. Similarly to what as been done in Refs.~\cite{Maltoni:2018ttu,Liu:2018peg, Bizon:2018syu}, in this work we adopt as starting point an agnostic attitude towards the values that $\kt$ and $\kf$ can assume, in order to cover the sensitivity that future colliders can probe. We will later comment on bounds on $\kt$ and $\kf$ making different UV assumptions.

\medskip

In this work we calculate the effects of anomalous cubic and quartic couplings in double Higgs production at hadron colliders. While $\lt$ affects the $gg \to HH$ amplitude already at the Born level, $\lf$ enters only via NLO EW corrections, {\it i.e.}, at the two-loop level. Before discussing the details of the calculation it is convenient to anticipate what are the quantities that enter in our phenomenological predictions. 
 \begin{figure}
 \centering
 \includegraphics[scale=0.7]{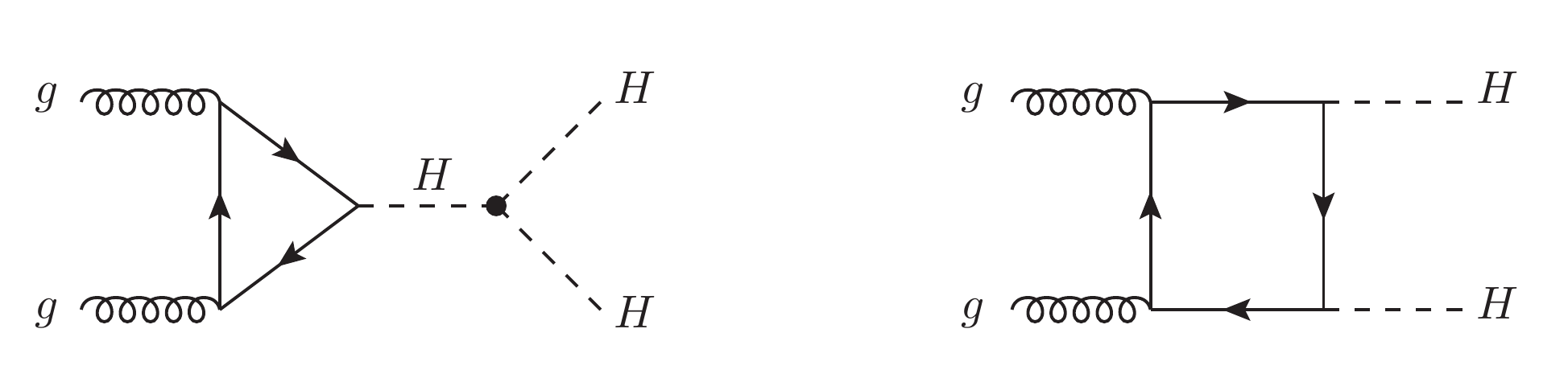} 
 \caption{Double Higgs production at LO in SM. The triangle diagram is sensitive to the cubic coupling.}
 \label{fig:HH1L}
 \end{figure}
In fig.~\ref{fig:HH1L} we display the one-loop diagrams of the Born amplitude in $HH$ production. While the triangle (left diagram) depends on $\lt$, the box (right diagram) does not. Moreover, it is well known that the interference effects between the two diagrams leads to large cancellations. QCD corrections have been computed up to next-to-next-to-LO \cite{deFlorian:2015moa,Grazzini:2018bsd}  and, besides reducing the scale dependence, they increase the LO cross section by roughly a factor of 2.  In this work we will assume that QCD corrections factorise from the two-loop EW effects that we calculate. While the accuracy of this assumption has been directly tested only in very few cases \cite{Dittmaier:2015rxo, Bonetti:2018ukf, Anastasiou:2018adr}, it has been often employed in the past, both due to the difficulty of calculating QCD-EW mixed corrections and due to the theoretical arguments supporting its validity.  
\begin{figure}[h]
\centering
 \includegraphics[scale=0.75]{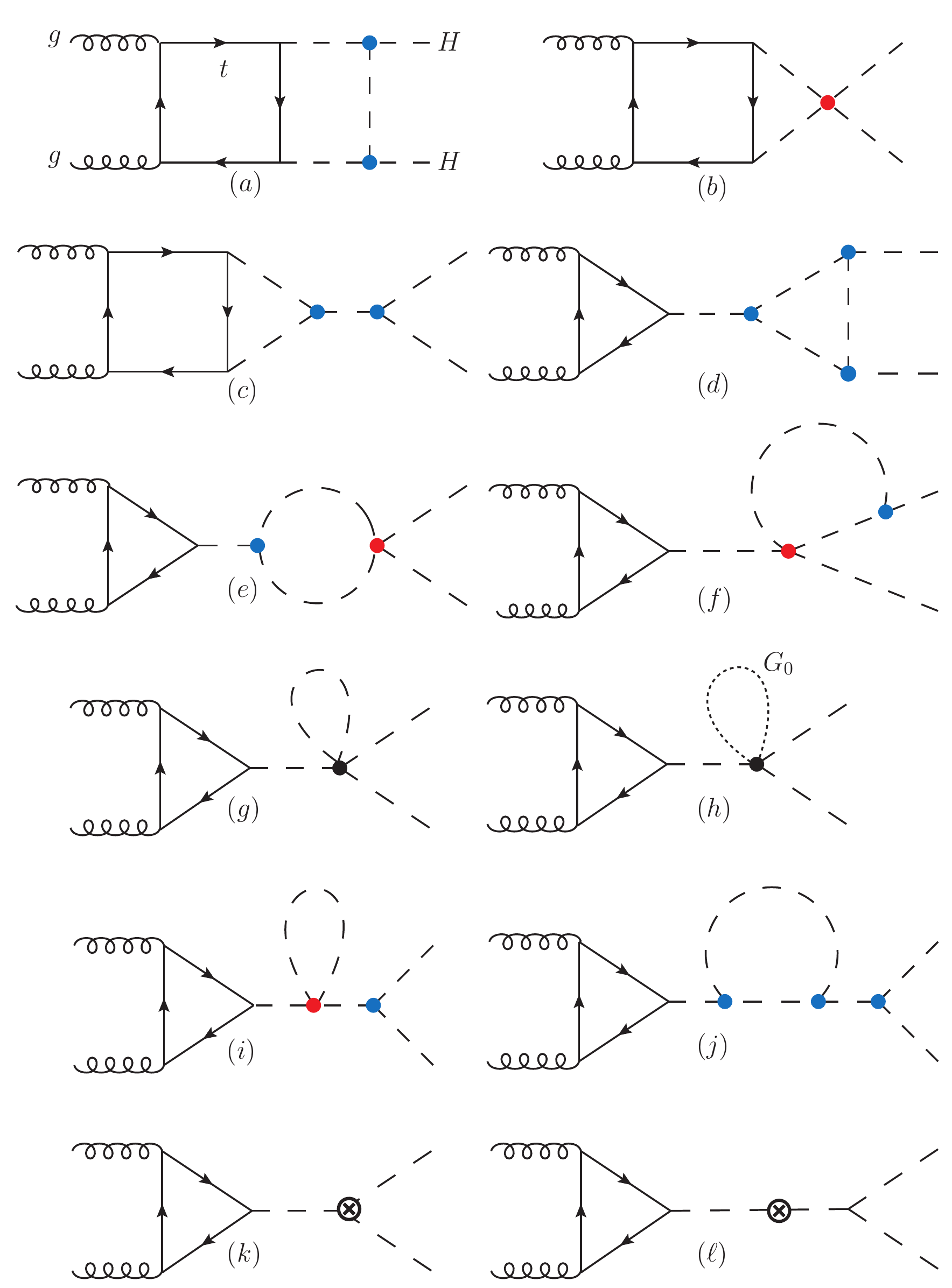}
 \caption{Two-loop topologies involving $\cbs$ and $\cbe$ effects on Higgs self coupling in $gg \to HH$. Except diagrams $(g)$ and $(h)$,  all topologies are present in the SM. We have marked with a  blob all the vertices involving $\cbs$ and $\cbe$; cubic vertices are  in blue while quartic ones are in red. Diagrams $(a)$-$(c)$ are non-factorisable two-loop topologies. Diagrams $(d)$-$(h)$, together with the counterterm $(k)$, can be evaluated via the one-loop form factor $V[HHH]$, while $(i)$,$(j)$ and $(l)$ with the $P[HH]$ one.}\label{fig:hh2l}
\end{figure} 
Two-loop corrections to $HH$ production involve further $\lt$ effects and introduce a $\lf$ dependence, as can be seen in fig.~\ref{fig:hh2l}. All the contributions arising from the two-loop topologies depicted in fig.~\ref{fig:hh2l} have been evaluated and renormalised via UV counterterms; more details concerning the calculation are given in Sec.~\ref{sec:details}.

Following the approach presented in Ref.~\cite{Maltoni:2018ttu} for $e^+e^-$ collisions, we define the quantity to be used in phenomenological investigations as
\begin{eqnarray}
    \sigma^{\rm pheno}_{\rm NLO}  &=& \sigma_{\rm LO} + \Delta\sigma_{\bar c_6}+ \Delta\sigma_{\bar c_8}\, \label{sHHNLOpheno}\,,
\end{eqnarray}
where 
\begin{eqnarray}
\sigma_{\rm LO} &=& \sigma_{0} +\sigma_{1} \bar c_6 + \sigma_{2} \bar c_6^2\, , \label{sHHLO} 
\end{eqnarray}
is the LO prediction. In eq.~\eqref{sHHLO}, $\sigma_{0}$ is the SM prediction, $\sigma_1$ corresponds to the leading contribution in the EFT expansion, being of order $(v/\Lambda)^2$, while
$\sigma_2$ is of order  $(v/\Lambda)^4$ and corresponds to the squared EFT term. Clearly, no contribution proportional to $\cbe$ appears at LO.
The NLO corrections are included through the terms
\begin{eqnarray}
    \Delta\sigma_{\bar c_6}&=&  \bar c_6^2   \Big[ \sigma_{30}  \bar c_6 +\sigma_{40} \bar c_6^2 \Big] +    \tilde \sigma_{20} \bar c_6^2\, , \label{Dscs}\\
    \Delta\sigma_{\bar c_8}&=&\bar c_8 \Big[\sigma_{01} +\sigma_{11} \bar c_6 +\sigma_{21}\bar c_6^2\Big]  \label{Dsce}\,,
\end{eqnarray}
which
are  the loop corrections induced by $\cbs$ on top of $\cbs$ and the two-loop $\cbe$-dependent part, respectively. They both  
originate from the topologies shown in Fig.~\ref{fig:hh2l}. In the following we explain the rationale behind these formulae and the meaning of the different $\sigma_{i(j)} \cbs^i \cbe^j$ terms entering them.

First of all it is important to note that we organise the different contributions in terms of $\cbs$ and $\cbe$ and not $\lt$ and $\lf$. As explained in Ref.~\cite{Maltoni:2018ttu} this organisation reflects the necessary EFT expansion that has to be performed in order to renormalise UV divergences
and obtain gauge invariant predictions. We recall that $\cbs$  can be directly related to $\lt$, while $\cbe$ captures the violation of the relation $\kqual  = 6  \ktre -5 $, which holds if only $\cbs$ is present, {\it cf.}~eqs.~\eqref{k3inv} and \eqref{k4inv}.

Our goal is not to determine the ultimate precision that can be achieved at future colliders on $\cbs$ and $\cbe$. Rather, we want to perform the first sensibility study on the determination of the cubic and quartic Higgs self couplings via double Higgs production at future hadron colliders. For this reason, SM EW corrections on top of $\sigma_{\rm LO}$ are not taken into account. Since we are agnostic about the possible size of $\cbs$, large cubic couplings are possible and lead to sizable enhancements via topologies such as $(d)$ in fig.~\ref{fig:hh2l} \cite{DiLuzio:2017tfn,Maltoni:2018ttu}. For this reason, in $\Delta\sigma_{\bar c_6}$ we take into account all the contributions of order $\cbs^3$ and $\cbs^4$. These two contributions are relevant only for large $\cbs$, since otherwise they are suppressed w.r.t. the contributions appearing at LO. We remind that in Refs.~\cite{DiLuzio:2017tfn,Maltoni:2018ttu} it has been shown that $\Delta\sigma_{\bar c_6}$, and therefore  $\sigma^{\rm pheno}_{\rm NLO}$, in general makes sense only in the range $|\cbs|<5$. Outside this range perturbativity is violated for any prediction involving the bulk of $HH$ production. We will comment more on this point in Sec.~\ref{sec:num}. At variance with  Ref.~\cite{Maltoni:2018ttu}, we include also the term $ \tilde \sigma_{20} \bar c_6^2$ in eq.~\eqref{Dscs}. This term includes only part of the two-loop contributions of order $\cbs^2$ and its purpose is to preserve the large cancellations that are present in $\sigma_{\rm LO}$ between different $\sigma_i \cbs^i$ terms, since also in $\Delta\sigma_{\bar c_6}$ these cancellations are distributed among different $\sigma_{i0}$ terms. On the other hand, it is relevant only for $\cbs \sim 2$ where the cross section reaches the smallest value and the cancellations are the largest.\footnote{We have verified that the inclusion of the corresponding term in the $e^+e^-$ studies in Ref.~\cite{Maltoni:2018ttu} would lead to negligible differences.}  
      
The quantity $\Delta\sigma_{\bar c_8}$ is the most relevant part of our computation and it solely induces the sensitivity on $\cbe$. At variance with Ref.~\cite{Bizon:2018syu}, where only the topology $(b)$ has been considered, in this term we take into account also all the contributions originating from topologies  $(e)$-$(i)$, which contribute at the same level and therefore cannot be ignored in any gauge-invariant calculation.\footnote{ Note that the topology $(g)$ involves a $H^5$ interaction which in principle depends also on the $\bar c_{10}$ Wilson coefficient form the dimension-10 operator $\left(\Phi^\dag\Phi -\frac{1}{2}v^2 \right)^5$. As discussed in Ref.~\cite{Maltoni:2018ttu}, the effect of this diagram can be redefined as a constant shift on $\cbs$ and therefore our calculation is sensitive on a linear combination of $\cbs$ and $\bar c_{10}$, which we set equal to zero. Nevertheless, the $\cbs$ and $\cbe$ contributions emerging from this diagrams are taken into account. See Ref.~\cite{Maltoni:2018ttu} for more details.} 
Also for the case of $\cbe$, a theoretical bound based on the perturbativity requirement can be set \cite{Maltoni:2018ttu} and corresponds to  $|\cbe|<31$.

\subsection{Organisation of the calculation}
\label{sec:details}
In this section we give more details about our computational framework.
Let us first consider the origin of the contributions in eqs.~\eqref{Dscs} and ~\eqref{Dsce}, in particular the presence of the term $\tilde \sigma_{20}$.
Using the same notations as for the $\sigma_{i(j)}$ terms, we define the different contributions of order $\cbs^i \cbe^j$ entering the $\M(gg \to HH)$ amplitude as $\M_{i(j)}$. Denoting by $\M^{1 \rm L}$ and $\M^{2 \rm L}$ the one-loop and two-loop amplitudes, we define
\beqn
\M^{1 \rm L}&=& \M^{1 \rm L}_0 + \cbs \M^{1 \rm L}_1\, , \label{M1L} \\
\M^{2 \rm L}&=&\sum_{i+2j\le 3} \cbs^i\cbe^j\M^{2 \rm L}_{ij}\, . \label{M2L}
\eeqn
The SM term $\M^{1 \rm L}_0$ receives contributions from both the one-loop triangle and box diagrams in fig.~\ref{fig:HH1L}. The relation between eqs.~\eqref{Dscs} and \eqref{Dsce} and the $\M_{i(j)}$ terms is: 
\begin{eqnarray}
    \Delta\sigma_{\bar c_6}&\propto& 2\Re\left[(\M^{1 \rm L}_0 + \cbs\M^{1 \rm L}_1)(\cbs^2\M^{2 \rm L}_{20} + \cbs^3\M^{2 \rm L}_{30})^*\right] \, , \label{DscsM}\\
    \Delta\sigma_{\bar c_8}&\propto&2\Re \left[(\M^{1 \rm L}_0 + \cbs\M^{1 \rm L}_1)(\cbe\M^{2 \rm L}_{01} + \cbs \cbe\M^{2 \rm L}_{11})^*\right]\, .  \label{DsceM}
\end{eqnarray}
In other words, $\Delta\sigma_{\bar c_6}$ and $\Delta\sigma_{\bar c_8}$ originate from the interference of $\M^{1 \rm L}$ with the terms with the largest dependence on $\cbs$, $(\cbs^2\M^{2 \rm L}_{20} + \cbs^3\M^{2 \rm L}_{30})$, and all the terms that depend on $\cbe$, $(\cbe\M^{2 \rm L}_{01} + \cbs \cbe\M^{2 \rm L}_{11})$. However, while the perturbative orders in $\Delta\sigma_{\bar c_8}$ and the interference terms emerging from the r.h.s.~of eq.~\eqref{DsceM} are in one-to-one correspondence, this is not true for $\Delta\sigma_{\bar c_6}$. The term $2\Re\left[(\M^{1 \rm L}_0 )(\M^{2 \rm L}_{20})^*\right]$ from the r.h.s.~of 
   eq.~\eqref{DscsM}, which gives rise to $\tilde \sigma_{20}$, multiplies the same $\cbs$ powers as the term $2\Re\left[(\M^{1 \rm L}_1 )(\M^{2 \rm L}_{10})^*\right]$, which we do not include in our computation. As already mentioned, we include the (formally subleading) term $\tilde \sigma_{20}$ because of the large cancellations among the triangle and box topologies at LO, and the fact that they contribute to different $\cbs$ powers; these cancellations are expected to be not substantially spoiled by NLO corrections. By keeping at the same level the entire $\M^{1 \rm L}$ amplitude of eq.~\eqref{M1L} in the interference leading to $  \Delta\sigma_{\bar c_6}$ we avoid that similar cancellations in NLO corrections are truncated by the $\cbs$ expansion. As already mentioned, this is relevant only for $\cbs\sim2$, where $\sigma$ has the minimum value, precisely due to the aforementioned cancellations. We remark, however, that this does not change the formal accuracy of our NLO corrections, which  is of order $\cbs^3$ and $\cbs^4$.
   
\medskip

The two-loop contributions entering the different $\M_{ij}$ sub-amplitudes can be further classified into three types:
\begin{itemize}
\item{Factorisable two-loop contributions ($\mathcal{F}$),}
\item{Non-factorisable two-loop contributions ($\mathcal{N}$), }
\item{Higgs wave-function counterterms ($\mathcal{W}$).}
\end{itemize} 
This classification is based on Feynman diagrams and can be easily understood from the topologies in fig.~\ref{fig:hh2l}. The first category $\mathcal{F}$ corresponds to the factorisable topologies $(d)$-$(j)$, together with the vertex counterterms in topologies $(k)$ and $(l)$.
 Their contributions are separately UV divergent, but their sum is finite, also for each  separate $\cbs^i\cbe^j$ order considered in this work. In particular, the topologies $(i)$, $(j)$ and $(l)$  can be evaluated together via the UV-finite $P[HH]$ form factor given in Ref.~\cite{Maltoni:2018ttu}, while all the remaining topologies from category $\mathcal{F}$ via the UV-finite $V[HHH]$ form factor given in the same reference. We remind the reader that topology $(d)$ is UV finite.

The non-factorisable two-loop contributions correspond to the topologies $(a)$-$(c)$ which are not available.  From a technical point of view, their computation is the most difficult and important part of this work. Details are given in sec.~\ref{sec:nonfac}. Moreover, we find that numerically their phenomenological impact is non-negligible w.r.t.~the factorisable ones.

We remind the reader that the Higgs wave-function renormalisation constant involves a quadratic dependence on $\ktre$ and therefore both a quadratic and linear dependence on $\cbs$ \cite{Gorbahn:2016uoy,Degrassi:2016wml}. Moreover, its contribution is UV-finite. Similarly to Ref.~\cite{Maltoni:2018ttu}, its contribution is not included in the $P[HH]$ and $V[HHH]$ form factors and has to be separately added. The third category $\mathcal{W}$ corresponds to these additional contributions, which can be easily calculated via the LO diagrams and the SM contribution of $\lt$ to the Higgs wave-function, namely,
		\begin{align}
			\delta Z_H^{\textrm{SM},\lambda_3}=-\frac{9\lambda \mh^2}{16\pi^2}B_0^{\prime}(\mh^2,\mh^2,\mh^2)\, ,
		\end{align} 
where $B_0^{\prime}(\mh^2,\mh^2,\mh^2)$ is the derivative of the $B_0(p^2,\mh^2,\mh^2)$ scalar integral evaluated at $p^2=\mh^2$ 

Based on the classifications we have just introduced, the different $\M_{ij}^{\rm 2L}$ terms can be further divided into
\begin{eqnarray}
	\M_{20}^{\rm 2L}&=&\M_{20}^{\mathcal{W}}+\M_{20}^{\mathcal{F}}+\M_{20}^{\mathcal{N}}\, , \nonumber\\
	\M_{30}^{\rm 2L}&=&\M_{30}^{\mathcal{W}}+\M_{30}^{\mathcal{F}}\, , \nonumber\\
	\M_{01}^{\rm 2L}&=&\phantom{\M_{30}^{\mathcal{W}}+}~\M_{01}^{\mathcal{F}}+\M_{01}^{\mathcal{N}}\, , \nonumber\\
	\M_{11}^{\rm 2L}&=&\phantom{\M_{30}^{\mathcal{W}}+}~\M_{11}^{\mathcal{F}}\, .
\end{eqnarray}

It will be useful to subdivide the amplitude further according to the spin exchanged in the $s$-channel.
In view of the description of the calculation of two-loop non-factorisable diagrams, it is important to note that only the topology ($a$) includes both a spin-0 and spin-2 component; all the other topologies in fig.~\ref{fig:hh2l} are solely spin-0. In the case of one-loop diagrams, the triangle is also solely spin-0, while the box includes both a spin-0 and spin-2 component. Thus, the spin-2 contribution of the box diagram interferes only with the spin-2 component of the topology ($a$), while the spin-0 part of the box diagram and the triangle diagram interfere with all the two-loop topologies. 

 Since the diagrams in the topology ($a$), which involves both spin-0 and spin-2 components, lead to contributions of order $\cbs^2$, we can further define
\begin{eqnarray}
	\M_{0,20}&=&\M_{0,20}^{\mathcal{W}}+\M_{0,20}^{\mathcal{F}}+\M_{0,20}^{\mathcal{N}}\, , \nonumber \\
	\M_{2,20}&=&\M_{2,20}^{\mathcal{W}}+\phantom{\M_{0,20}^{\mathcal{F}}}+\M_{2,20}^{\mathcal{N}}\,,
\end{eqnarray}
where the first lower index denotes the spin component. With this notation we can directly express the $\M_{0,20}^{\mathcal{W}}$, $\M_{2,20}^{\mathcal{W}}$ and also $\M_{30}^{\mathcal{F}}$ terms as
\begin{eqnarray}
			\M_{0,20}^{\mathcal{W}}&=&\delta Z_H^{\rm \textrm{SM},\lambda_3} (2\M_{0,1}^{\rm 1L}+\M_{0,0}^{\rm 1L})\, , \label{WF1}\\
			\M_{2,20}^{\mathcal{W}}&=&\delta Z_H^{\rm \textrm{SM},\lambda_3}\M_{2,0}^{\rm 1L} \, ,  \label{WF2} \\
		        \M_{30}^{\mathcal{W}}=\M_{0,30}^{\mathcal{W}}&=&\delta Z_H^{\rm \textrm{SM},\lambda_3}\M_{0,1}^{\rm 1L}\,. \label{WF3}
\end{eqnarray}
Therefore, thanks to eqs.~\eqref{WF1}-\eqref{WF3} and the formulae for the $P[HH]$ and $V[HHH]$ form factors provided in Ref.~\cite{Maltoni:2018ttu}, both the $\mathcal{F}$ and $\mathcal{W}$ contributions can be calculated. The only missing component in our calculation are the non-factorizable ($\mathcal{N}$) contributions, which are discussed in the next section.

\begin{figure}
\centering
 \includegraphics[scale=0.6]{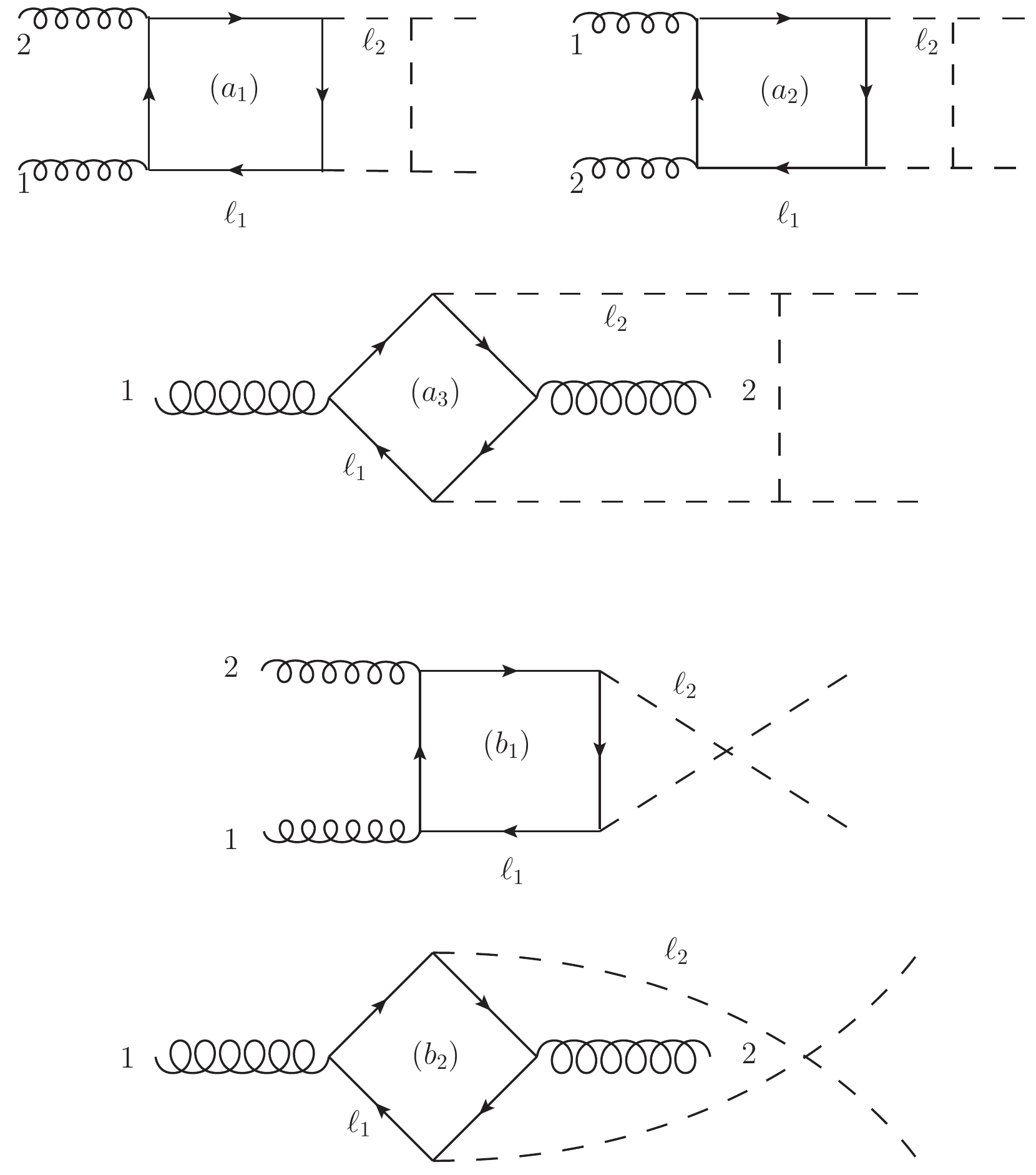}
 \caption{Non-factorizable two-loop diagrams of classes $(a)$ and $(b)$.}\label{fig:hh2l-dbx-bxtr}
\end{figure}

 \subsection{Two-loop non-factorisable terms}
 \label{sec:nonfac}
 \subsubsection{Reduction to form factors}
All the non-factorisable ($\mathcal{N}$) contributions originate from the topologies $(a)$, $(b)$ and $(c)$ in fig.~\ref{fig:hh2l}, these topologies can be further divided in sub-topologies; we show them for $(a)$ and $(b)$ in fig.~\ref{fig:hh2l-dbx-bxtr}, those for $(c)$ can be trivially obtained adding an $H$ propagator in $(b)$. In the topology $(a)$ (double-box) there are in total 6 diagrams
of which 3 are 
independent due to charge conjugation property of the fermion loop: 2 planar, ($a_1$) and ($a_2$), and 1 non-planar, ($a_2$). In the topology ($b$) (box-triangle) there are in total 3 diagrams, 2 of them are planar and charged conjugate, leading to $(b_1)$, the remaining diagram is instead non-planar, $(b_2)$. The case of $(c)$, is analogous to $(b)$, including an $H$ propagator.
 Topology $(a)$, as we already said,  contributes to both $\M_{0,20}^{\mathcal{N}}$ and $\M_{2,20}^{\mathcal{N}}$, while topology $(c)$ is in one-to-one correspondence with $\M_{10}^{\mathcal{N}}$. Topology $(b)$ contributes also to  $\M_{0,20}^{\mathcal{N}}$, which is therefore the only non-factorisable term receiving contributions from two different topologies. We can schematically summarise all this information as
\begin{eqnarray}
\M_{2,20}^{\mathcal{N}}, ~\M_{0,20}^{\mathcal{N}} \Longleftarrow \M_a &=& 2 (\M_{a_1} + \M_{a_2} + \M_{a_3})\label{eqn:amp-dbx} \, ,\\
 \M_{0,10}^{\mathcal{N}}=\M_{10}^{\mathcal{N}}\Longleftarrow\M_b &=& 2 \M_{b_1} + \M_{b_2}\label{eqn:amp-bxtr} \, , \\
\M_{0,20}^{\mathcal{N}} \Longleftarrow \M_c &=& \M_b \times \frac{6 v^2}{\lambda_4}  \frac{\lambda_3^2}{p_{12}^2-\mh^2}\, ,
\end{eqnarray}
where the $ \Longleftarrow$ arrow should be understood as ``contributes to'' and we have further remarked that $\M_{10}^{\mathcal{N}}$ is all spin-zero. It is important to note that the sums of diagrams in each topology $(a)$, $(b)$ and $(c)$ are separately finite and  gauge invariant.

\medskip 

The calculation of all the non-factorisable two-loop diagrams is performed via numerical methods. 
As a first step, two-loop diagrams are generated with {\sc\small  QGRAF} \cite{Nogueira:1991ex} and the amplitudes are written in {\sc \small FORM}~\cite{Vermaseren:2000nd} in $d=4-2\epsilon$ dimensions. Then, the amplitudes are projected onto spin-0 and spin-2 form factors.\footnote{ In this work, this projection has been used also for the evaluation of $\mathcal{F}$ and $\mathcal{W}$ contributions.} Assigning the following on-shell $p_i$ momenta to the external particles,
 \begin{equation}
  g(p_1) + g(p_2) \to H(p_3) + H(p_4)\, ,
 \end{equation}
 where all the $p_i$ are considered as incoming, 
 both $\M^{\rm 1L}$ and $\M^{\rm 2L}$, and any of their  gauge-invariant sub-amplitudes,  can be projected onto two spin-0 and spin-2 bases ${\cal A}_1$ and ${\cal A}_2$~\cite{Degrassi:2016vss,Borowka:2016ypz}, and expressed via corresponding form factors denoted as $F_0$ and $F_2$. Specifically,
 \begin{eqnarray}\label{eqn:Mmunu}
	 \mathcal{M}^{\mu_1\mu_2} \epsilon_{1,\mu_1}\epsilon_{2,\mu_2}= \delta^{c_1c_2}\mathcal{A}_0^{\mu_1\mu_2} \epsilon_{1,\mu_1} \epsilon_{2,\mu_2}F_0 + \delta^{c_1c_2}{\cal A}_2^{\mu_1\mu_2} \epsilon_{1,\mu_1}\epsilon_{2,\mu_2}F_2\, .
 \end{eqnarray}
 In eq.~\eqref{eqn:Mmunu}  $\epsilon_{1}$ and $\epsilon_{2}$ are the (transverse) polarisation vectors for the two incoming on-shell gluons, while $\mu_1$ and $\mu_2$ ($c_1$ and $c_2$) are their corresponding Lorentz(colour) indices.
 The tensor bases $\mathcal{A}_{0}^{\mu_1\mu_2}$ and $\mathcal{A}_{2}^{\mu_1\mu_2}$ can be arbitrarily chosen and we decided to use the orthonormal ones, which satisfy the relation\footnote{The inner product stands for the contraction with polarisation vectors and summation over all physical polarisations.}
 \begin{align}
	\mathcal{A}_0\cdot \mathcal{A}_0=\mathcal{A}_2\cdot \mathcal{A}_2=1,\mathcal{A}_0\cdot \mathcal{A}_2=0\, .\label{eq:orth-proj}
\end{align}
 The two tensor bases $\mathcal{A}_{0}^{\mu_1\mu_2}$ and $\mathcal{A}_{2}^{\mu_1\mu_2}$ in $d$-dimensions read \footnote{The expression for the second 
 projector in Ref.~\cite{Degrassi:2016vss} contains a typo that is corrected here.} 
 \begin{align}
 {\cal A}_0^{\mu_1\mu_2} =& \frac{1}{\sqrt{d-2}} \left( g^{\mu_1\mu_2} - \frac{p_1^{\mu_2} p_2^{\mu_1}}{p_1\cdot p_2} \right  )\, ,\\
 {\cal A}_2^{\mu_1\mu_2} =& \frac{1}{2}\sqrt{\frac{d-2}{d-3}} \Big( 
 - \frac{d-4}{d-2} \Big[ g^{\mu_1\mu_2} - \frac{p_1^{\mu_2} p_2^{\mu_1}}{p_1\cdot p_2} \Big]+ g^{\mu_1\mu_2} \nn \\
  +& \frac{(p_3\cdot p_3)p_1^{\mu_2}p_2^{\mu_1} 
 + (2 p_1\cdot p_2) p_3^{\mu_1}p_3^{\mu_2}
 - (2 p_1\cdot p_3) p_2^{\mu_1}p_3^{\mu_2}
 - (2 p_2\cdot p_3) p_3^{\mu_1}p_1^{\mu_2}
      }{p_T^2 (p_1\cdot p_2)} \Big)\, ,
	  \label{eq:proj}
 \end{align}
where $p_T^2 = (s_{13}s_{23}-\mh^4)/s_{12}$ denotes the square of the Higgs-boson transverse momentum w.r.t.~the gluons in the center-of-mass rest frame, using the convention $s_{ij}=(p_i+p_j)^2$.

After the above projection, we obtain the spin-dependent non-factorisable amplitudes $\M_{0,20}^{\mathcal{N}}$, $\M_{2,20}^{\mathcal{N}}$  and $\M_{0,01}^{\mathcal{N}}$ written in terms of form factors, {\it i.e.},  $F_{0,20}^{\mathcal{N}}$, $F_{2,20}^{\mathcal{N}}$ and $F_{0,01}^{\mathcal{N}}$ where
\beq
F_{0,20}^{\mathcal{N}}=F_{0,a}+F_{0,c},~~~ F_{2,20}^{\mathcal{N}}=F_{2,a},~~~  {\rm and} ~~~F_{0,01}^{\mathcal{N}}=F_{0,b}\, .\label{ffactors}
\eeq
In other words, all the the $\mathcal{N}$ contributions entering our calculation can be expressed via the $F_{0,20}^{\mathcal{N}}$, $F_{2,20}^{\mathcal{N}}$ and $F_{0,01}^{\mathcal{N}}$ form factors, which in turn depend on the non-vanishing spin-0 and spin-2 projections of the $(a)$-$(c)$ topologies, $F_{0,a}$, $F_{0,b}$ $F_{0,c}$ and $F_{2,a}$.

\subsubsection{Numerical evaluation of the form factors}
 The form factors $F_{0,a}$, $F_{0,b}$ $F_{0,c}$ and $F_{2,a}$ are computed with {\sc\small pySecDec} 
\cite{Borowka:2017idc,Borowka:2017esm}, a toolbox for the 
numerical evaluation of multi-loop integrals. We remind the reader that  {\sc\small pySecDec}  can readily compute 
loop integrals with massive internal lines and/or off-shell legs. Moreover, compared to 
its predecessor \textsc{SecDec 3} \cite{Borowka:2015mxa}, it facilitates  the creation of integral 
libraries, allowing for a direct incorporation of the code into the 
calculation of the full amplitude. 

Before using  {\sc\small pySecDec}, we simplify the numerators of the the loop integrals in the form factors, in order to obtain tensor integrals that optimise the speed of the computation. It is important to note that the form factors $F_{0,a}$ and $F_{2,a}$ involve 7-propagator diagrams while $F_{0,b}$ and $F_{0,c}$ 6-propagator ones.
Using propagator identities in {\sc \small FORM}, 
we obtain  a total of 11 integral expressions for $F_{0,a}$,  24 for $F_{2,a}$ and  9 for $F_{0,b}$ and $F_{0,c}$. The corresponding topologies are depicted in Appendix \ref{app:topo}.
For simplicity, the overall 
coupling factors, colour factors ($\delta^{a_1a_2}/2$) and factor of (-1) due to fermion loop are removed from the tensor integrals.
In particular, the quantities directly calculated via {\sc\small pySecDec} are $\tilde F_{0,a}$, $\tilde F_{2,a}$, $\tilde F_{0,b}$ and $\tilde F_{0,c}$, where
we define
\begin{eqnarray}
 F_{0,a}&=& g_s^2 \frac{m_t^2}{v^2} (6\lambda v)^2 \Big(\frac{i}{16\pi^2}\Big)^2 \frac{\delta^{a_1a_2}}{2} \tilde F_{0,a}\, , \label{eq:tilderule} \\ 
  F_{0,b}&= &g_s^2 \frac{m_t^2}{v^2} (6\lambda) \Big(\frac{i}{16\pi^2}\Big)^2 \frac{\delta^{a_1a_2}}{2} \tilde F_{0,b}\, . \label{eq:tilderule2}
\end{eqnarray}
$\tilde F_{2,a} $ is related to $ F_{2,a}$ like in eq.~\eqref{eq:tilderule} and $\tilde F_{0,c}$ is related to $ F_{0,c}$ like in eq.~\eqref{eq:tilderule2}.

In order to improve on the speed and convergence of the numerical evaluation, further measures are taken. First, only the finite parts are evaluated. To do this 
correctly, the integrands generated by {\sc\small pySecDec} are multiplied with their prefactors, containing $\mathcal{O}(\epsilon)$ terms, before the integration. Nevertheless, we have cross-checked for specific phase-space points  that UV- divergencies cancel for each diagram, although individual integral expressions can be separately UV-divergent.

Second, all integrals with the same denominator structure are added together before numerical 
integration. We have checked that the summation of several denominator structures prior to numerical integration does not lead to a faster convergence.

Third, different integrators were chosen 
for different integrals. A deterministic integrator like {\sc\small Cuhre} \cite{Berntsen:1991:ADA}, which is part of the
 {\sc\small Cuba} library \cite{Hahn:2004fe} and linked to \textsc{pySecDec}, is generally very fast and accurate for 
integrals with up to 5 integral dimensions. Beyond 5 dimensions, the integrator {\sc\small Vegas} \cite{Lepage:1977sw} is chosen. 
Furthermore, both {\sc\small Vegas} and {\sc\small Cuhre} give a $\chi^2$ estimate, stating the probability that 
the uncertainty associated to the result is accurate. Tests have repeatedly shown that the 
{\sc\small Cuhre} results can be trusted only if $\chi^2$ is well below 1. Therefore, a routine was 
included to reperform the numerical integration with the more adaptive but generally slower 
integrator  {\sc\small Vegas} when the $\chi^2$ value is too high. 
With this procedure we minimise cancellations and make sure that our numerical result is stable.

We have already mentioned that the UV finiteness of the form factors has been explicitly verified. Further tests have also been performed in order to ensure the correctness of the calculation. 
We have cross-checked the large $\mt$ limits for the $(b)$ and $(c)$ topologies (box-triangle) against analytical results. By setting $s_{12}=\mh^2$ we have found perfect agreement with the expression given in Ref.~\cite{Degrassi:2016wml}. Also, for the $(a)$ topology (double-box), we have numerically tested that by artificially setting to $m_X$ the mass in the Higgs propagator connecting the two final-state Higgs,  denoting the amplitude as $\M_{a,X}$, we obtain $\M_{a,X}\rightarrow \M_b [ -2 \lambda_3^2/(\lambda_4 m_X^2)]$ in the limit $m_X\rightarrow\infty$. In other words, by integrating out the heavy state $X$, the $(a)$ topology reduces to the $(b)$ topology where as expected the quartic coupling is an effective coupling  $\lambda_4 = -2 \lambda_3^2/m_X^2$. The factor of 2 originates from the number of diagrams
contributing to the double-box amplitude, which is twice the number of diagrams contributing to
the box-triangle amplitude.

\subsubsection{Grids for phase-space integration}
\label{sec:interpolation}

Up to this point we have discussed the strategy used for the evaluation of non-factorisable terms for a given phase space point. However, in this work we are interested in phenomenological predictions at colliders. Thus, the partonic squared matrix-elements have to be integrated over the phase-space and convoluted with parton-distribution-functions (PDFs). To this purpose, given the limited speed in the evaluation of the non-factorisable factors, it is helpful to build a grid that can be interpolated and quickly integrated over the relevant phase-space. In the following we explain  how we have generated these grids, which have then been used together with an in-house Montecarlo for obtaining the phenomenological results of Sec.~\ref{sec:constraints}. 

Let us start by discussing the spin-0 component at two loops.
The box-triangle diagrams, topologies $(b)$ and $(c)$, depend on only one kinematic
variable $s_{12}$, hence a one-dimensional grid  for $F_{0,b}$ and  $F_{0,c}$
is sufficient and with enough sampled values of $s_{12}$ a linear interpolation can be used. On the contrary, the double box diagrams, topology (a), depend on both $s_{12}$ and the angle $\theta$ between $p_1$ and $p_3$. However,  the dependence of  $F_{0,a}$ on $\theta$ is actually small  and it can be approximated by the first few terms in the 
partial wave expansion \cite{Jacob:1959at} of $F_{0,a}$ as
         
\begin{align}
	{\tilde F}_{0,a}(s,\theta) = \sum_{i=0}^{\infty}a_{i}^{\prime}(s)d_{0,0}^{i}(\theta)=\sum_{i=0}^{\infty}a_{i}(s)P_{i}(\cos\theta)\,.\label{eq:dbx-exp}
\end{align}
We truncate the expansion in order to approximate the full results. We find that the $\theta$ dependence is weak, especially for $s_{12}<4\mh^2$, {\it i.e.}, below the top-pair threshold in the loops. In this phase-space region the top-quark loop can be integrated out, obtaining an effective $HHgg$ coupling among the Higgs bosons and the gluons.
 With such an EFT description in the $\mt \to \infty$ limit there is no $\theta$ dependence.
Thus, the dominant contribution originates from the term without $\theta$ dependence, namely, the $a_0(s)$ term.
In order to have the $\theta$ dependence under control and to test the validity of the partial wave expansion,
we do not only include the first term but also the second term,\footnote{Since $gg\to HH$ is by definition symmetric, the $a_i(s)$ coefficients are  zero for odd values of $i$.}
\begin{align}
	{\tilde F}_{0,a}(s,\theta) \approx a_0(s)+a_2(s)P_2(\cos\theta)\label{eq:dbx-reg}\, .
\end{align}
 \begin{figure}
 \includegraphics[width=0.45\textwidth]{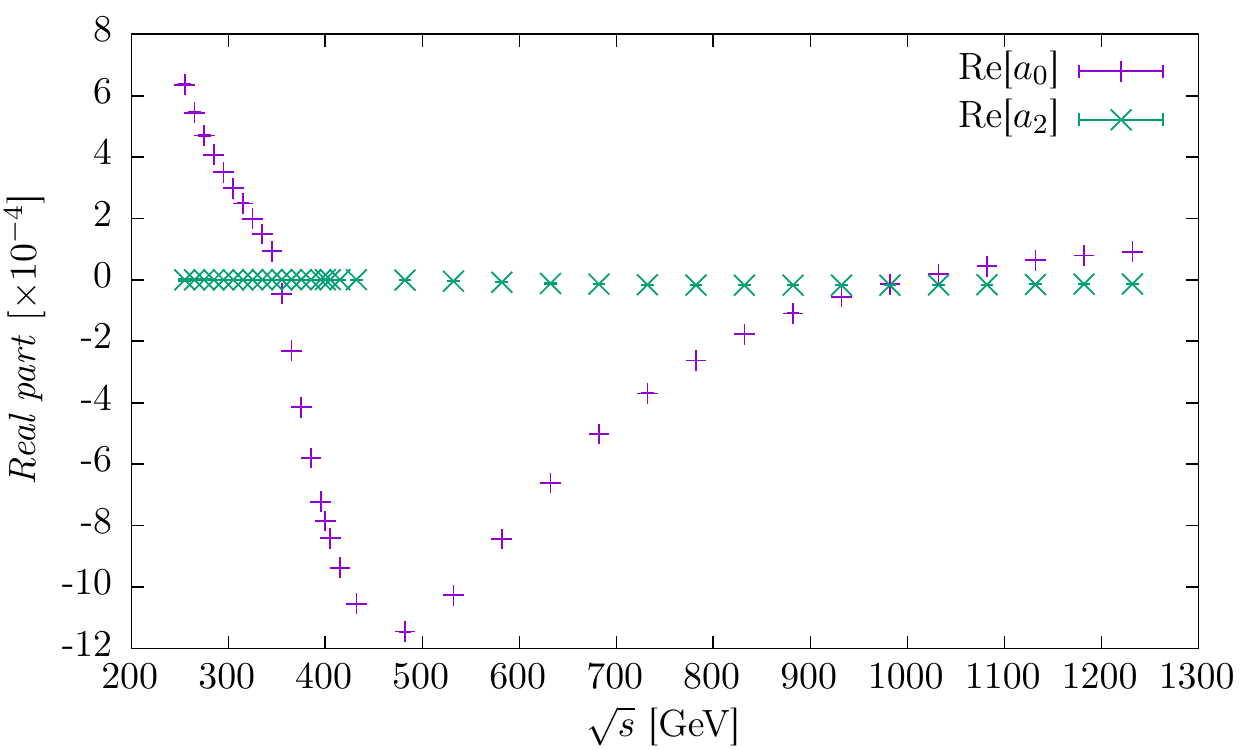}
\includegraphics[width=0.45\textwidth]{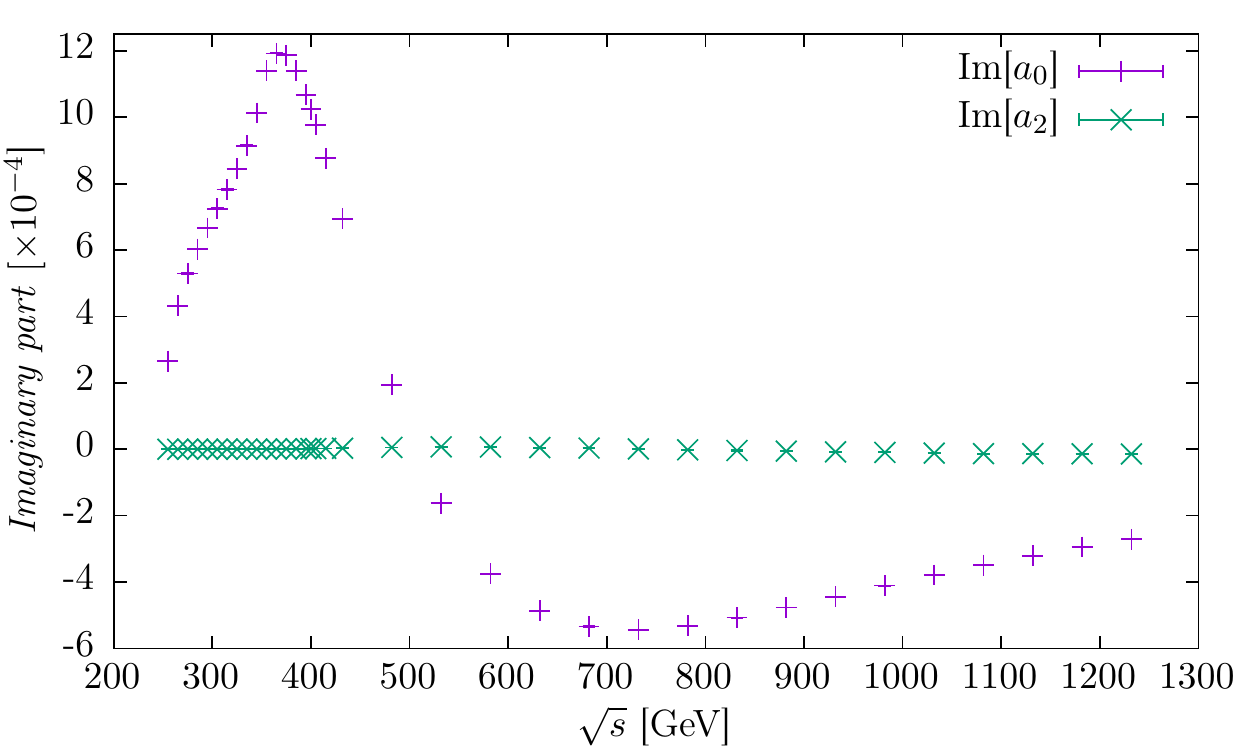}
 \caption{Fit results:  real (left) and imaginary (right) parts of $a_0(s)$ and $a_2(s)$}\label{fig:dbox-a0a2}
 \end{figure}

         For each value of $s$, different values of $\theta$ have been sampled in order to perform a linear regression of $a_0(s)$ and $a_2(s)$. Afterwards, a linear interpolation is separately performed on both the values of $a_0(s)$ and $a_2(s)$. 		 
		 The validity of the truncation of the partial-wave expansion at $a_2(s)$ has also been investigated. First of all, we found that both the real and imaginary parts of $a_2(s)$ are substantially smaller than those of $a_0(s)$, as can be seen in fig.~\ref{fig:dbox-a0a2}. Thus, contributions from higher-order $a_i(s)$ terms are expected to be even smaller than $a_2(s)$. Moreover we have estimated their contribution by comparing the value obtained with the approximation in eq.~\eqref{eq:dbx-reg} after the regression and the actual value obtained. We can conclude that the truncation uncertainty is at the $\mathcal{O}(1\%)$ level.

Let us conclude this section by commenting on the spin-2 contribution  $F_{2,a}$. Although there is a large dependence on $\theta$, we have verified that its contribution is strongly suppressed w.r.t.~the spin-0 contribution. For this reason we safely ignore this contribution in our phenomenological study of Sec.~\ref{sec:constraints}.

\section{Numerical Results}
\label{sec:num}

In this section we discuss the numerical results obtained for the $m(HH)$ distribution and the total rates at different collider energies. The phenomenological analyses of Sec.~\ref{sec:constraints} are based on these results.

In our calculation, we have used the following input parameters for the masses of the heavy SM particles,
\begin{equation}
m_t = 173.2 \text{ GeV}\, , \quad m_W = 80.385 \text{ GeV} \, , \quad m_Z = 91.1876 \text{ GeV}\, , \quad m_H = 125.09 \text{ GeV} \,,
\end{equation}
whereas all the other masses are set equal to zero. Similarly to Ref.~\cite{Maltoni:2018ttu}, we renormalise $\alpha$ in the $G_\mu$-scheme and we use as input parameter
\begin{equation}
G_\mu = 1.1663787 \cdot 10^{-5} \text{ GeV}^{-2} \,.
\end{equation}
The renormalisation scale for $\alpha_s$ and factorisation scale are set to be $\mu_R=\mu_F=\frac{1}{2}m(HH)=\frac{1}{2}\sqrt{\hat s}$, and we have used the Parton-Distribution-Functions (PDF) set {\sc\small CT14LO} \cite{Dulat:2015mca}. 
We remind the reader that in our calculation we renormalise $\cbs$ in the $\MSbar$ scheme and we set the renormalisation scale to  $\mu_{\textrm{EFT}}=2m_H$. Moreover, we assume both the Wilson coefficients $\cbs$ and $\cbe$  at the scale $\mu_{\textrm{EFT}}$.

\begin{table}
\footnotesize
    \centering
    \begin{tabular}{|c||c|c|c|c|c|c|c|c|c|}
        \hline
      $\sqrt{s}$ [TeV] & $\sigma_{0}$ [fb] & $\sigma_{1}$ [fb] & $\sigma_{2}$ [fb]  \\
        \hline
        \hline
       14 & 19.49 & -15.59 & 5.414  \\
        & - & (-80.0\%)& (27.8\%) \\
        \hline        
       27 & 78.30 & -59.39 & 19.58  \\
               & - & (-75.8\%)& (25.0\%) \\
       \hline
       100 & 790.8  & -556.8 & 170.8 \\
                      & - & (-70.5\%)& (21.6\%) \\
        \hline
    \end{tabular}
	\caption{LO contributions to $ \sigma^{\rm pheno}_{\rm NLO} $. We show for every entry the ratio with $\sigma_0$ at the same energy.\label{tab:totxsLO}
   }
\end{table}

\begin{table}
\footnotesize
    \centering
    \begin{tabular}{|c||c|c|c||c|c|c|c|c|c|}
        \hline
      $\sqrt{s}$ [TeV] &  $\tilde \sigma_{20}$ [fb] & $\sigma_{30}$ [fb] & $\sigma_{40}$ [fb] & $\sigma_{01}$ [fb] & $\sigma_{11}$ [fb] & $\sigma_{21}$ [fb] \\
        \hline
        \hline
       14 &   0.7112 & -0.5427 & 0.0620 & 0.3514 & -0.0464 & -0.1433 \\
             & (3.6\%) &  (-2.8\%)& (0.3\%) & (1.8\%)&  (-0.2\%)&  (-0.7\%)\\
        \hline        
       27 &  2.673 & -1.936 & 0.2102 & 1.3552 & -0.137 &  -0.5127 \\
             & (3.4\%) &  (-2.5\%)& (0.3\%) & (1.7\%)&  (-0.2\%)&  (-0.7\%)\\       
       \hline
       100 &  24.55 & -16.53  & 1.663  & 12.932 & -0.88 & -4.411 \\
                    & (3.1\%) &  (-2.1\%)& (0.2\%) & (1.6\%)&  (-0.1\%)&  (-0.6\%)\\
        \hline
    \end{tabular}
	\caption{Two-loop contributions to $ \sigma^{\rm pheno}_{\rm NLO} $. We show for every entry the ratio with $\sigma_0$ at the same energy.\label{tab:totxsNLO}
     }
\end{table}

In table \ref{tab:totxsLO}, we list the three different $\sigma_i$ contributions entering the LO part of $\sigma^{\rm pheno}_{\rm NLO}$ at 14, 27 and 100 TeV proton--proton collisions.  Similarly, in table \ref{tab:totxsNLO} we list all the two-loop $\sigma_{ij}$ contributions entering $\sigma^{\rm pheno}_{\rm NLO}$. We display in parentheses also their ratio with the LO prediction in the SM, $\sigma_0=\sigma_{\rm LO}^{\rm SM}$.  As can be seen in tables \ref{tab:totxsLO} and \ref{tab:totxsNLO}, cross sections considerably grow with the energy, while all the contributions induced by $\cbs$ and $\cbe$ mildly decrease in comparison with $\sigma_{\rm LO}^{\rm SM}$. Indeed, at large energies, the one-loop box diagrams is dominant w.r.t.~the one with a triangle, which is the only one leading to $\cbs$ contributions at LO and to $\cbe$ contributions via loop corrections.

\begin{figure}
 \centering
 \includegraphics[scale=0.5]{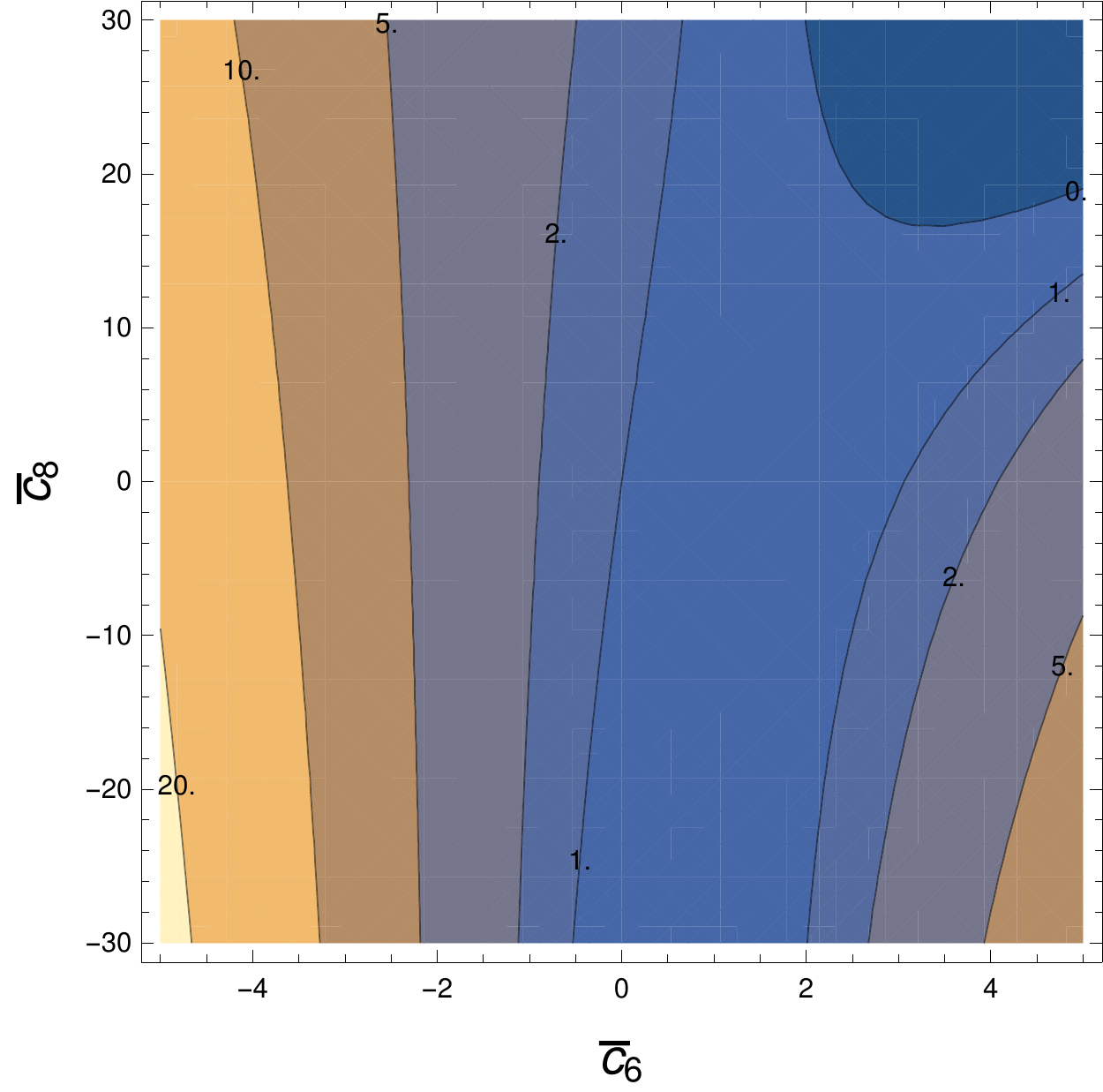}
 \hspace{0.5cm}
 \includegraphics[scale=0.5]{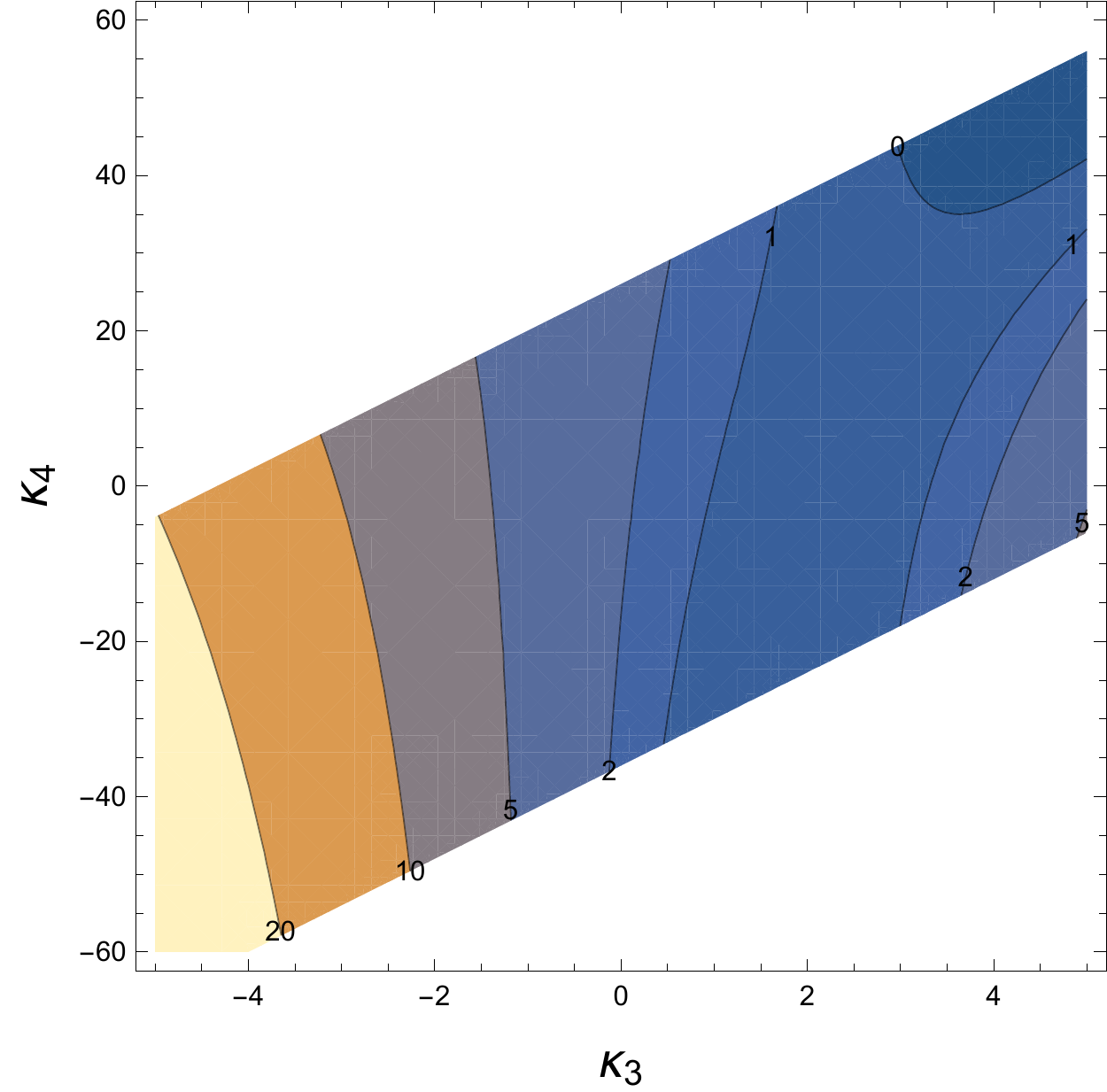}
 \includegraphics[scale=0.5]{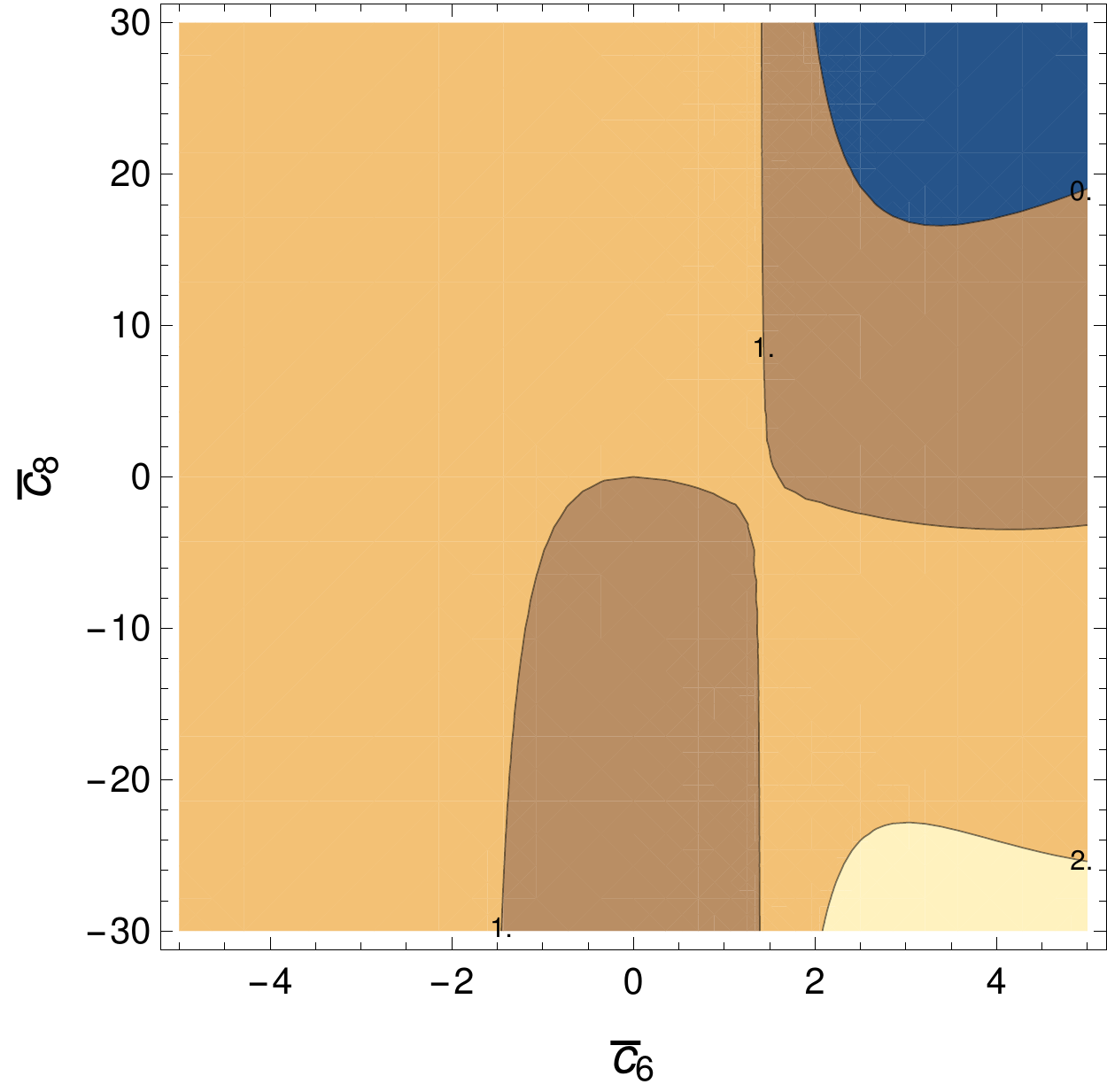}
 \hspace{0.5cm}
 \includegraphics[scale=0.5]{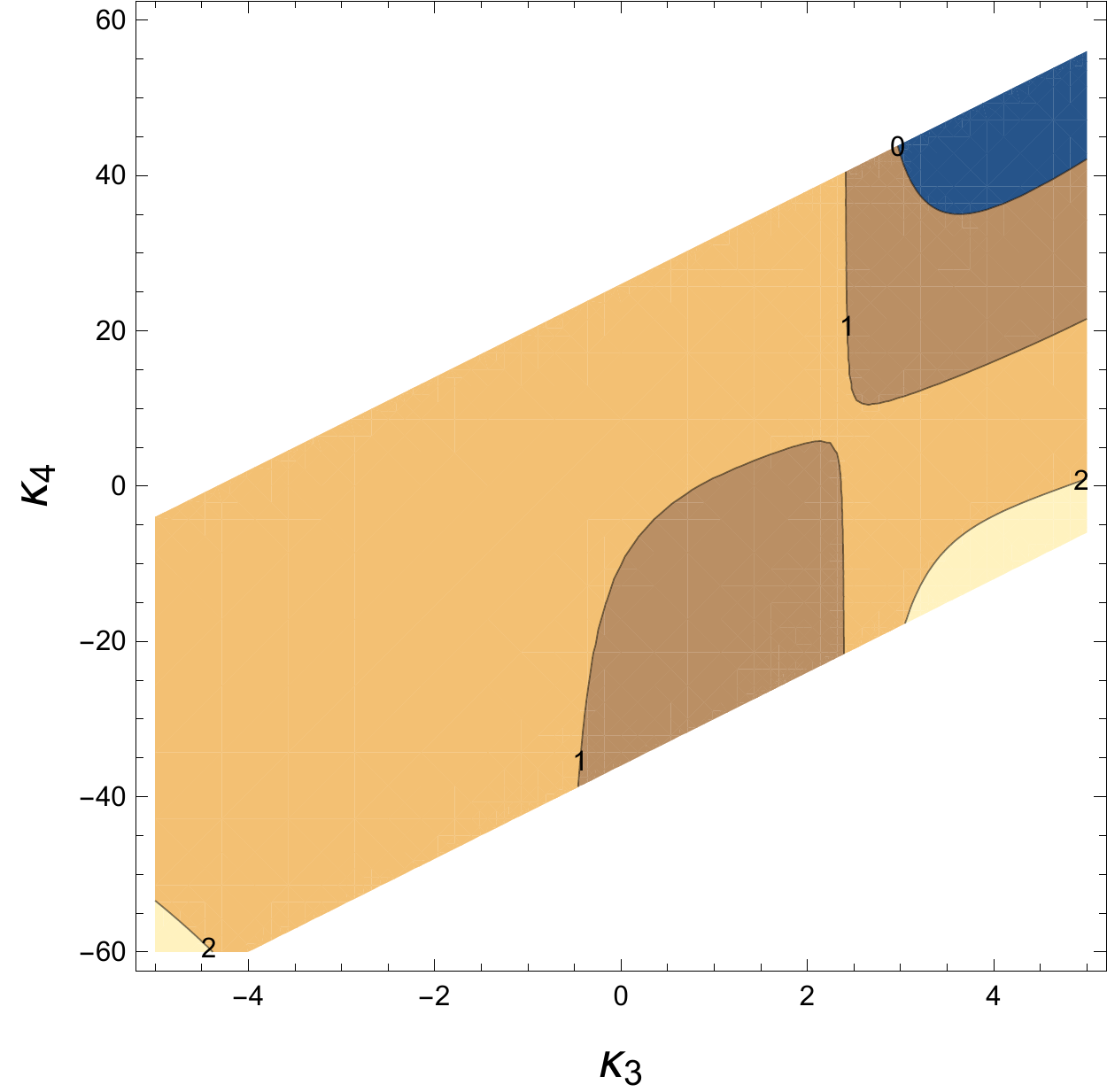}

 \caption{Contour plots at 14 TeV for $\sigma^{\rm pheno}_{\rm NLO}/\sigma^{\rm SM}_{\rm LO}$ (top) and $\sigma^{\rm pheno}_{\rm NLO}/\sigma_{\rm LO}$ (bottom). Left plots show results in the ($\cbs$,$\cbe$) plane, while right plots in the ($\ktre,\kqual$) plane. \label{fig:contour}}
\end{figure}

\begin{figure}[h!]
\centering
 \includegraphics[scale=0.7]{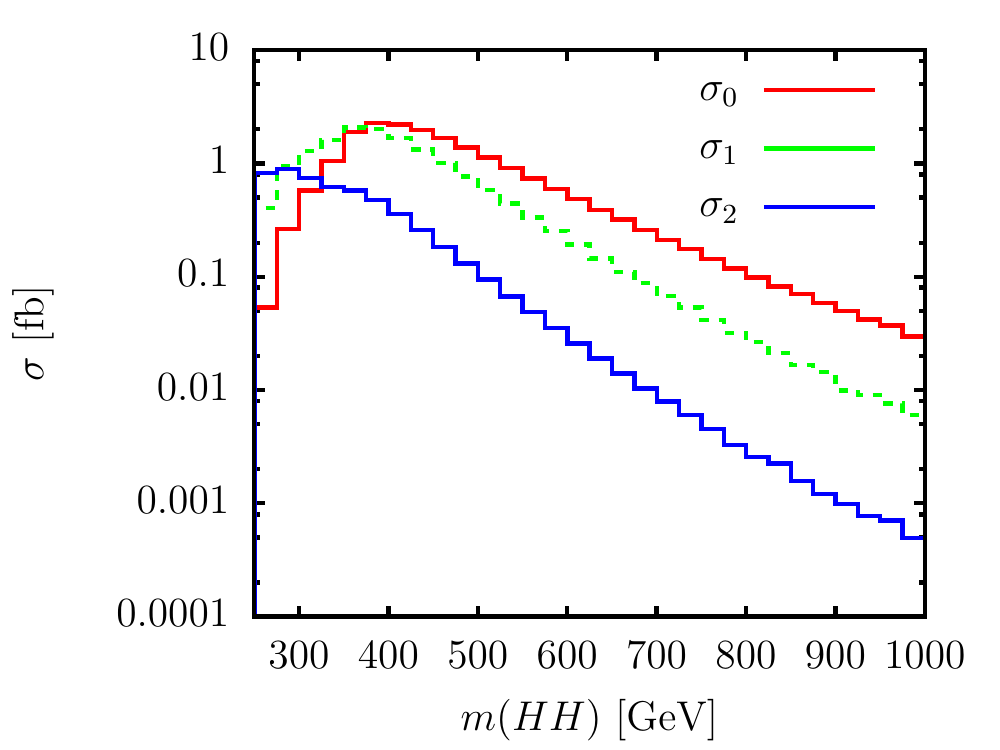}
 \\
  \includegraphics[scale=0.7]{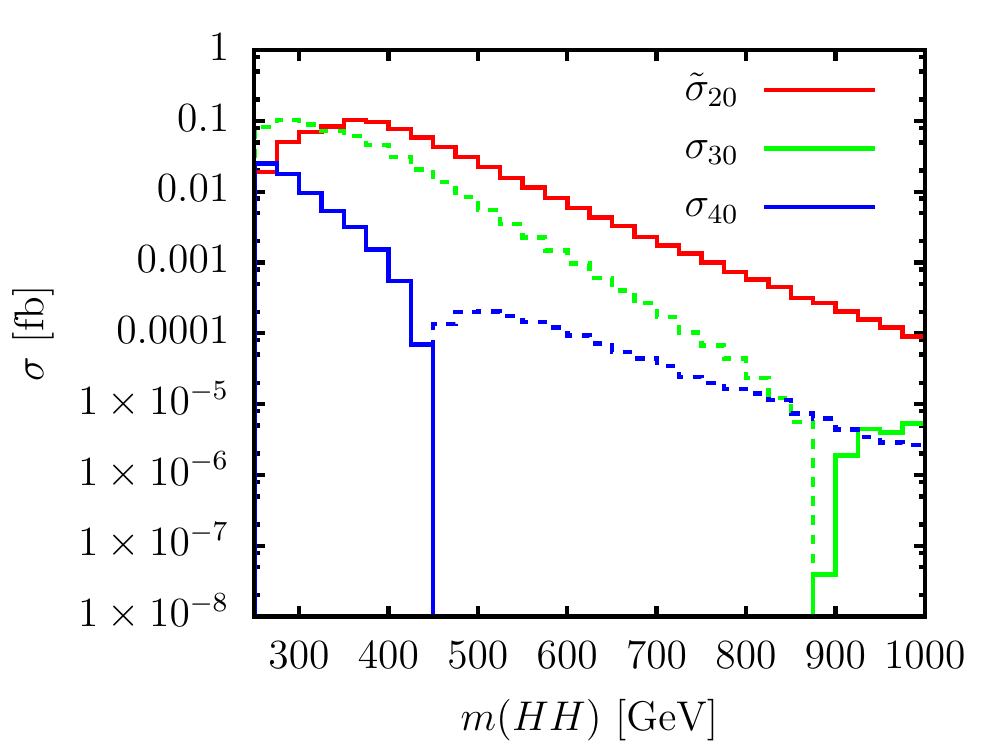}
   \includegraphics[scale=0.7]{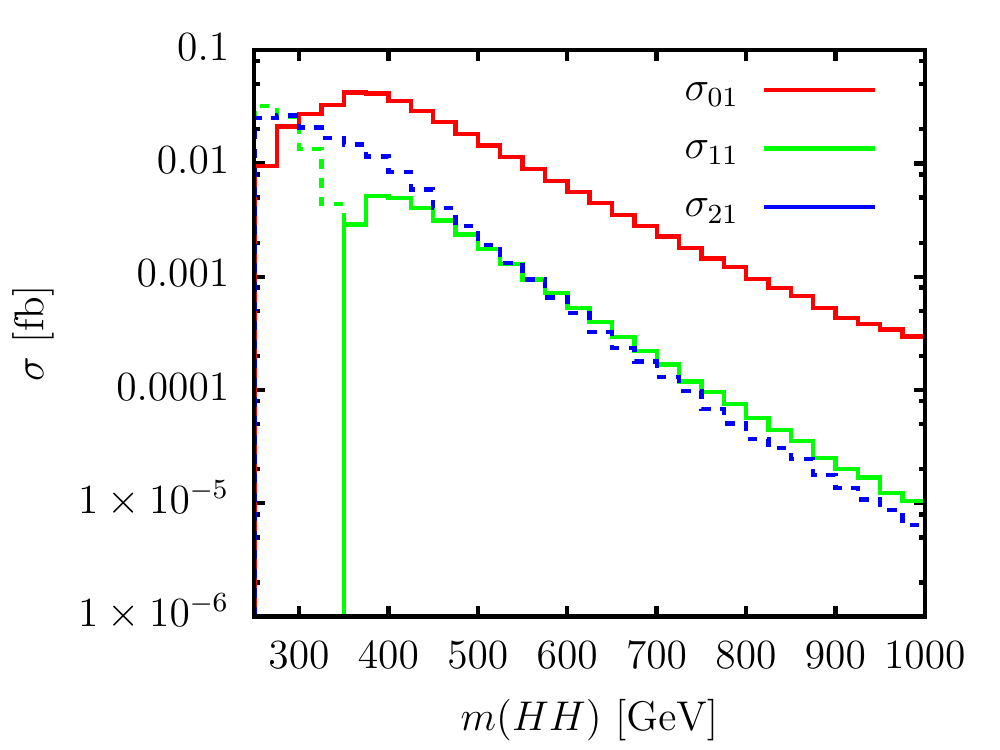}
 \caption{Individual $\sigma_{i(j)}$ contributions at 14 TeV as function of $m(HH)$. The upper plot display contributions to $\sigma_{\rm LO}$, while the lower plots those to $ \Delta\sigma_{\bar c_6}$ (left) and to  $\Delta\sigma_{\bar c_8}$ (right). \label{fig:logscale}}
\end{figure}

In fig.~\ref{fig:contour} we show four different contour plots for the 14 TeV energy. The upper plots show the ratio $\sigma^{\rm pheno}_{\rm NLO}/\sigma^{\rm SM}_{\rm LO}$, {\it i.e.}, the ratio between our phenomenological prediction and the SM one, while the lower plots show the ratio  $\sigma^{\rm pheno}_{\rm NLO}/\sigma_{\rm LO}$, which corresponds to the $K$-factor from two-loop corrections in our calculations. The left plots display these ratios in the $(\cbs,\cbe)$ plane, while the right plots in the $(\ktre,\kqual)$ one. In the  plots we consider the perturbativity regime $|\cbs|<5$ and $|\cbe|<31$, which leads to values of $|\kqual|$ up to $\sim$ 60.  The upper plots show that large values of $\kt$ can considerably enhance the value of the total cross section. For $\cbs<0$ there is only a small dependence on $\cbe$, while for  $\cbs>0$ the dependence is sizable, and it even leads to negative cross sections for both large and positive $\cbs$ and $\cbe$. These effects are induced by the loop corrections; the LO predictions cannot be negative since they originate from a squared amplitude. It can be seen also in the lower plots where the contour line for $\sigma^{\rm pheno}_{\rm NLO}/\sigma_{\rm LO}=0$ is the same  of $\sigma^{\rm pheno}_{\rm NLO}/\sigma^{\rm SM}_{\rm LO}=0$ in the upper plots. For negative values our prediction is unphysical, so it cannot be used for phenomenological studies. This is caused by the sum of $\cbs$ and $\cbe$ two-loop effects, which is large in absolute value. For the same reason also a region with $\sigma^{\rm pheno}_{\rm NLO}/\sigma_{\rm LO}>2$ is present for large and positive(negative) $\cbs(\cbe)$. However, we do not exclude it since it is simply denoting a large one-loop $K$-factor. 

\begin{figure}[t]
\centering
 \includegraphics[scale=0.5]{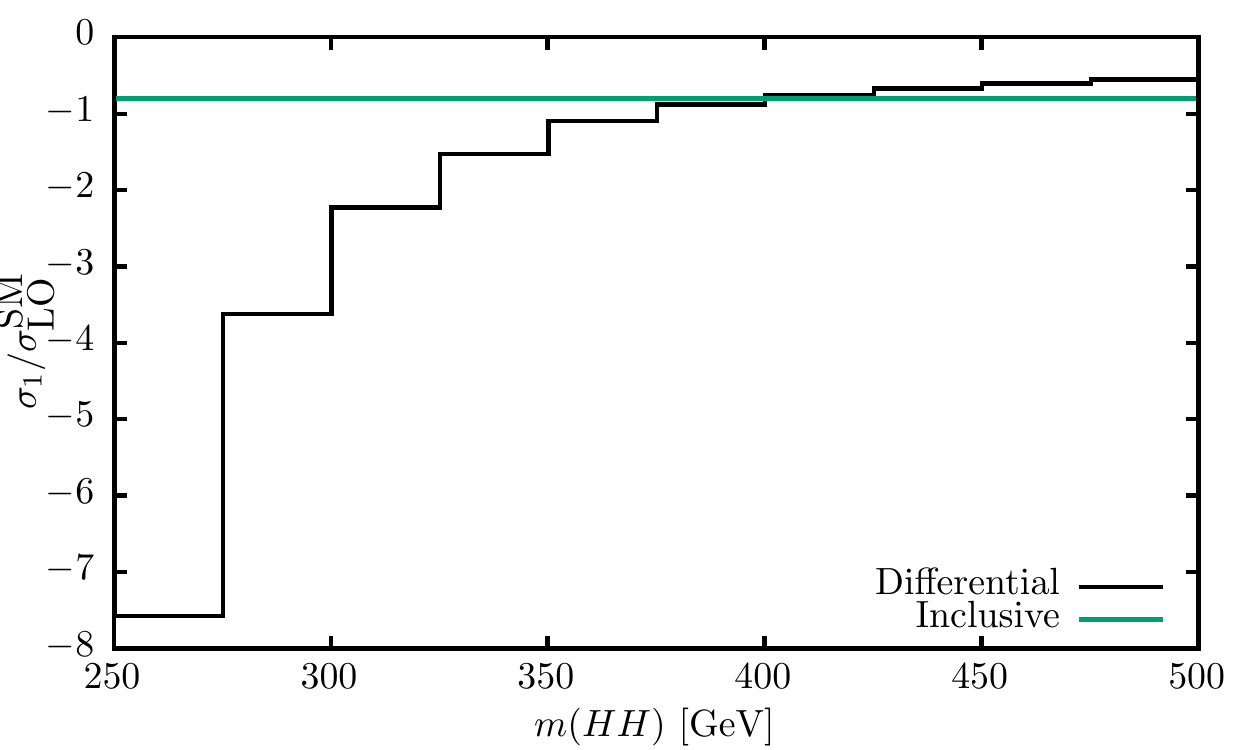}
 \includegraphics[scale=0.5]{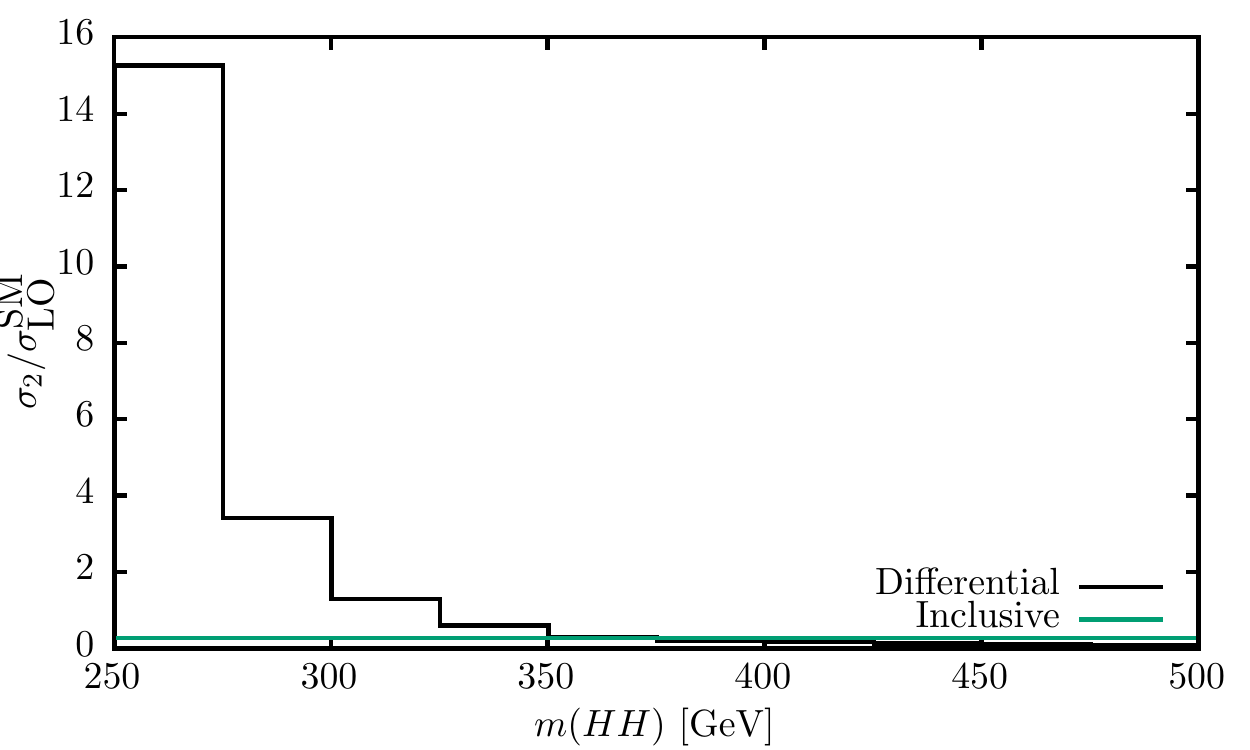}
  \includegraphics[scale=0.5]{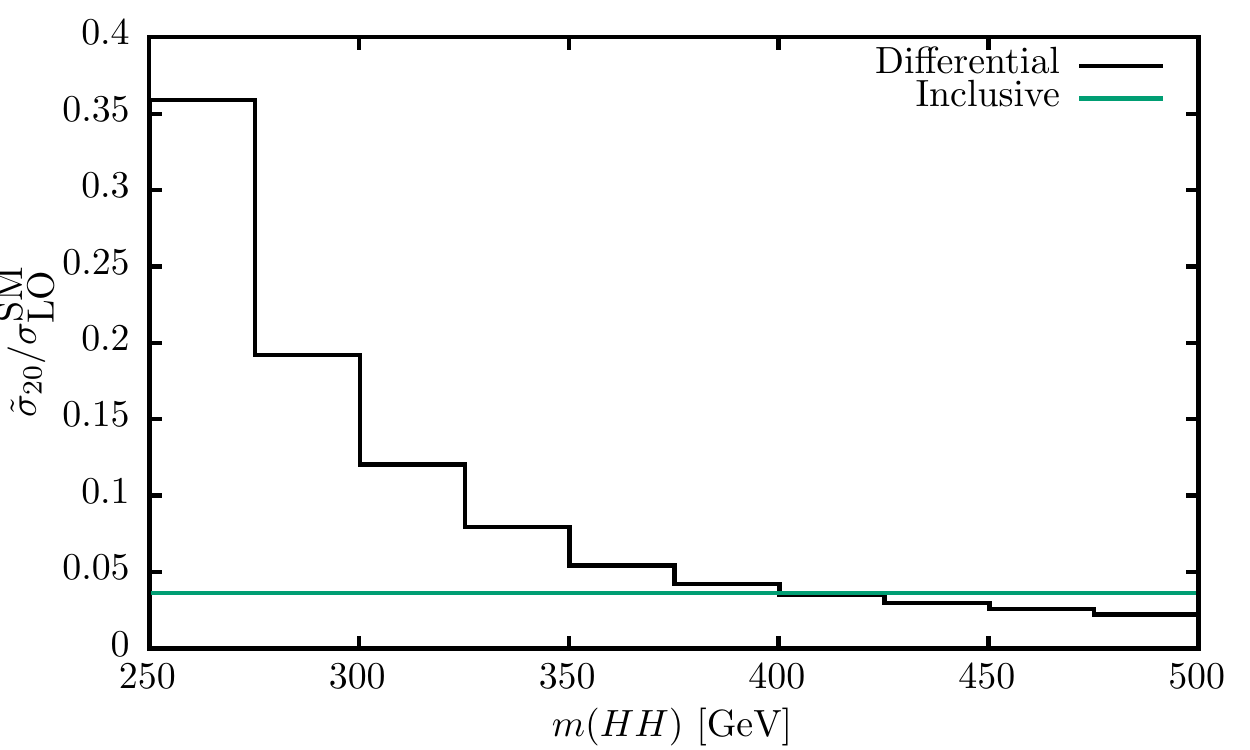}
 \includegraphics[scale=0.5]{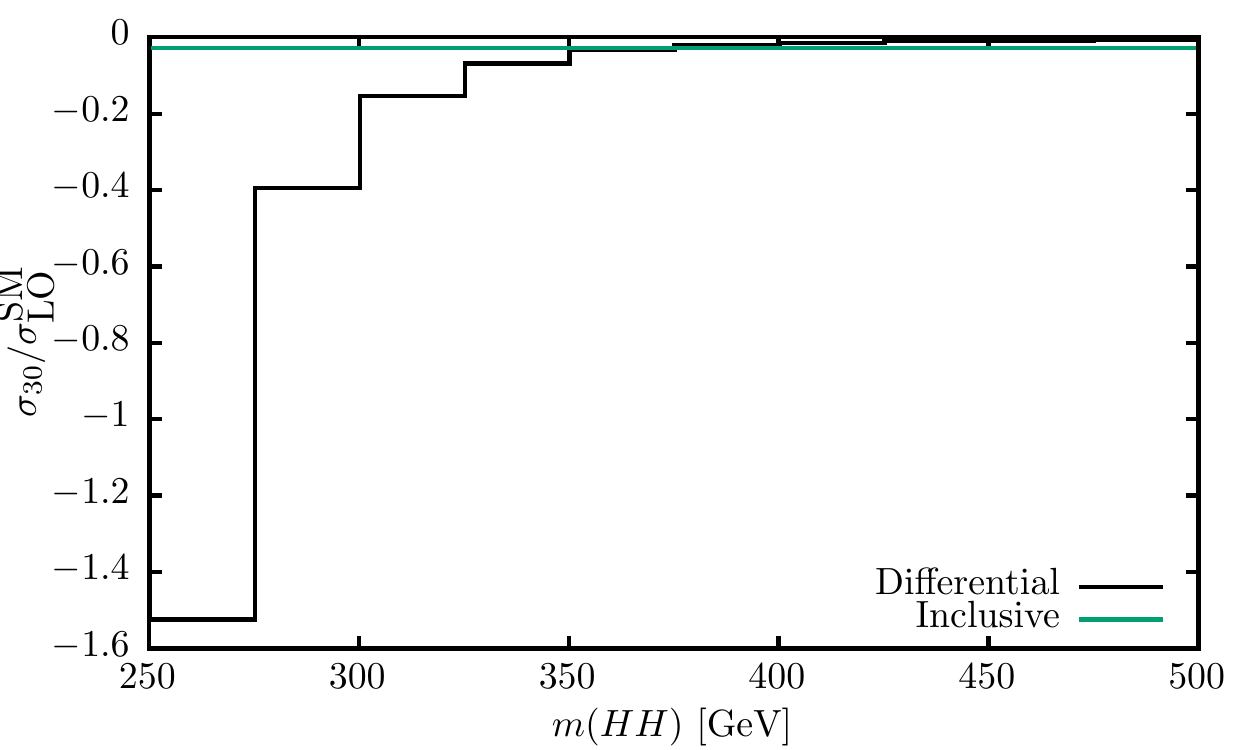}
  \includegraphics[scale=0.5]{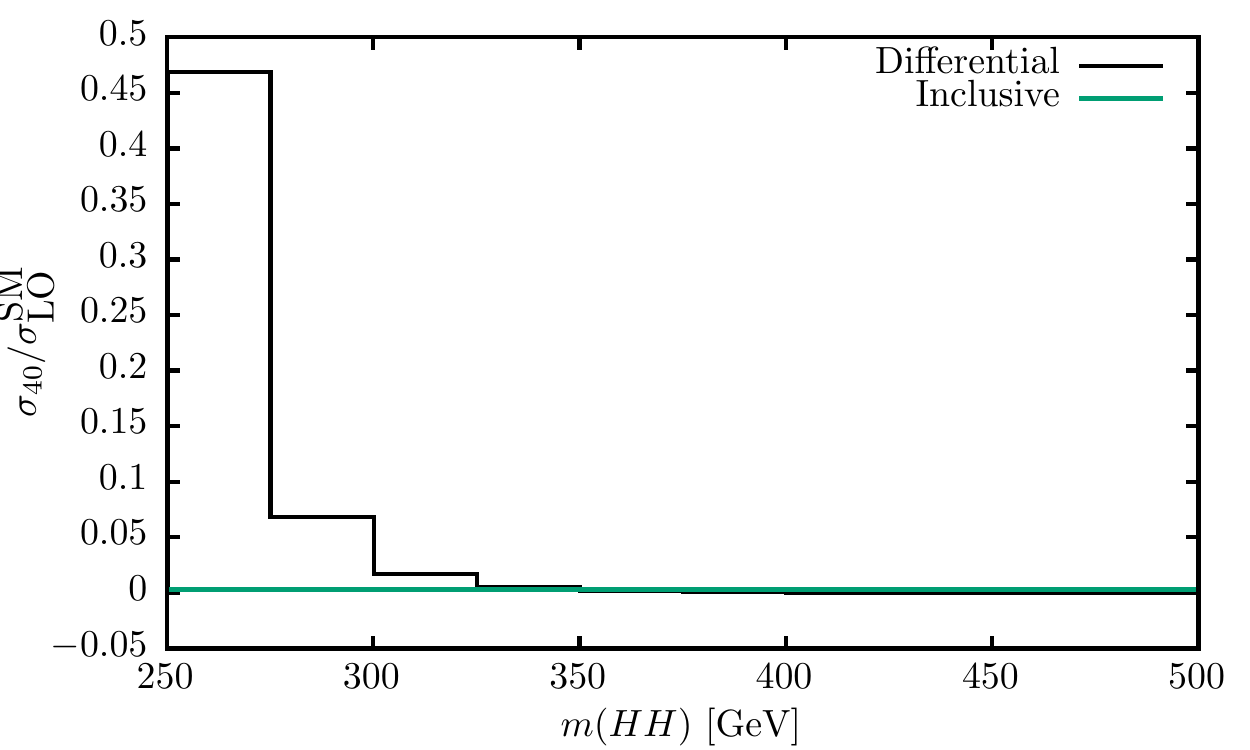}
 \includegraphics[scale=0.5]{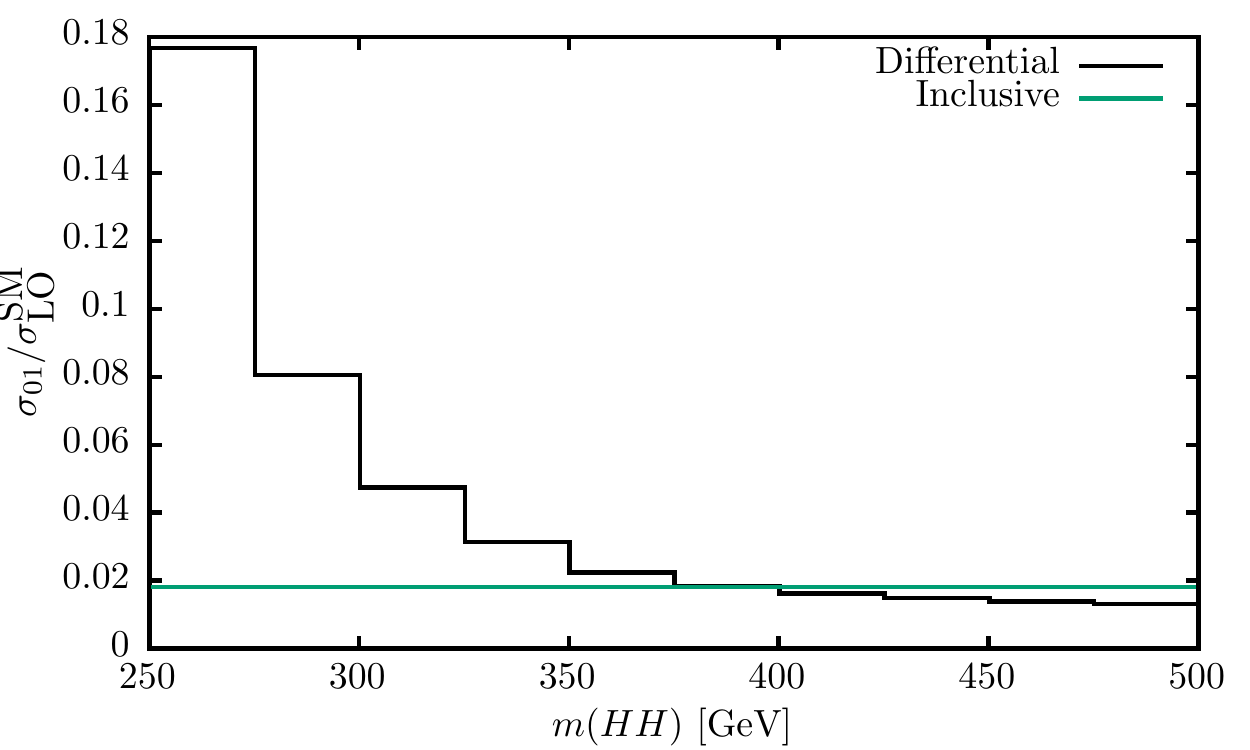}
 \includegraphics[scale=0.5]{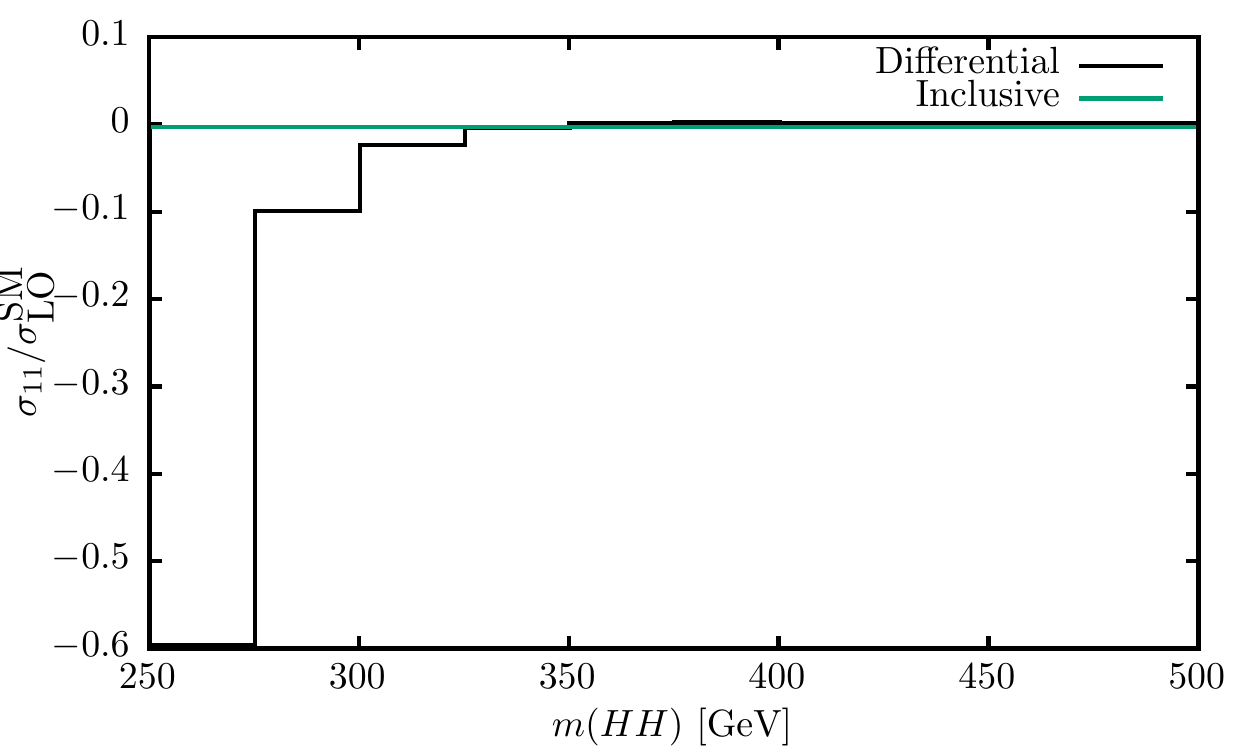}
 \includegraphics[scale=0.5]{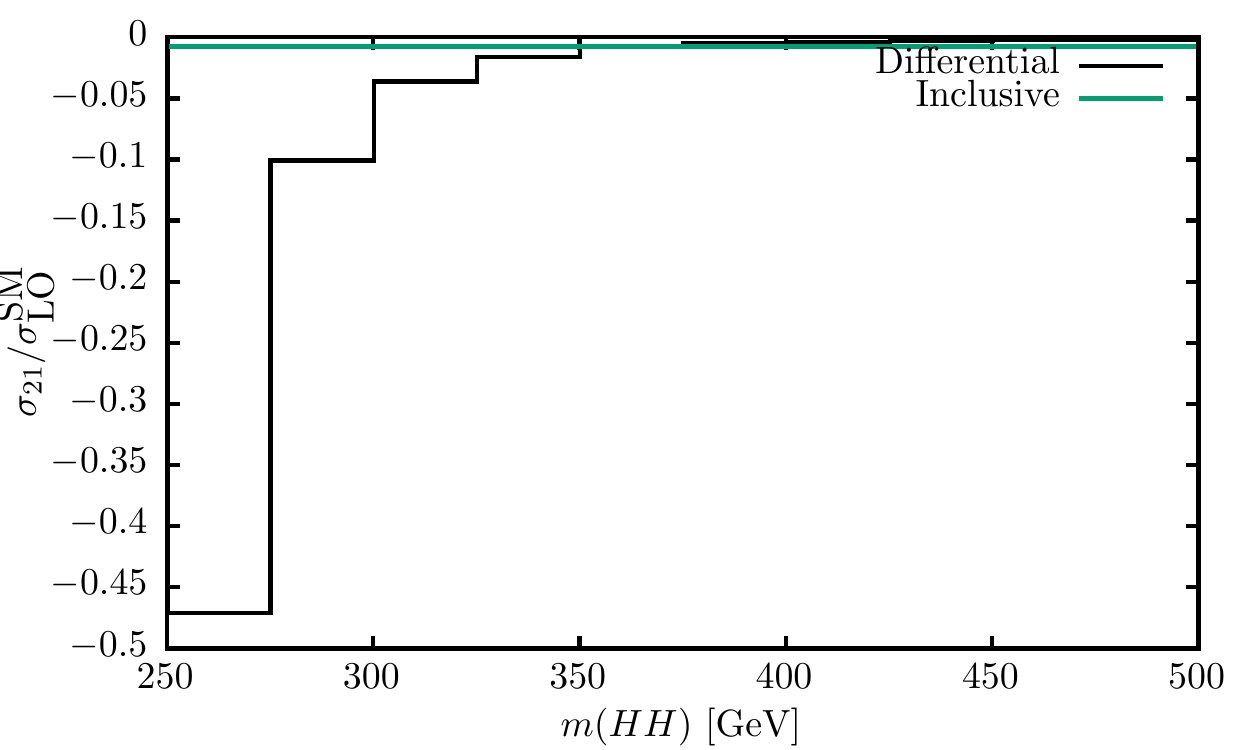}
 \caption{Relative impact of the $\sigma_{i(j)}$ contributions to the $m(HH)$ distribution at 14 TeV (black)
        compared to the same quantity at total cross section level (green).}\label{fig:diff-ratio}
\end{figure}

\begin{figure}
\centering
 \includegraphics[scale=0.71]{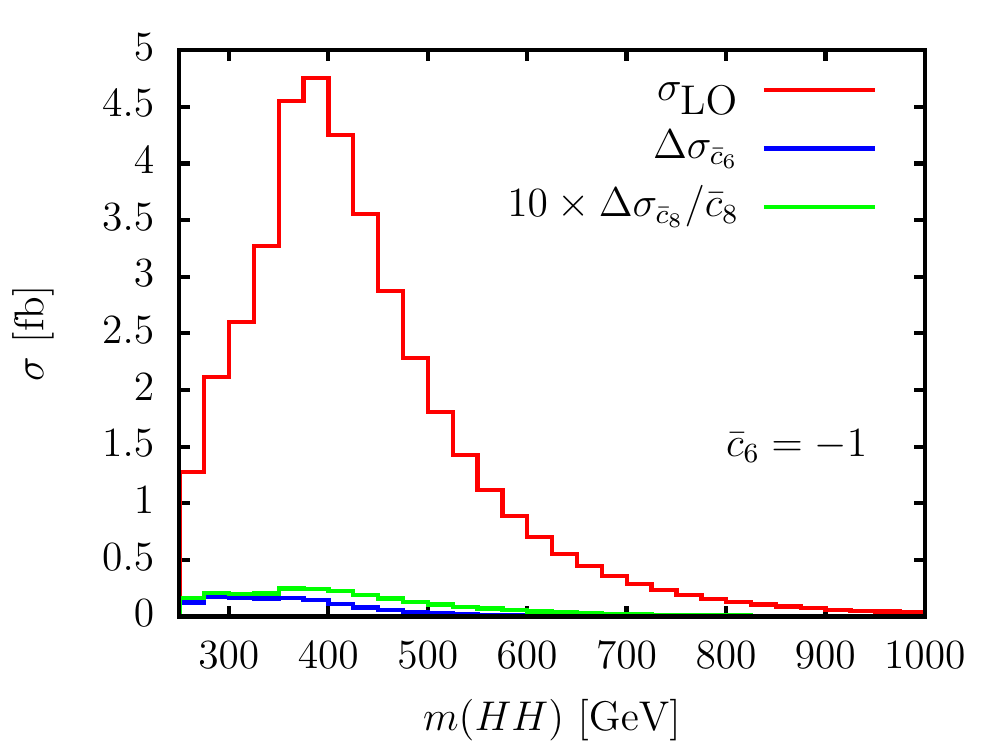}
 \includegraphics[scale=0.71]{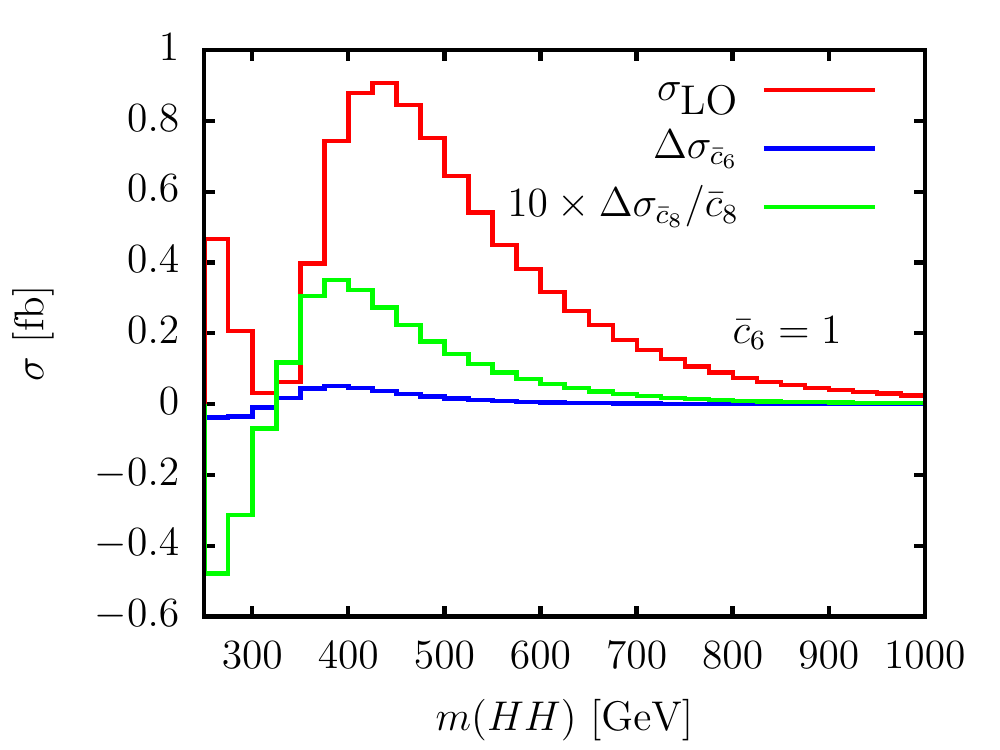}
 \includegraphics[scale=0.71]{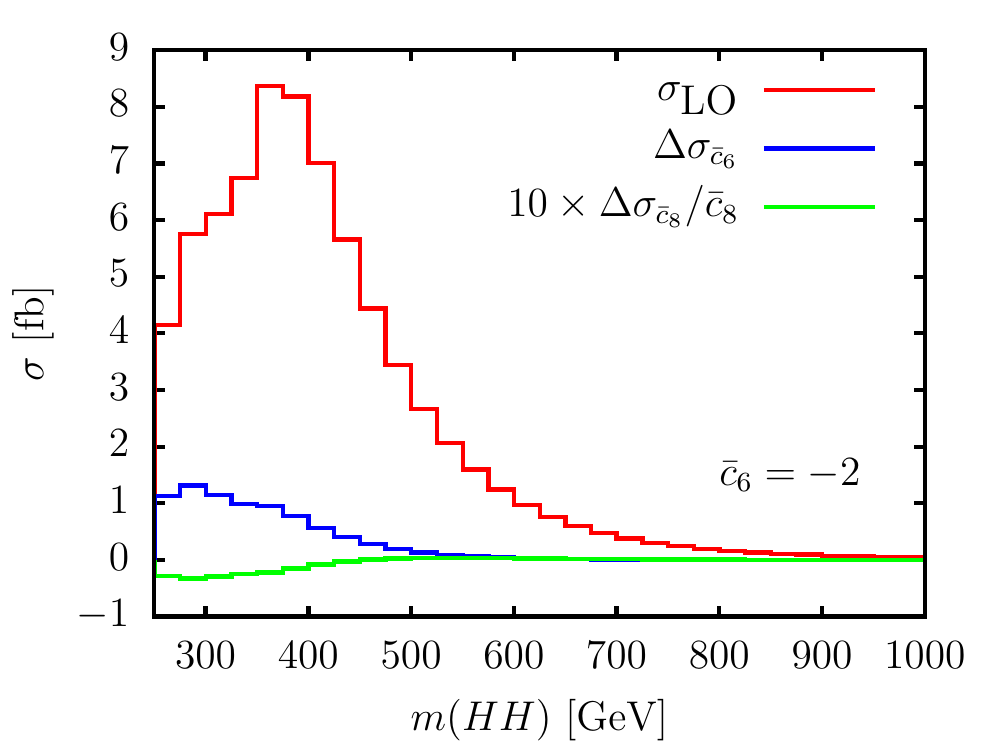}
 \includegraphics[scale=0.71]{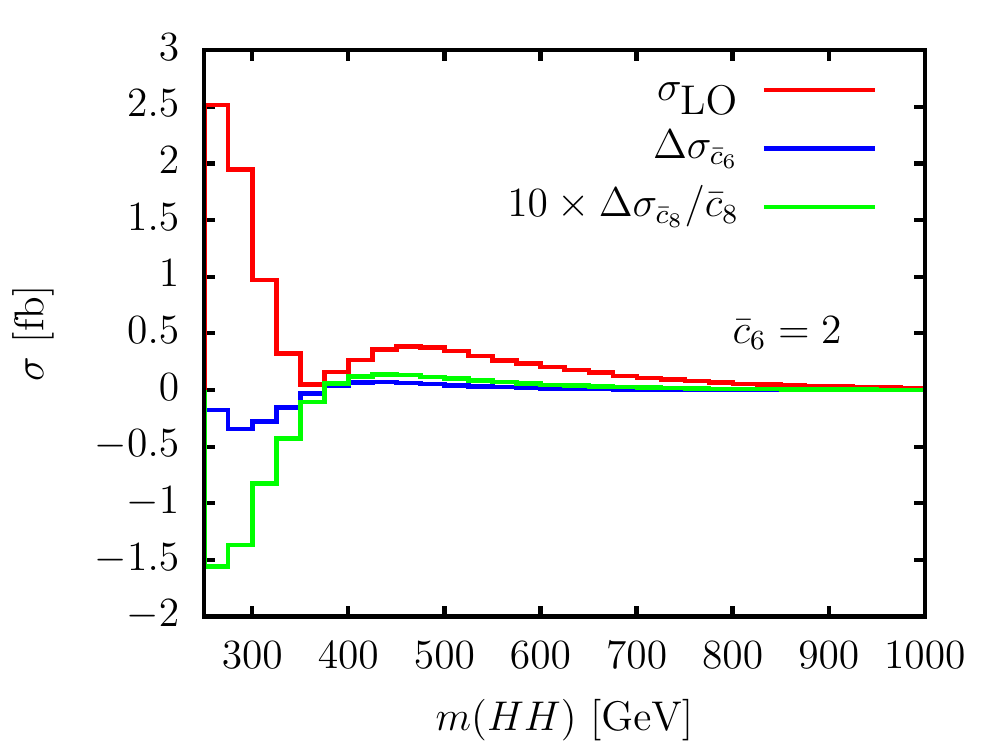}
 \includegraphics[scale=0.71]{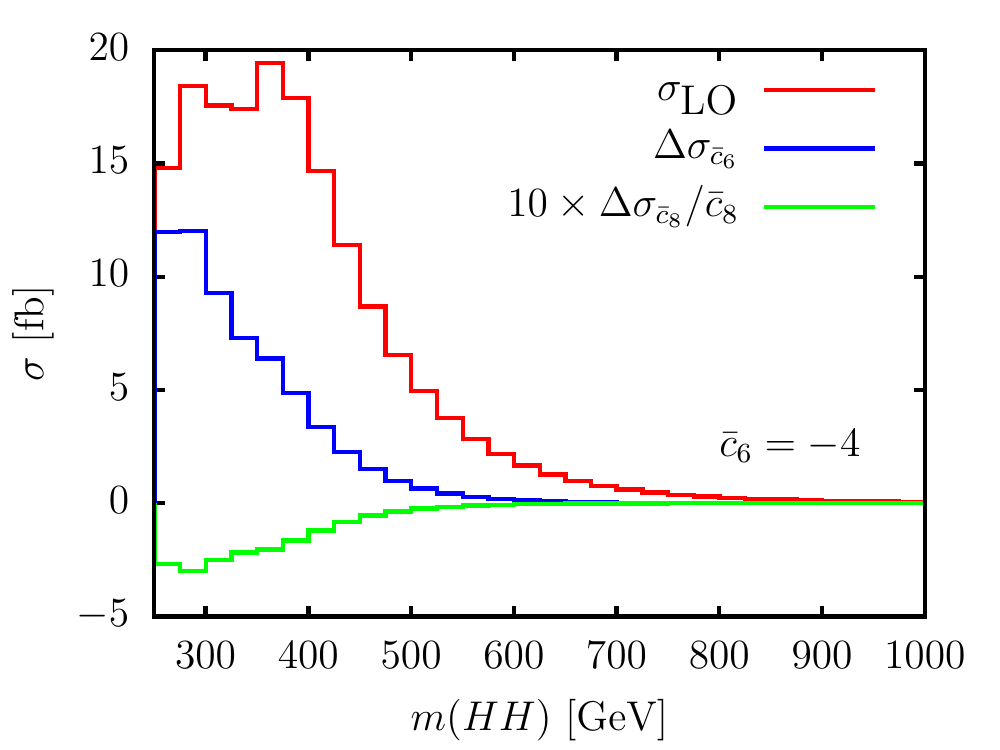}
 \includegraphics[scale=0.71]{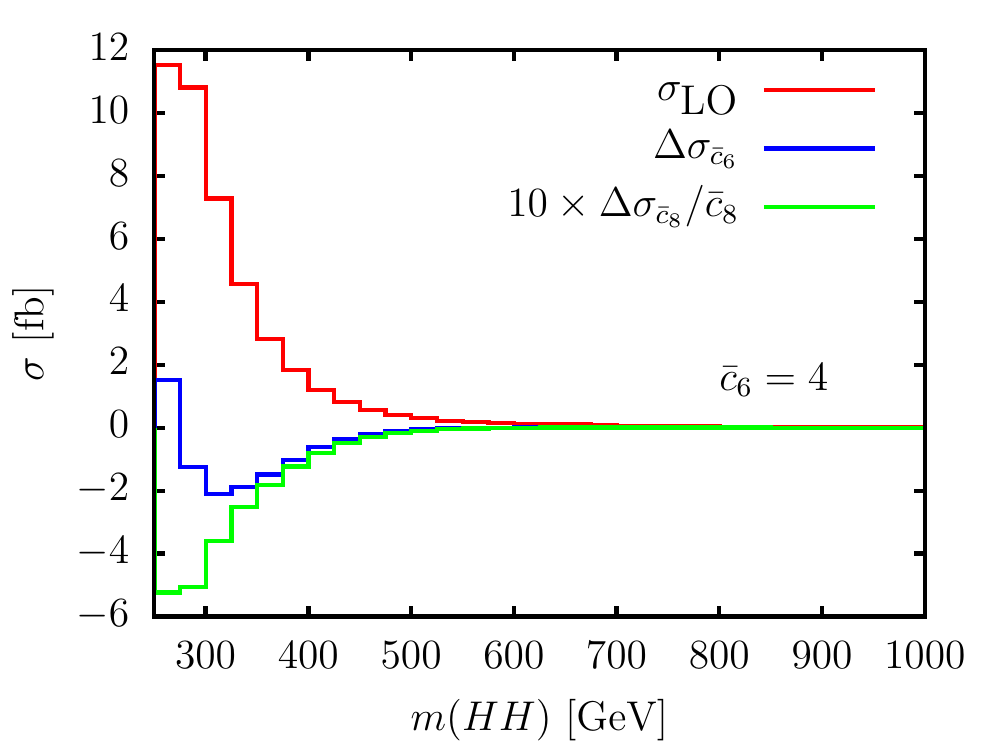}
 \caption{Different contributions ($\sigma_{\rm LO}$, $ \Delta\sigma_{\bar c_6}$ and  $\Delta\sigma_{\bar c_8}/\cbe$) to the $m(HH)$ distribution at 14 TeV for different $\cbs$ values.  \label{fig:distc6}}
\end{figure}

We move now to the differential distributions. In fig.~\ref{fig:logscale} we show the individual $\sigma_i$ (upper plot) and $\sigma_{ij}$ contributions (lower plots) to the  $m(HH)$ distribution at 14 TeV.\footnote{Besides an overall rescaling of the normalisation, distributions are very similar at 100 TeV so we do not show them.} In the case of negative values we plot their absolute values and display the result as a dashed line. Moreover,  we show in fig.~\ref{fig:diff-ratio} the ratio of any $\sigma_i$ and $\sigma_{ij}$ contribution over $\sigma_{\rm LO}^{\rm SM}$. In any plot this ratio is displayed as a black line, while we show in green the same result at the inclusive level, {\it i.e.}, the values in parentheses in tables \ref{tab:totxsLO} and \ref{tab:totxsNLO}.  
We observe that the $\cbs$- and $\cbe$-induced contributions are most important close to threshold. Moreover, the quantities $\sigma_1$, $\sigma_{30}$ and $\sigma_{21}$ are negative. Therefore, large cancellations are present and shapes strongly depend on the value of $\cbs$ and loop corrections also on $\cbe$.  
In order to better show this point, in fig.~\ref{fig:distc6} we plot $\sigma_{\rm LO}$ for representative values of $\cbs$, namely, $\cbs=\pm1,\pm2, \pm4$. Moreover we plot the quantities $\Delta\sigma_{\bar c_6}$ and $\Delta\sigma_{\bar c_8}/\cbe$ from eqs.~\eqref{Dscs} and \eqref{Dsce}. As already explained, $\Delta\sigma_{\bar c_6}$ and $\Delta\sigma_{\bar c_8}/\cbe$ correspond to the loop corrections induced by $\cbs$ on top of $\cbs$ and the two-loop $\cbe$-dependent part, respectively. The normalisation and shape of $\sigma_{\rm LO}$ strongly depend on $\cbs$. The difference in shape is crucial in order to discriminate $\cbs$ values leading to the same total cross section and it is exploited in our work, which is based on the analysis of the $m(HH)$ distribution.  The  $\Delta\sigma_{\bar c_6}$ corrections grow for large $|\cbs|$ and the impact of $\Delta\sigma_{\bar c_8}/\cbe$ is larger for negative values of $\cbs$. In both cases, the largest effects are close to the threshold, as expected.

\section{Constraints on the Higgs self couplings}
\label{sec:constraints}
\subsection{General set up}
In this section we discuss the $\cbs$ and $\cbe$ ($\kt$ and $\kf$) constraints  that can be derived from the measurements of double Higgs production in proton--proton collisions at the LHC and a 100 TeV future collider. We consider the $b \bar b \gamma \gamma$ signature, which  has been identified as  the most promising channel
 and allow for the reconstruction of the di-Higgs invariance mass $m(HH)$. In order to be close to a realistic experimental analysis, we  follow the study of Ref.~\cite{Azatov:2015oxa} for the case of HL-LHC and 100 TeV collisions with 30 ab$^{-1}$ of luminosity.\footnote{In principle, also the analysis in Ref.~\cite{Goncalves:2018yva} can be used. However, the amount of details provided by the authors is not sufficient for performing our study. For the same reason, we do not show results at 27 TeV in our paper, although may be extracted performing the analysis in Ref.~\cite{Goncalves:2018yva}.} We use the same selection cuts for the $b \bar b \gamma \gamma$  signature, we divide the reconstructed $m(HH)$ distribution in the same six bins and for each bin we take directly from Ref.~\cite{Azatov:2015oxa} the predictions for the background and for the signal in the SM. Results in Ref.~\cite{Azatov:2015oxa} take into account higher-order QCD corrections for both the signal and the background and also showering, hadronisation and detector effects. In our analyses we assume that $\cbs$ and $\cbe$ effects factorise QCD corrections and we compute the effects of selection cuts (see Appendix \ref{app:cuts}) adding $H$ decays at the parton level. Thus, we also assume that  showering, hadronisation and detector effects factorise the effect of selections cuts on the  $b \bar b \gamma \gamma$  signature.

In order to set limits on $\cbs$ and $\cbe$ we perform a $\chi^2$ fit on the $m(HH)$ distribution. For simplicity, as done in Ref.~\cite{Azatov:2015oxa}, we will include statistical uncertainties only. The impact of theoretical uncertainties and experimental systematic uncertainties is expected to be much smaller than statistical ones \cite{Azatov:2015oxa, Bizon:2018syu}, therefore they would not in general lead to significant differences; some caveats are present for the 100 TeV case and will be discussed afterwards. On the other hand, we have found that assuming $\cbs$ and $\cbe$ effects as flat within each of the six bins of the reconstructed $m(HH)$ distributions can strongly distort the results. Indeed, in each $m(HH)$ bin, $\cbs$ and $\cbe$ effects are not flat over the full $b \bar b \gamma \gamma$ phase-space. Thus, selection cuts have an impact not only on the total number of events observed but also on the ratio $ \sigma^{\rm pheno}_{\rm NLO}  / \sigma^{\rm SM}_{\rm LO}  $. More details about the fit procedure can be found in Appendix \ref{app:fit}.

Similarly to what has been in done in Ref.~\cite{Maltoni:2018ttu}, we consider two different scenarios for setting bounds on Higgs self couplings:
\begin{enumerate}
    \item  {\bf Scenario 1: Well-behaved EFT ($\ktre\ne1, \kqual\sim 6 \ktre -1$).} \\  The contribution from $\bar c_8$ is suppressed w.r.t.~the one from $\cbs$, hence we can safely set $\bar c_8=0$.
        We do not assume only $\cbs\sim0$, {\it i.e.}, an SM-like configuration, but also allow for large BSM effects ($|\cbs|\gtrsim 0$). 
        
    \item {\bf Scenario 2: General parameterisation allowing for $\ktre\ne1$ and $\kqual\ne 6 \ktre -1$.}\\
    Effects from $\bar c_8$ are not negligible and therefore we consider $\bar c_8\ne 0$. Also in this case, we consider $\cbs\sim0$ or $|\cbs|\gtrsim 0$, allowing for large BSM effects.
\end{enumerate}

In Scenario 1, assuming that Nature corresponds to $\cbs=\cbstr$, we will analyse the constraints that can be set on  $\cbs$. In the Scenario 2, setting $\cbetr=0$, we explore the constraints that can be set on the $(\cbs,\cbe)$ plane, for different value of $\cbstr$. One may be tempted to study also a ``Scenario 3'', as done in Ref.~\cite{Bizon:2018syu}, where $\cbs=0$ and $\cbe\ne 0$. However, this configuration is unstable. Indeed, it is easily spoiled by the running of $\cbs$ and $\cbe$ at different scales,\footnote{As can be easily derived by the counterterm for $\cbs$ given in Ref.~\cite{Maltoni:2018ttu}, the one-loop $\beta$-function for $\cbs$ contains terms proportional to $\cbe$ and independent on $\cbs$.} since it is not protected by any symmetry and not emerging from an EFT expansion. For this reason we refrain from considering this scenario.

\subsection{Scenario 1}

\begin{figure}
 \centering
 \includegraphics[scale=0.5]{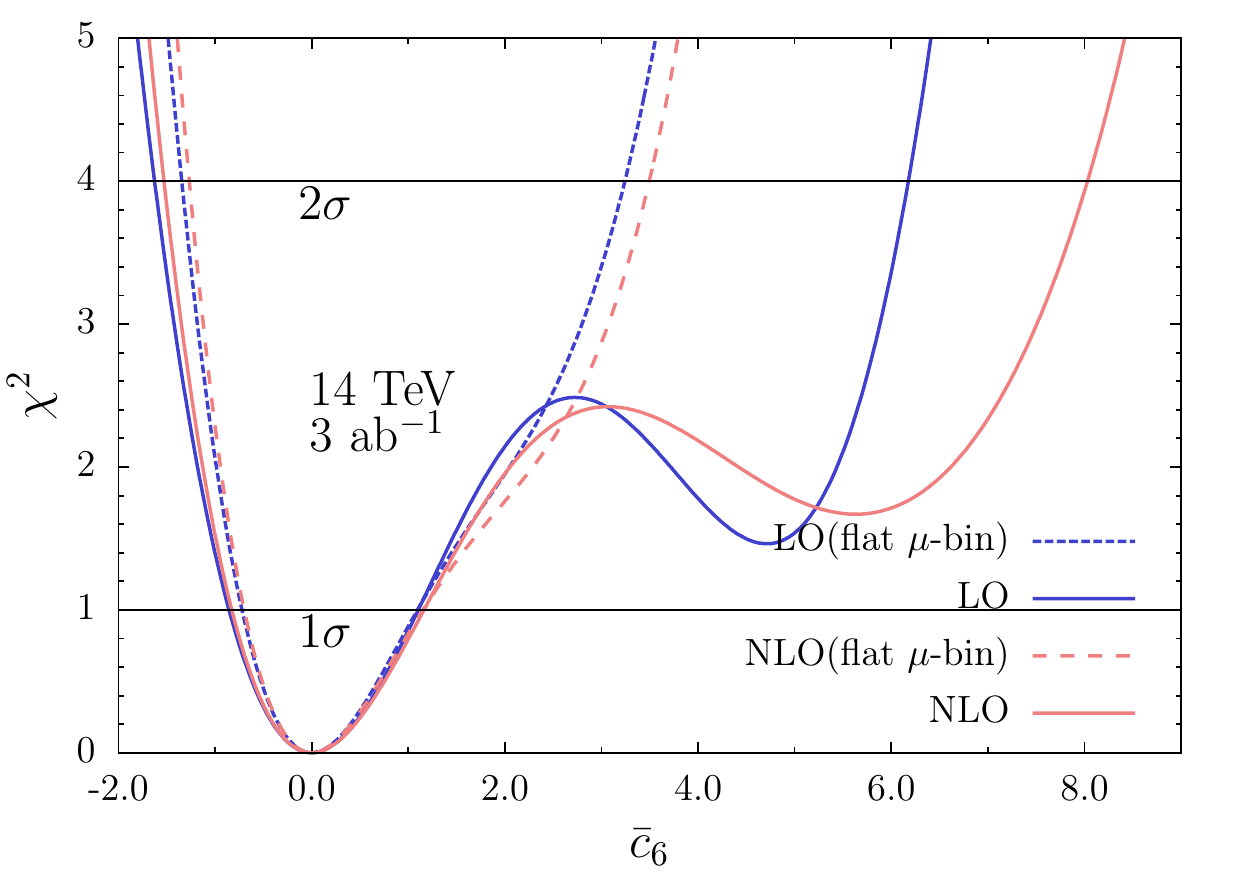}
 \includegraphics[scale=0.5]{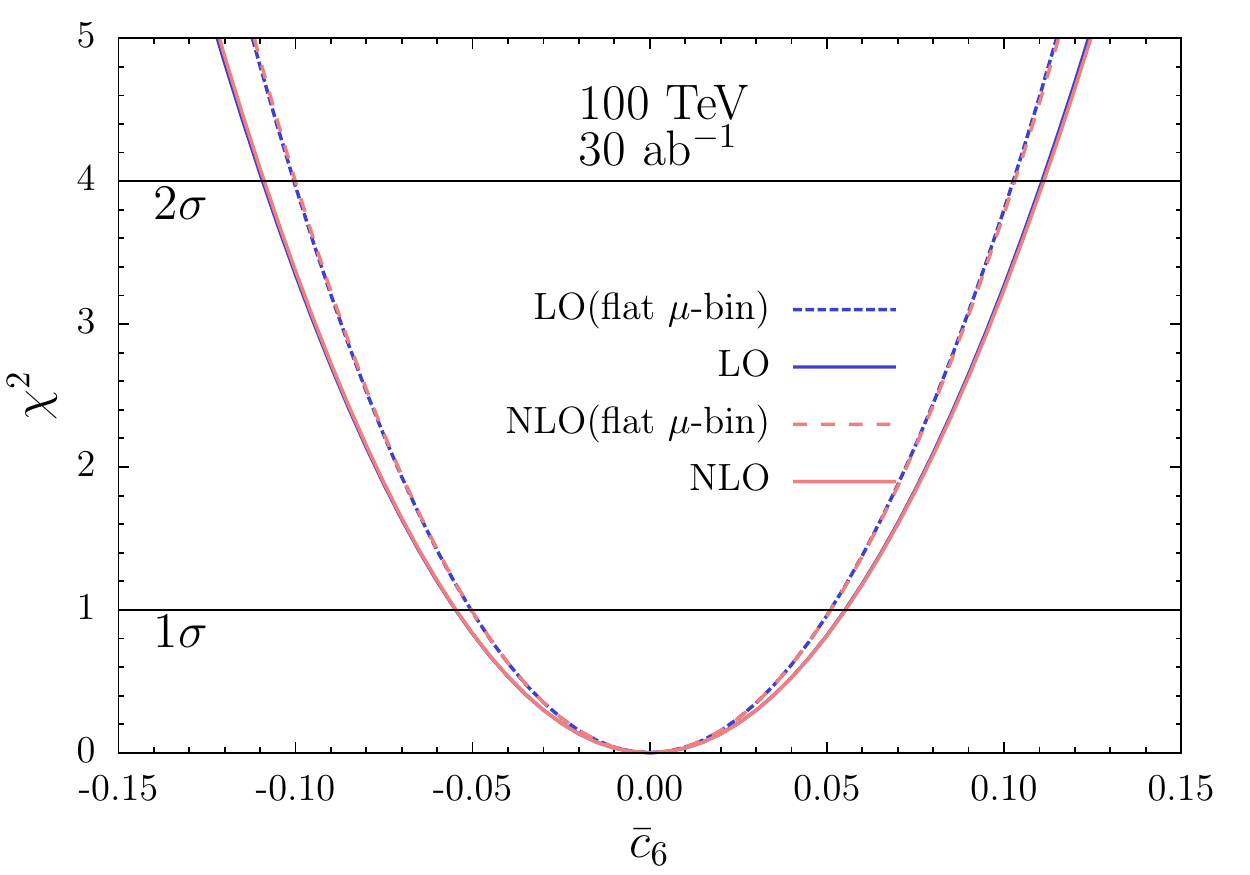}
 \caption{$\chi^2$ as a function of ${\bar c_6}$ for ${\bar c_8}=0$ at 14 (left) and 100 (right) TeV. \label{fig:chi2}}
\end{figure}

We start considering the $\chi^2$ function and the $1\sigma$ and $2\sigma$ bounds that can be obtained for $\cbs$ assuming $\cbstr=0$, at the HL-LHC and at a future 100 TeV collider. In fig.~\ref{fig:chi2} we plot the $\chi^2$ function\footnote{In fact, the plots display the quantity $\chi^2 -\min(\chi^2)$. For brevity we will refer to it as $\chi^2$.} using $\sigma^{\rm pheno}_{\rm NLO}$ or $\sigma_{\rm LO}$ in the fit. Moreover, we show the relevance of fully differential information in the treatment of $\cbs$ and $\cbe$ effects. In the case denoted  as ``flat $\mu$-bin'' in the plot, we assume that for each $m(HH)$-bin the impact of $\cbs$ effects can be evaluated via the ratio $\sigma/\sigma_{\rm LO}$ without taking into account the selection cuts on the $b \bar b \gamma \gamma$ final state, where $\sigma$  can be either $\sigma^{\rm pheno}_{\rm NLO}$ or $\sigma_{\rm LO}$. We remark that both in the ``flat $\mu$-bin'' and  normal cases, selection cuts are taken into account for  the SM signal; the ``flat $\mu$-bin'' concerns only the modelling of $\cbs$ and $\cbe$ effects for the $m(HH)$-binning of the fit. More details are given in Appendix \ref{app:fit}.
As can be seen  in fig.~\ref{fig:chi2}, NLO effects, which in Scenario 1 corresponds to $\Delta\sigma_{\bar c_6}$ only, are relevant only for large values of $\cbs$. On the contrary, the ``flat $\mu$-bin'' assumption strongly distorts the $\chi^2$ profile, especially for positive values of $\cbs$. Indeed, as can be seen from the dashed lines, with this assumption the $2\sigma$ bounds at 14 TeV would be artificially improved. This effect is due to the fact that for  $\cbs \gtrsim 2$ the bulk of events is in the first bin(s) of the $m(HH)$ distribution (see fig.~\ref{fig:distc6}) and selection cuts strongly depend on $m(HH)$ especially close to the threshold (see Appendix \ref{app:cuts}). 

Consistently taking into account the selection cuts in our analysis, we find the following $2\sigma$ intervals, 
\begin{align}
-0.5<\ktre&=1+\cbs<8 \qquad &{\rm at~14~TeV~with~3~ab^{-1}}\,, \label{eq:142s}\\
0.9<\ktre&=1+\cbs<1.1 \qquad &{\rm at~100~TeV~with~30~ab^{-1}\label{eq:1002s}}\, .
\end{align} 

\begin{figure}
 \centering
 \includegraphics[scale=0.5]{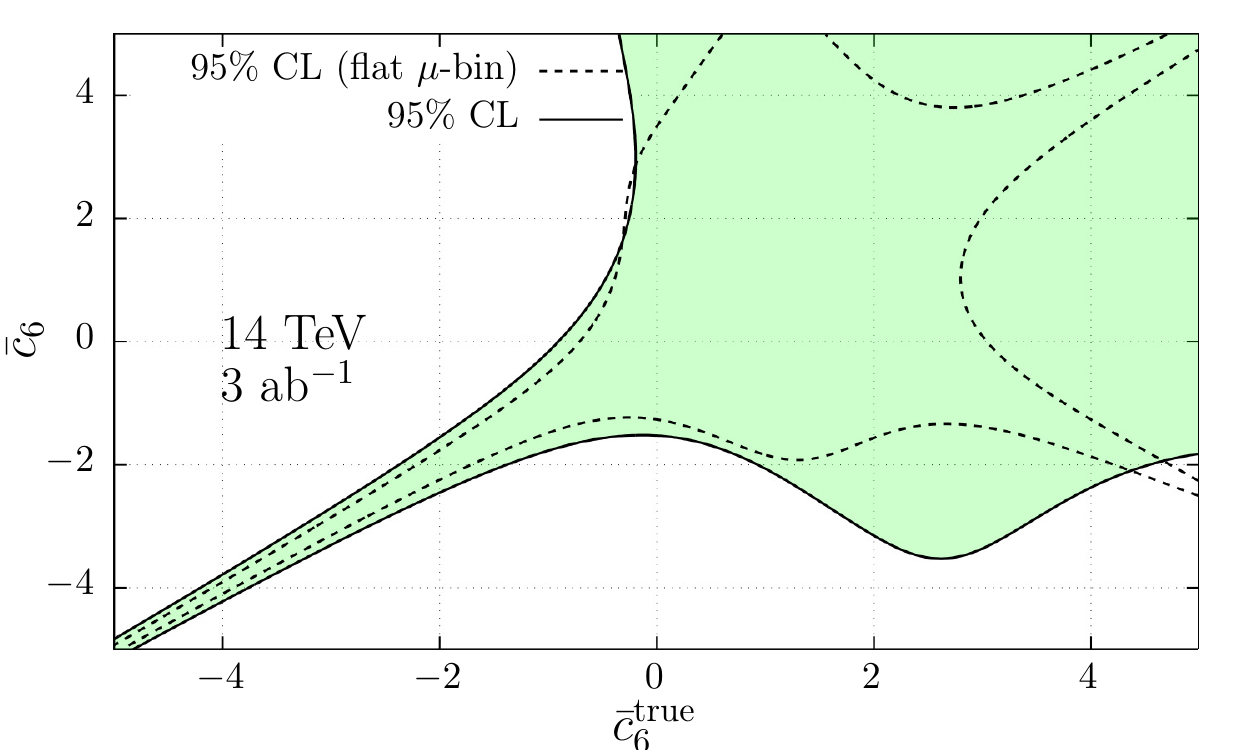}
 \includegraphics[scale=0.5]{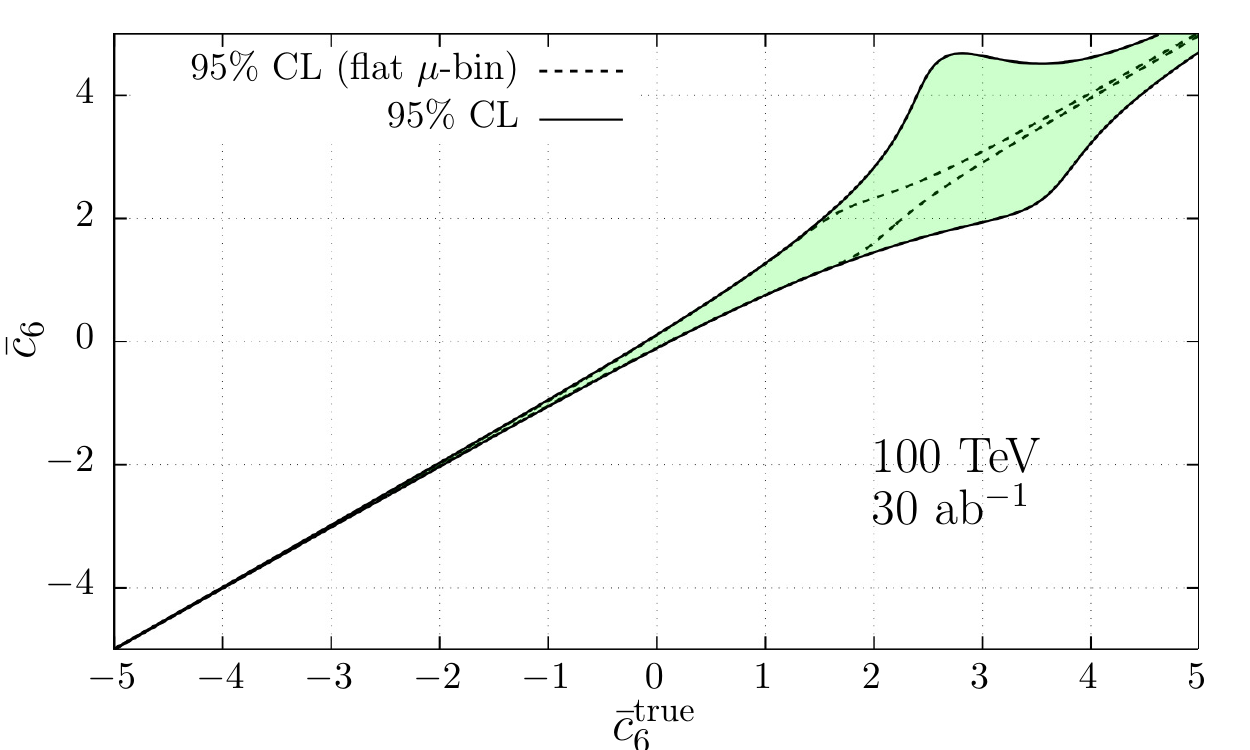}
 \caption{Sensitivity ($2\sigma$ bounds) on ${\bar c}_6$ as function of $c_6^{\rm true}$ for 14 TeV (left) and 100 TeV (right).  \label{fig:c6lims}}
\end{figure}

We now move to the case where $\cbstr$ can be different from zero. In fig.~\ref{fig:c6lims} we show $2\sigma$ bounds for $\cbs$ as a function of $\cbstr$. It turns out that if $\cbs$ is negative, bounds can be sizeably stronger. For instance, assuming $\cbstr=-2$ a limit $-1.5.<\ktre=1+\cbs<-0.5$ can be obtained at HL-LHC, which is remarkably more stringent than in the  $\cbstr=0$ case of \eqref{eq:142s}. In the case of 100 TeV, large and negative values of $\cbstr$ seem to lead to subpercent precision. This should be interpreted as indication that high precision may be reached in this scenario, but also that theory and systematic uncertainties have to be taken into account to estimate a realistic value. In both the plots of fig.~\ref{fig:c6lims} we show also results under the ``flat $\mu$-bin'' assumption as dashed
lines. As can be seen, this assumption would have a strong effect to the $\cbs$ bounds, especially for $\cbs \gtrsim 0$. 

\begin{figure}
 \centering
 \includegraphics[scale=0.7]{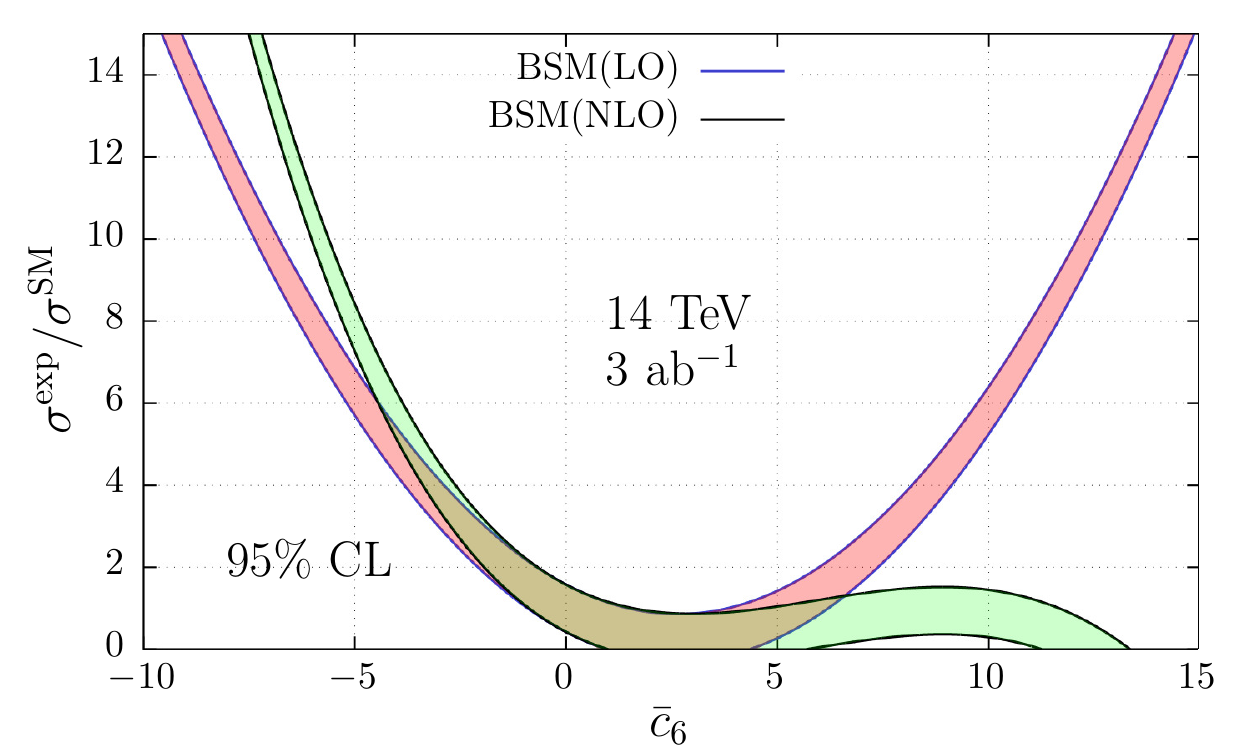}
 \caption{Bounds on $\cbs$ that can be set according to the supposedly double-Higgs measured cross section, normalised to the corresponding SM prediction. The red band is obtained considering only $\sigma_{\rm LO}$, while the green using $\sigma^{\rm pheno}_{\rm NLO}$. }
	\label{fig:ignorance}
\end{figure}

In this context we want also to stress an important point that has been somehow overlooked in both theory and experimental  studies on $\kt$-determination.
In Fig.~\ref{fig:ignorance} we plot the 2$\sigma$ constraints that can be obtained on $\cbs$ by varying of $\sigma^{\rm exp}/\sigma^{\rm SM}$, where $\sigma^{\rm exp}$ is the measured value and  $\sigma^{\rm SM}$ is the SM prediction. We derive the constraints using two different approximations: $\sigma^{\rm pheno}_{\rm NLO}$ and $\sigma_{\rm LO}$. As can be seen, for $|\cbs|\gtrsim5$, where perturbativity is violated, the constraints on $\cbs$ strongly depend on the choice between $\sigma^{\rm pheno}_{\rm NLO}$ and $\sigma_{\rm LO}$.  When data are fitted with $\sigma_{\rm LO}$ predictions, $\cbs$ or equivalently $\ktre$ is a parameter of ignorance that only for $|\ktre-1|=|\cbs|\lesssim5$ coincides to the quantity one is interested in. Outside this range,  $\cbs$(or $\kt$) is only suggesting how far from the SM predictions is the experimental result. The usage of $\sigma^{\rm pheno}_{\rm NLO}$ or any higher-order corrections in the place of $\sigma_{\rm LO}$ is not improving this situation, since the regime is not perturbative for $|\cbs|\gtrsim5$. In conclusion, one can set bounds outside the   
$|\ktre-1|=|\cbs|\lesssim5$ range, but only within this region they properly refer to the quantities we are interested in and defined via parameters in the Lagrangian.

\subsection{Scenario 2}

\begin{figure}
 \centering
  \includegraphics[scale=0.5]{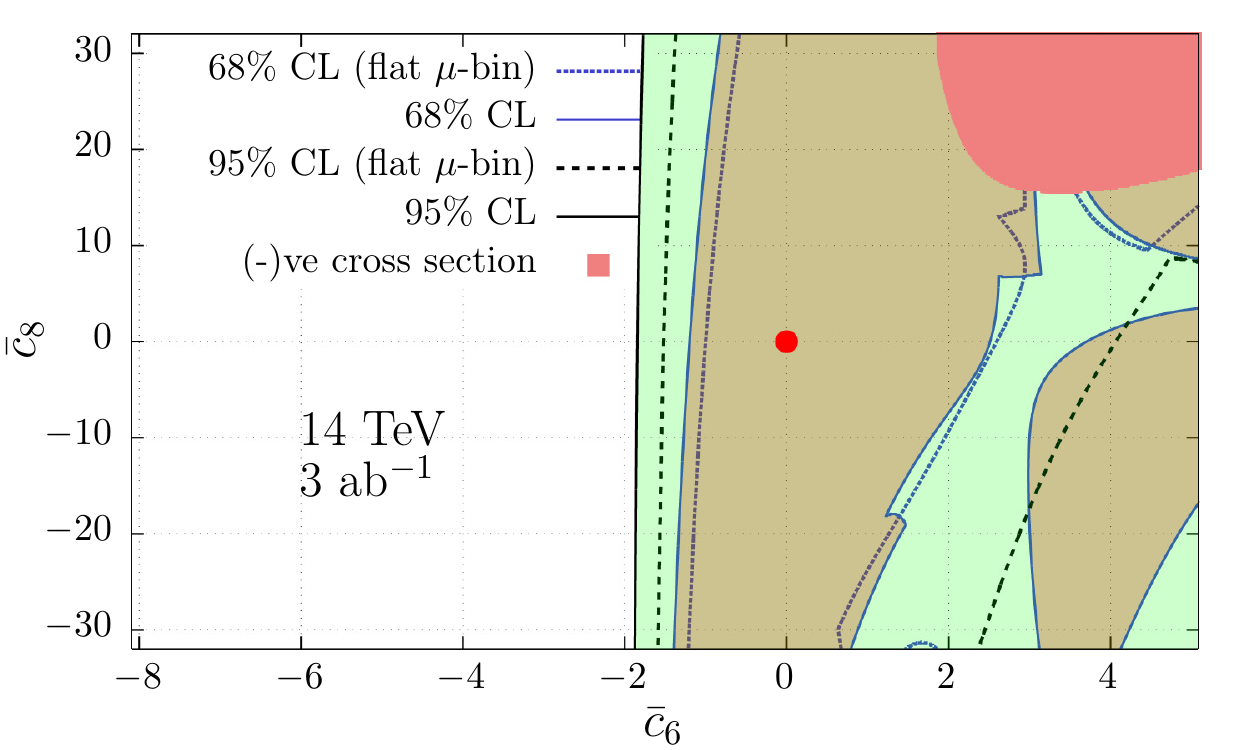}
    \includegraphics[scale=0.5]{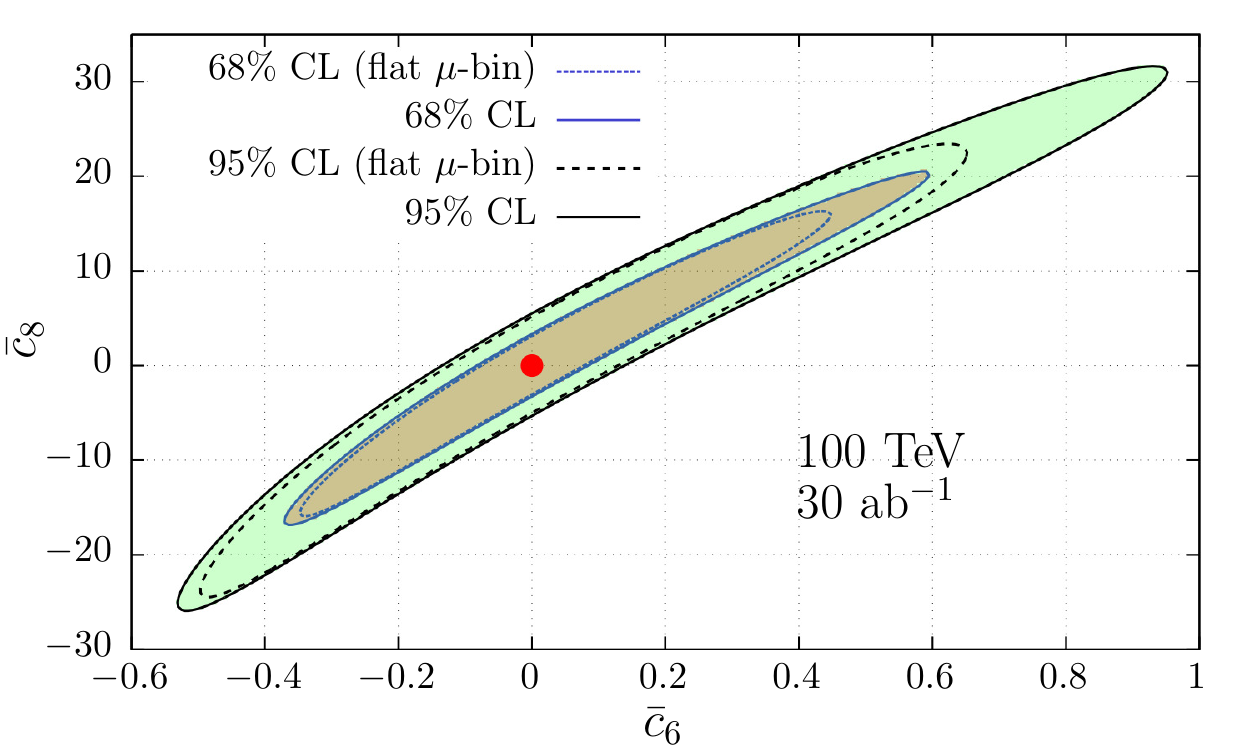}
	\caption{Expected 1$\sigma$ and 2$\sigma$ bounds in the ($\cbs$,$\cbe$) plane at 14 (left) and 100 TeV (right), assuming $\cbstr=\cbetr=0$ (denoted by red dots). \label{fig:c6c8lims}}
\end{figure}

This scenario allows us to discuss the most important phenomenological results of this work, {\it i.e.}, the expected constraints on $\cbs$ and $\cbe$ ($\kt$ and $\kf$) that can be obtained via double Higgs production at HL-LHC and a 100 TeV future collider.
Assuming $\cbetr=0$, these constraints  are shown in the left and right plot of Fig.~\ref{fig:c6c8lims}, respectively. We show $2\sigma$ results and again the effect due to the ``flat $\mu$-bin'' assumption, the red area corresponds to the region where the cross-section is negative ({\it cf.} left plots in Fig.~\ref{fig:contour}). As already mentioned, no phenomenological study can be performed in this configuration. Similarly, for a given $(\cbs,\cbe)$, predictions for some bins can be negative, while positive for others; we retain the information only for those bins where the cross-section is predicted to be positive. As can be seen from fig.~\ref{fig:c6c8lims}, at HL-LHC the presence of $\cbe$ contributions is not sizeably affecting the result in \eqref{eq:142s}, obtained under the assumption $\cbe=0$. On the other hand, no sensible constraints can be obtained at the HL-LHC on the $\cbe$ parameter. In other words, with a complete calculation and taking into account selection cuts and background effects, we find a  much less optimistic result than in Ref.~\cite{Bizon:2018syu}. 

Results at 100 TeV collisions are qualitatively very different than at the HL-LHC. The bounds on $\cbs$ are affected by the presence on $\cbe$. As can be seen from the right plot of Fig.~\ref{fig:c6c8lims}, the bounds are $0.4<\ktre=1+\cbs<2$, which is less precise than \eqref{eq:1002s}, obtained under the  assumption $\cbe=0$. Although most of the perturbativity $\cbe$ region is not excluded, there is a clear direction in the contours of the constraints in the $(\cbs,\cbe)$ plane.  

\begin{figure}
 \centering
 \includegraphics[scale=0.5]{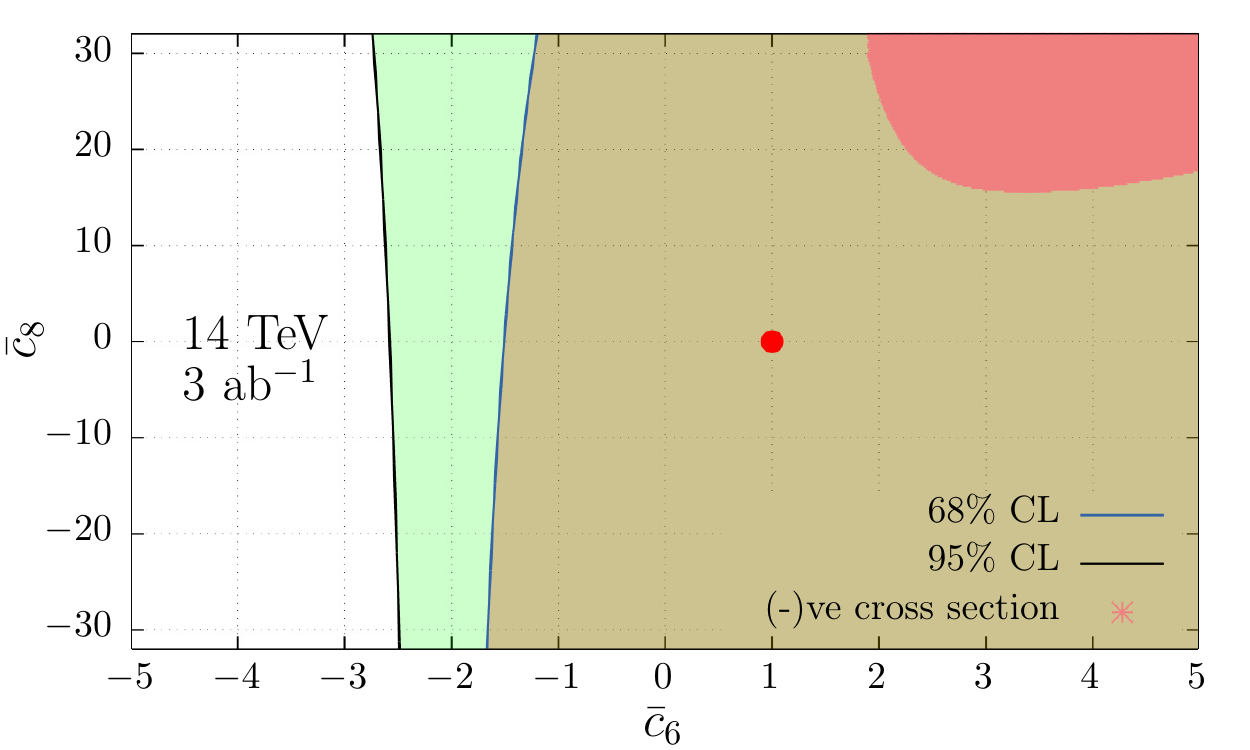}
  \includegraphics[scale=0.5]{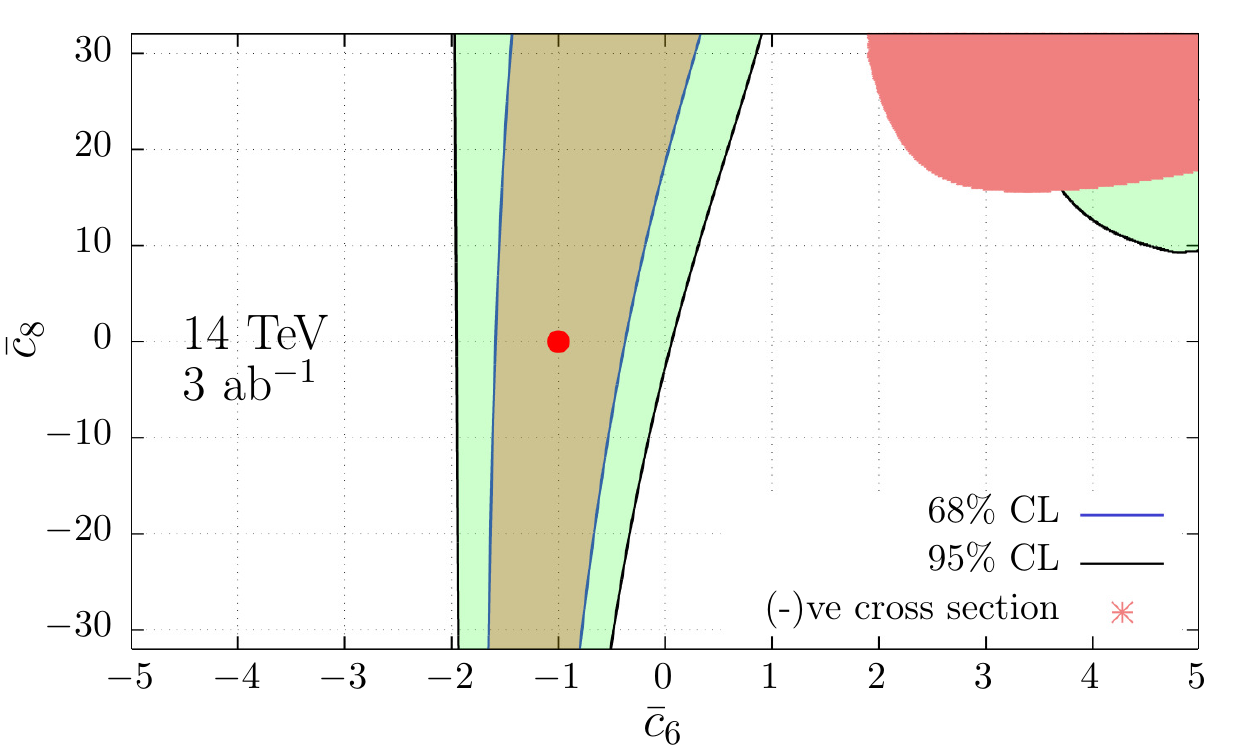}
 
 \includegraphics[scale=0.5]{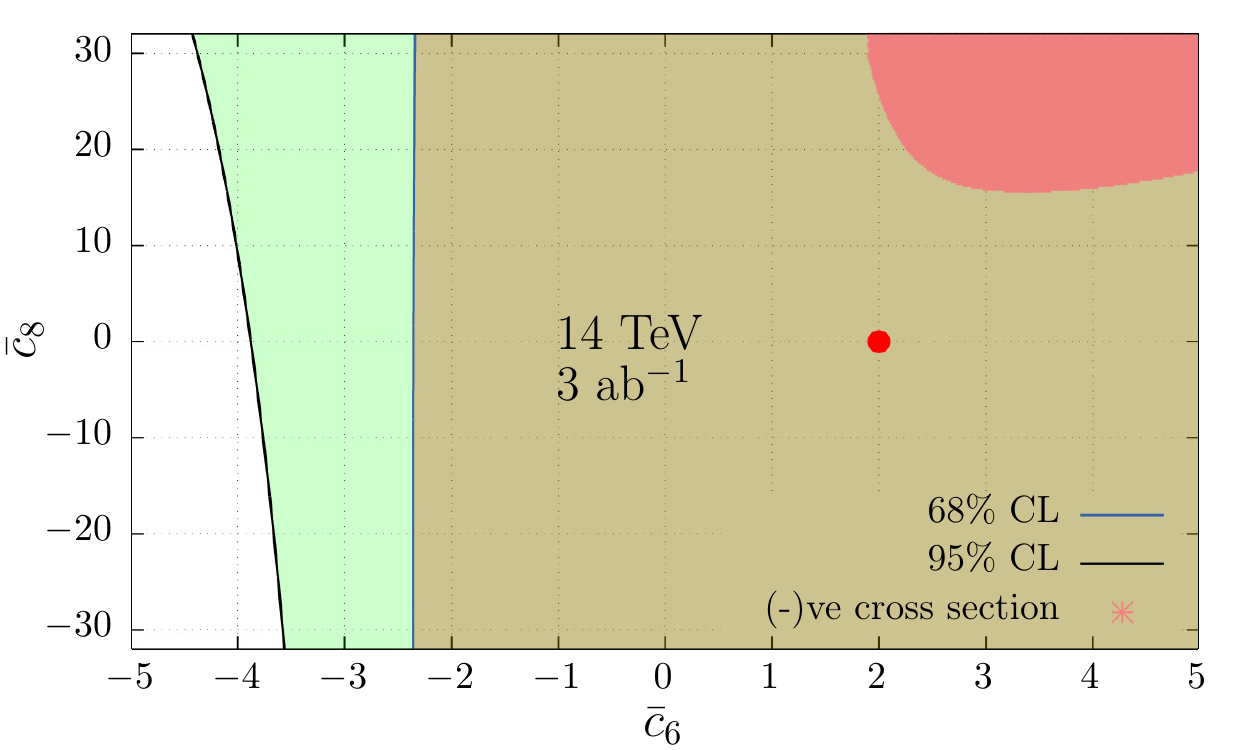}
  \includegraphics[scale=0.5]{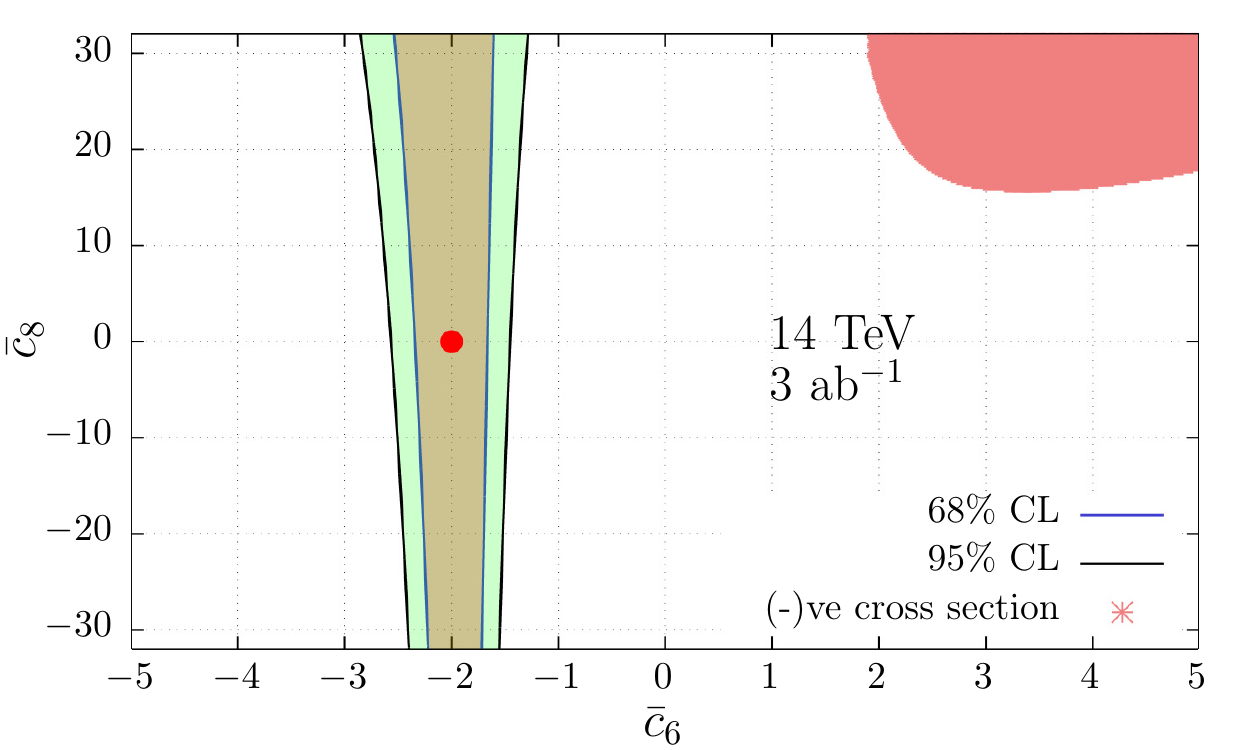}

 \includegraphics[scale=0.5]{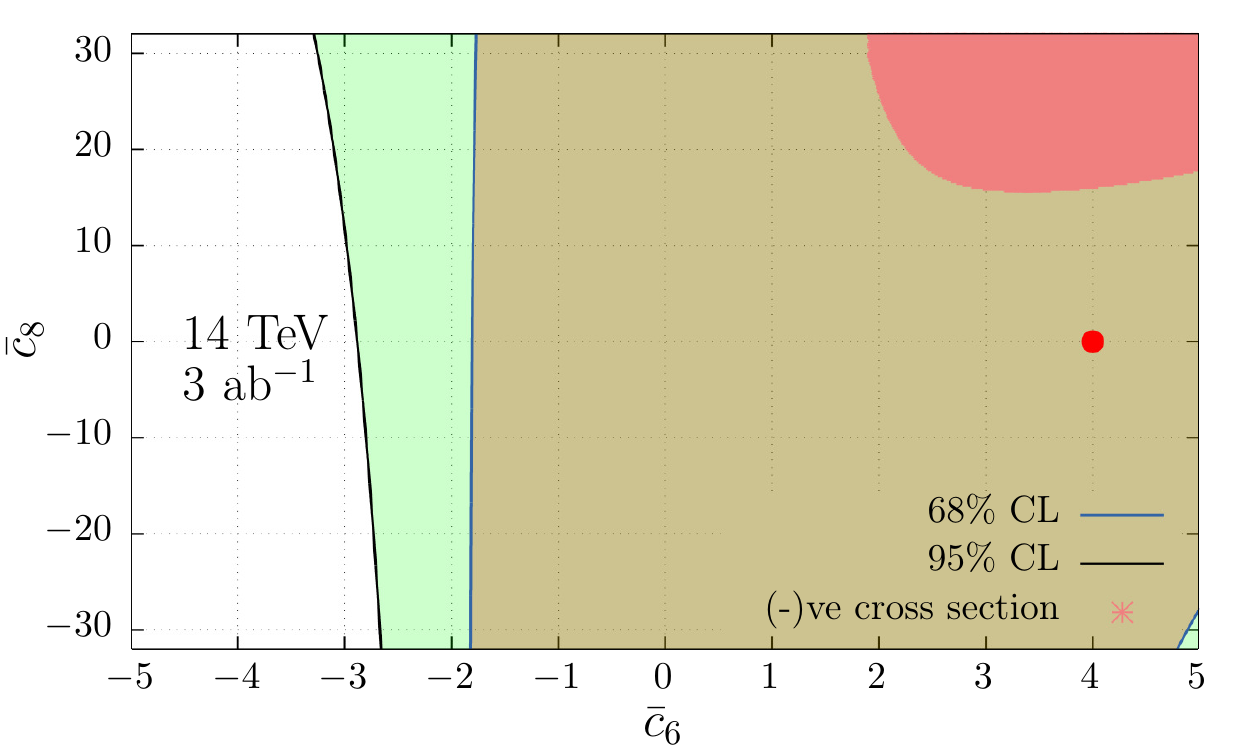}
  \includegraphics[scale=0.5]{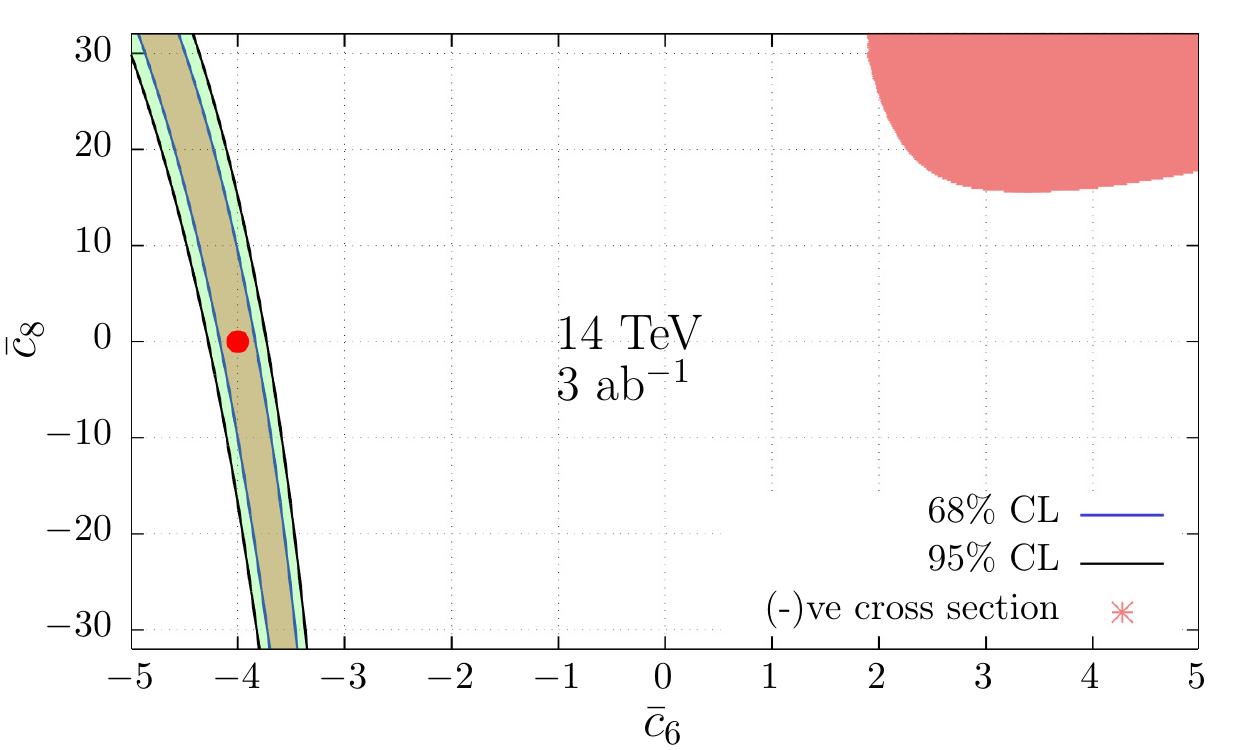}

	\caption{Expected 1$\sigma$ and 2$\sigma$ bounds in the ($\cbs$,$\cbe$) plane at 14 TeV, assuming $\cbstr=\pm1,\pm2,\pm4$ and $\cbetr=0$ (denoted by red dots). \label{fig:c6c814}}
\end{figure}

\begin{figure}
 \centering
 \includegraphics[scale=0.5]{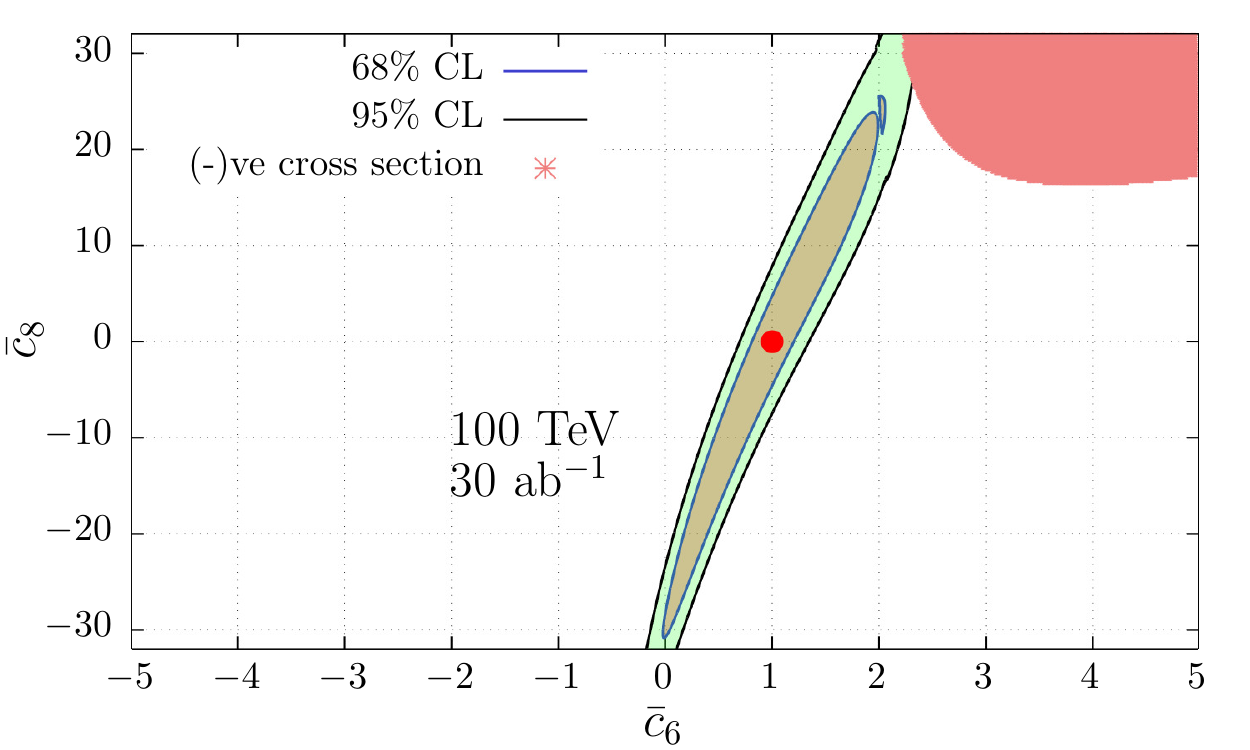}
  \includegraphics[scale=0.5]{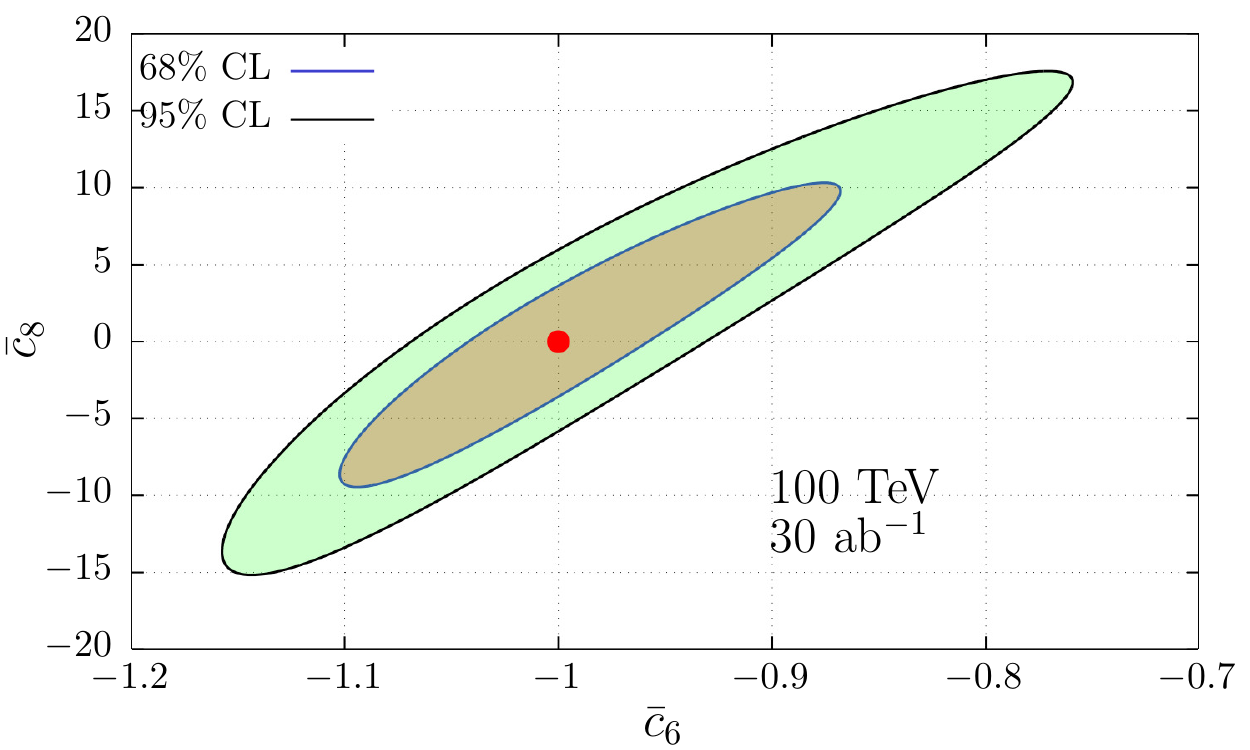}
 
 \includegraphics[scale=0.5]{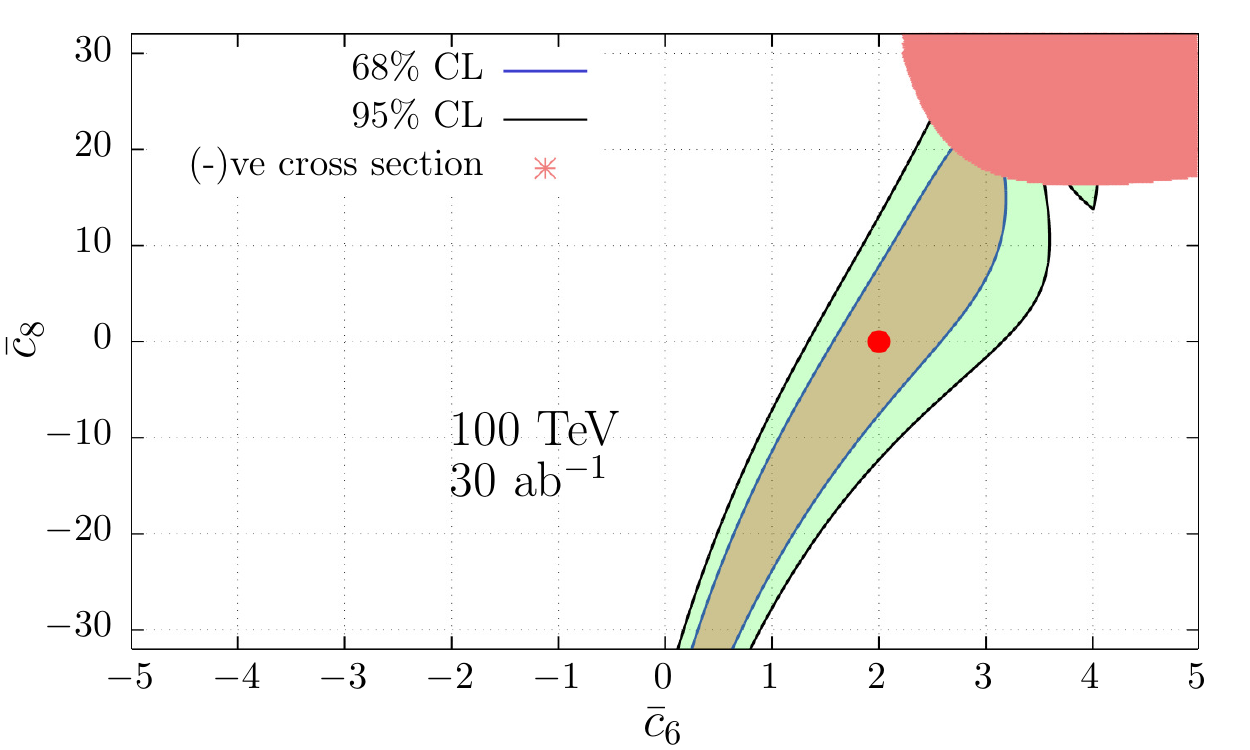}
  \includegraphics[scale=0.5]{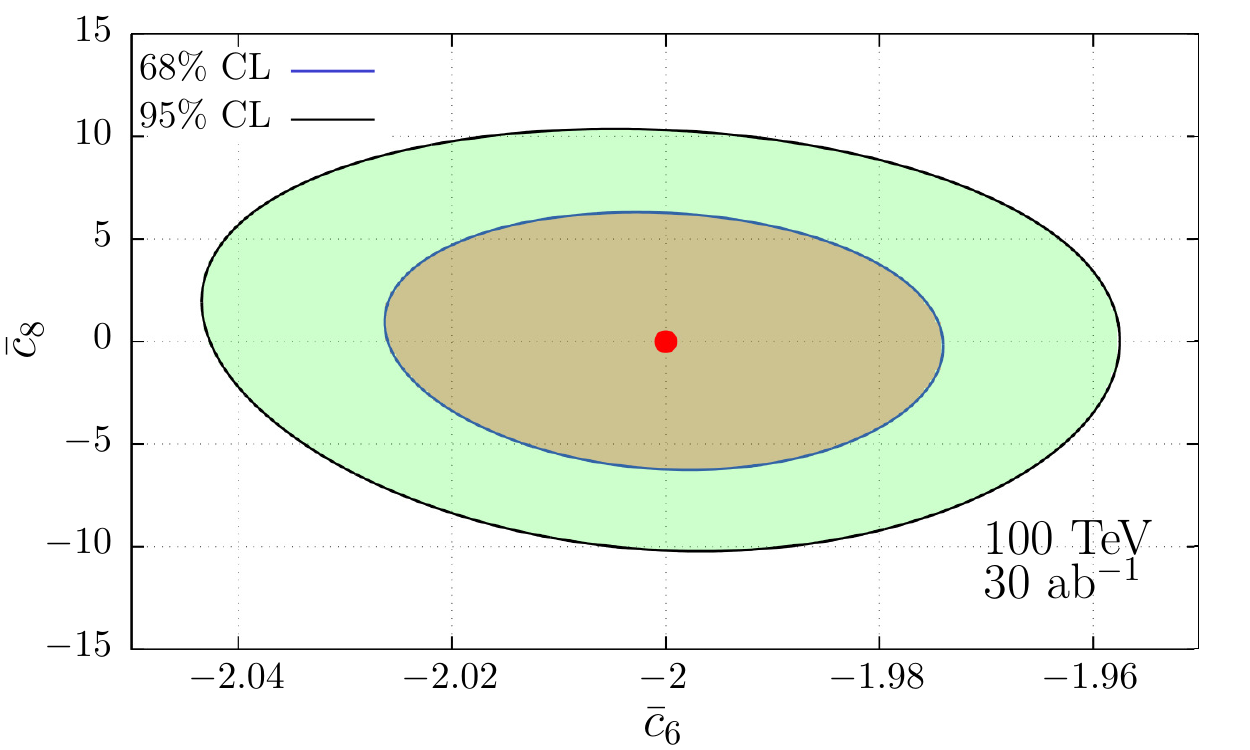}

 \includegraphics[scale=0.5]{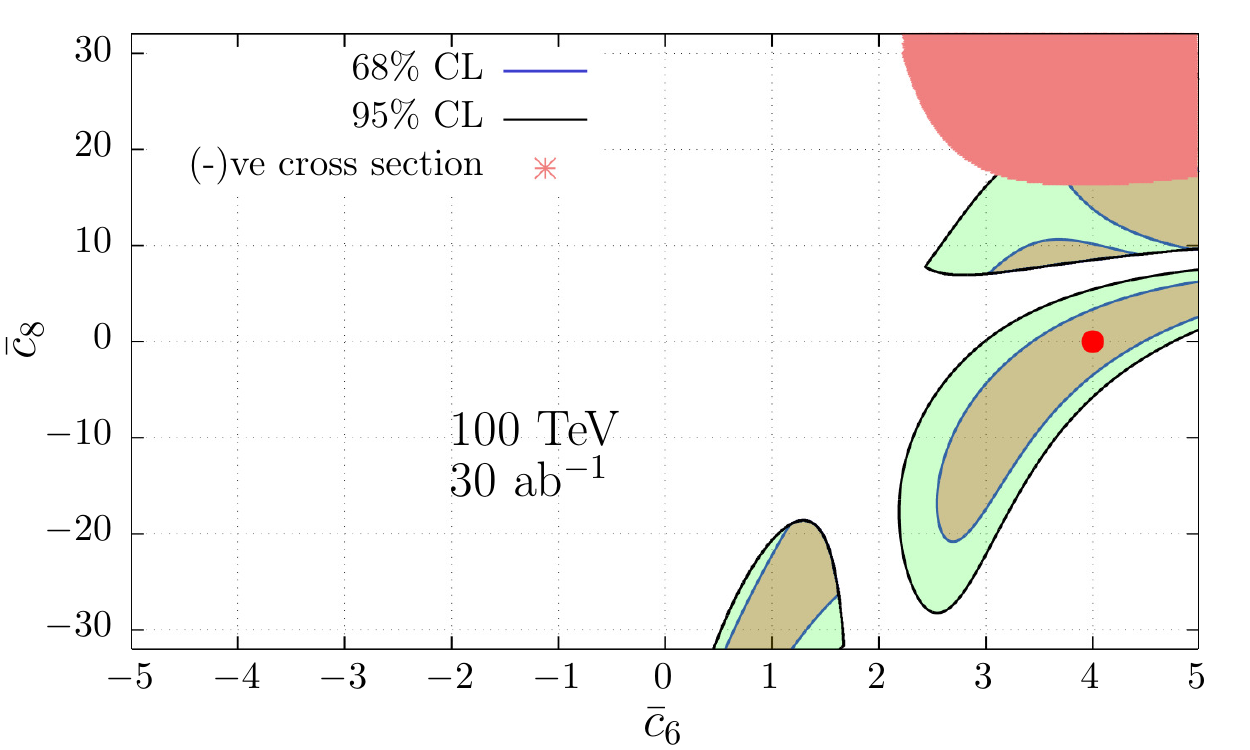}
  \includegraphics[scale=0.5]{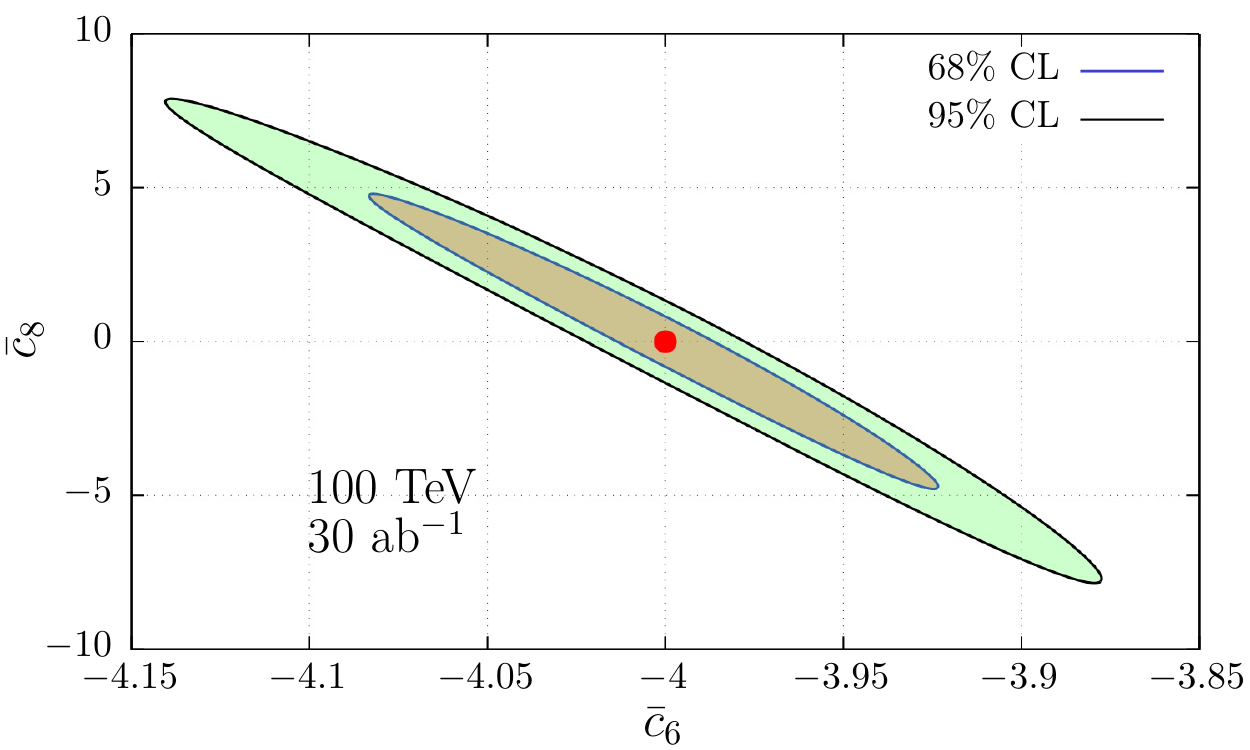}

 \caption{Expected 1$\sigma$ and 2$\sigma$ bounds in the ($\cbs$,$\cbe$) plane at  100 TeV, assuming $\cbstr=\pm1,\pm2,\pm4$ and $\cbetr=0$ (denoted by red dots).  \label{fig:c6c8100}}
\end{figure}

In Figs.~\ref{fig:c6c814} and \ref{fig:c6c8100} we show the constraints that can be set in the $(\cbs,\cbe)$ plane assuming $\cbetr=0$ and $\cbstr=\pm1,\pm2,\pm4$ for HL-LHC and a 100 TeV collider, respectively. 
As can be seen in Fig.~\ref{fig:c6c814}, at HL-LHC for large and positive values of $\cbstr$ we find results very close to  $\cbstr=0$. In general, including  $\cbstr$ negative values, we see that limits on $\cbs$ are not sizeably affected by the presence of $\cbe$. However,  sensible constraints on $\cbe$ cannot be obtained at the HL-LHC.  At 100 TeV, Fig.~\ref{fig:c6c8100}, (large) negative values of $\cbstr$ lead to strong constraints in the $(\cbs,\cbe)$ plane. However, we remind the reader that we do not take into account theory and experimental systematic uncertainties. As said for  the corresponding results in Scenario 1, these results may be affected by the aforementioned uncertainties.

\begin{figure}
 \centering
 \includegraphics[scale=0.5]{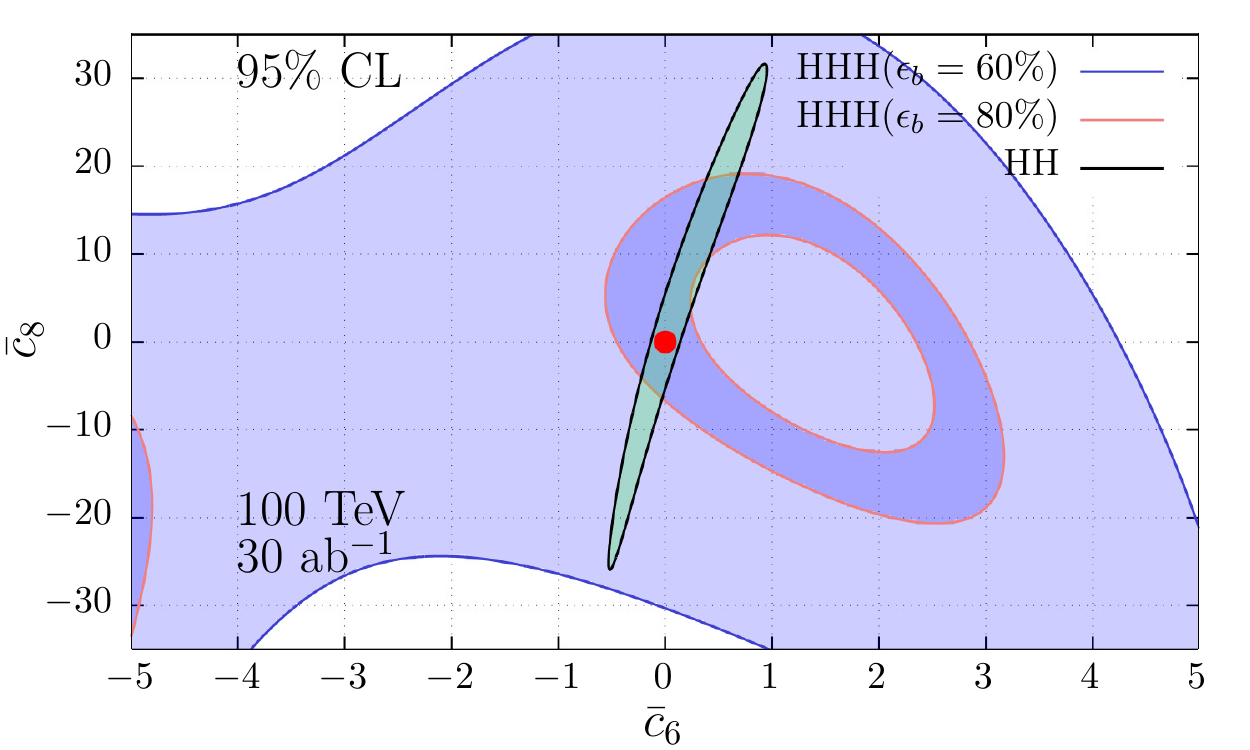}
 \includegraphics[scale=0.5]{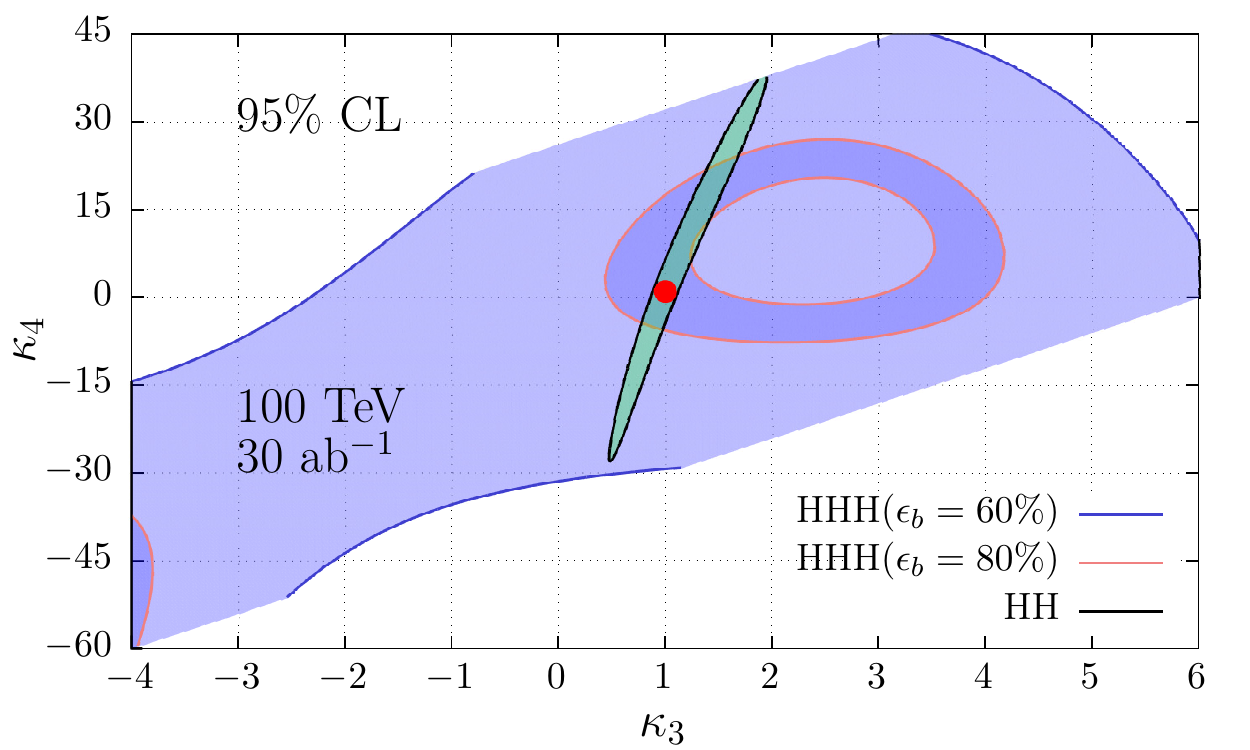}
	\caption{ Comparison between expected $2\sigma$-bounds from $HH(b \bar b\gamma\gamma)$ and $HHH(b \bar b b \bar b\gamma\gamma)$ at 100 TeV. The right plot  in ($\kappa_3,\kappa_4$) plane takes in account the perturbativity bounds on $\bar c_6$ and $\bar c_8$.  \label{fig:triple}}
\end{figure}

Last but not least, in Fig.~\ref{fig:triple} we compare the constraints obtained for $\cbstr = 0$ at 100 TeV (right plot of Fig.~\ref{fig:c6c8lims}) with the corresponding ones obtained following the analysis in Ref.~\cite{Papaefstathiou:2015paa, Contino:2016spe}, based on the $b \bar b b \bar b \gamma\gamma$ signature emerging from $pp \rightarrow HHH$ production.~\footnote{We have also looked at results from Ref.~\cite{Chen:2015gva}; following this analysis bounds are a bit stronger than in the case with 60\% $b$-tagging efficiency.} Triple Higgs bounds are derived via two different assumptions on $b$-tagging efficiency: optimistic (80\%) and conservative (60\%). As can be seen in Fig.~\ref{fig:triple}, double Higgs bounds are stronger than those from triple Higgs with the optimistic assumption. Especially, they are complementary to those from triple Higgs with the conservative assumption and their combination can lead to stronger results. We also show the corresponding comparison 
in $(\kappa_3,\kappa_4)$ plane taking into account the perturbative bounds on $\cbs$ and $\cbe$.

\clearpage

\section{Conclusion} \label{sec:conclusion}
The experimental determination of the Higgs potential and in particular of the Higgs self couplings is one the most far fetching goals of the HL-LHC and of future colliders. Its importance is matched only by the difficulty of such an endeavour: rates for multiple Higgs production which are directly sensitive to the self couplings, are very low making it hard to study distributions where most of the sensitivity actually lies. This is certainly true for the cubic coupling at the LHC, which can be accessed directly via $HH$ production, but becomes dramatic for the quartic coupling: its direct determination calls for measurements in the $HHH$ final state, whose production rate will be small even at a future 100 TeV pp colliders. 

The challenge on the one hand and the high-stakes on the other hand have provided strong motivation to the theoretical and experimental high-energy-physics community to devise alternative strategies. Among them, a new approach has emerged building up from the simple idea that single Higgs cross sections might display a sensitivity on the cubic coupling at higher orders.  Since the first proposal in the context of future $e^+e^-$ colliders~\cite{McCullough:2013rea}, the idea has been developed and extended to hadron colliders, eventually proving to be competitive with the direct determinations. A very first experimental analysis by CMS~\cite{CMS:2018rig} based on the proposal of refs.~\cite{Degrassi:2016wml, Maltoni:2017ims} has confirmed the expectations of the theoretical studies.  

Recently, some of us have proposed to extend the idea further and determine the (cubic and) quartic coupling exploiting the sensitivity coming from loop effects in $HH$ in the context of future $e^+e ^-$ colliders~\cite{Maltoni:2018ttu} . In this work we have moved one step further and explored the reach of hadron colliders by determining  the sensitivity to the (cubic and) quartic coupling of the main double Higgs production channel, $gg \to HH$, up to two loops. Being a technically challenging two-loop computation we have employed the most up-to-date numerical multi-loop techniques, providing for the first time a complete and consistent calculation of these effects.

We have considered two different scenarios, one ``EFT-like'' where the cubic and quartic couplings are related and one where they are varied independently. Our results clearly indicate that while the HL-LHC will have limited sensitivity, at the FCC-100 the precision on $HH$ differential measurements will be such that $HH$ will be more sensitive to independent  deviations in the self couplings than $HHH$ production itself. The best constraints on the quartic will therefore be obtained by combining $HH$ precision measurements with the direct determinations from $HHH$.  

\section*{Acknowledgements} 
We would like to thank Matthias Kerner
for a valuable comment on the input files for the numerical integration
of the two-loop amplitudes. S.B. would like to thank the hospitality of CP3 Louvain during the completion of the work. D.P.~is supported by the Alexander von Humboldt Foundation, in the framework of the Sofja Kovalevskaja Award Project ``Event Simulation for the Large Hadron Collider at High Precision''.  A.S. is supported by the MOVE-IN Louvain Cofund grant. The work of X.Z. is supported the European Union's Horizon 2020 research and innovation programme as part of the Marie Sk\l odowska-Curie Innovative Training Network MCnetITN3 (grant agreement no. 722104). 
This work has received funding from the ERC grant ``MathAm''  and from F.R.S.-FNRS under the `Excellence of Science' EOS be.h project n. 30820817. Computational resources have been provided by the supercomputing facilities of the Universit\'e catholique de Louvain (CISM/UCL) and the Consortium des \'Equipements de Calcul Intensif en F\'ed\'eration Wallonie Bruxelles (C\'ECI).

\clearpage
 
\appendix

 \section{Cut efficiency}
 \label{app:cuts}

In this section we explicitly write the cuts used in our analysis. 
The cuts are the same as in Ref.~\cite{Azatov:2015oxa}, on which our analysis is based. Specifically, at 14 TeV, we have,
\begin{align}
	p_T(b_1)>50\textrm{~GeV}&,\qquad p_T(b_2)>30\textrm{~GeV}\, , \nn\\
	p_T(\gamma_1)>50\textrm{~GeV}&, \qquad p_T(\gamma_2)>30\textrm{~GeV}\, , \nn\\
	|\eta(b)|<2.5&,\qquad |\eta(\gamma)|<2.5\, , \nn\\
	0.5<\Delta R(b,b)<2&,\qquad \Delta R(\gamma,\gamma)<2\, , \nn\\
	\Delta R(b,\gamma)>1.5\, &, 
\end{align}
while at 100 TeV, the $p_T$ cuts are replaced by:
\begin{align}
	p_T(b_1)>60\textrm{~GeV}&,\qquad  p_T(b_2)>40\textrm{~GeV}\, , \nn\\
	p_T(\gamma_1)>60\textrm{~GeV}&,\qquad  p_T(\gamma_2)>40\textrm{~GeV}\, .
\end{align}

\begin{figure}
	\includegraphics[width=0.45\textwidth]{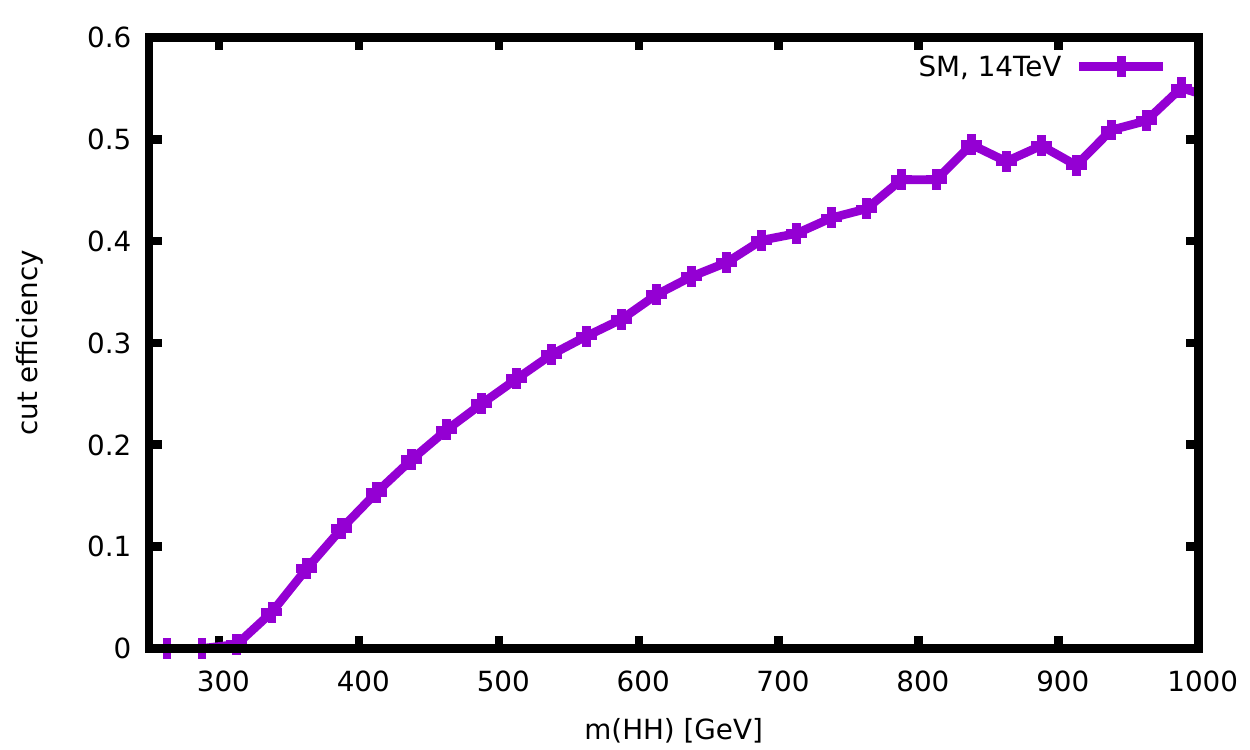}
	\includegraphics[width=0.45\textwidth]{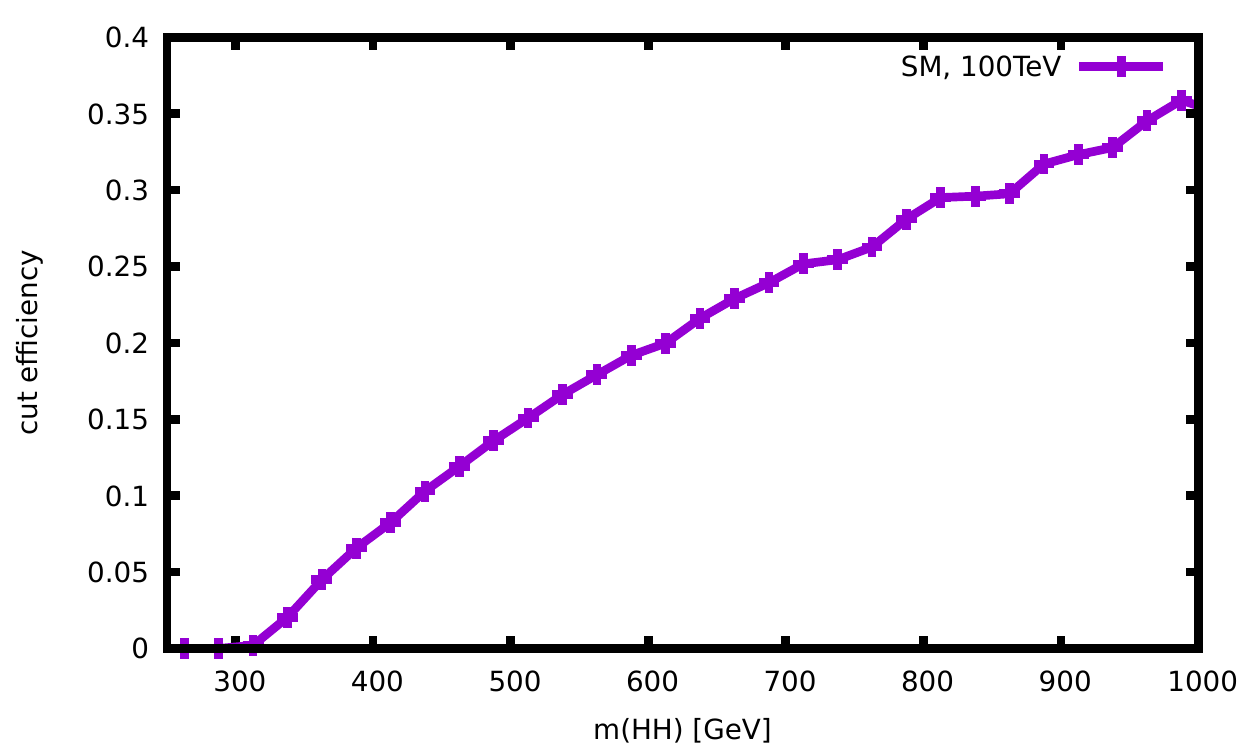}
	\caption{Differential cut efficiency for SM double-Higgs signal from the $b \bar b\gamma\gamma$ signature, at parton level.\label{fig:cuteff}}
\end{figure}

In fig. \ref{fig:cuteff}, we show the differential cut efficiency for the signal, assuming SM double Higgs production and narrow-width approximation.
In other words we plot the ratio between the number of events predicted in the SM with and without the cuts as function of $m(HH)$.
Since spin-0 contributions dominate for both SM and BSM cases,  cut efficiencies for BSM cases are very similar. 

The zero efficiency in the $250~{\rm GeV}<m(HH)<300~{\rm GeV}$ phase-space region is not a surprise;
when Higgs boson pairs are produced at the threshold, both the $b\bar b$ and $\gamma\gamma$ pairs from the Higgs decays are back-to-back and therefore rejected by the cuts $\Delta R(b,b)<2$ and $\Delta R(\gamma,\gamma)<2$. Increasing the energy, both Higgs can have non-vanishing transverse momentum and therefore their decay products can be not back-to-back and tend to be collimated for very high energies.

\section{Fit details}
\label{app:fit}
 In this Appendix we describe in detail the $\chi^2$ functions that have been used in this work for extracting  from $m(HH)$ distributions 1$\sigma$ and 2$\sigma$ bounds in the ($\cbs,\cbe$) parameter space. The general formula of the $\chi^2$ that has been exploited for our results has two degrees of freedom, $\bar{c}_6$ and $\bar{c}_8$, and reads
\begin{equation}
\chi^2=\sum_{i=1}^{n_{\rm bins}}\frac{[N_i^{HH}(\bar{c}_6,\bar{c}_8)-N_i^{HH}(\bar{c}_6^{\textrm{true}},\bar{c}_8^{\textrm{true}})]^2}{N_i^{HH}(\bar{c}_6,\bar{c}_8)+N_i^{\textrm{BKG}}}\theta(N_i^{HH}(\bar{c}_6,\bar{c}_8))\label{eq:chi2}
\end{equation}
where $\bar{c}_8^{\textrm{true}}$, and in some cases also $\cbe$, have  been set equal to zero. On the contrary $\bar{c}_6^{\textrm{true}}$ has been fixed to different values in the $-5<\cbs<5$ range and $\cbs$ has been kept always free.
In eq.~\eqref{eq:chi2}, $N_i^{HH}(\bar{c}_6,\bar{c}_8)$ is the number of signal events in each bin $i$  for the specific $\bar{c}_6$ and $\bar{c}_8$ values, while 
$N_i^{\textrm{BKG}}$ is the number of background events in the same bin, for a total of $n_{\rm bins}$. The $\theta$ function ensures that if the prediction for $N_i^{HH}(\bar{c}_6,\bar{c}_8)$ is negative, the information from the bin $i$ is discarded.

Bounds on $\cbs$ and $\cbe$ have been obtained following a fit procedure similar the one presented in Ref.~\cite{Azatov:2015oxa}, from which we have taken also the selection cuts (see  Appendix \ref{app:cuts}) and binning in the $m(HH)$ distribution. For this reason,  the value of $N_i^{\textrm{BKG}}$ is directly taken from Ref.~\cite{Azatov:2015oxa}. On the contrary, $N_i^{HH}(\bar{c}_6,\bar{c}_8)$ is derived from the value $N_i^{HH}(0,0)$, the SM prediction, from the same reference, which takes into account also higher-order QCD corrections. Assuming that these effect factorise with $\cbs$ and $\cbe$ corrections, the selection cuts of Appendix \ref{app:cuts} and the shower effects involved in the simulation of  $N_i^{HH}(0,0)$, we can obtain $N_i^{HH}(\bar{c}_6,\bar{c}_8)$ via the relation
\beq
N_i^{HH}(\bar{c}_6,\bar{c}_8)=N_i^{HH}(0,0) \mu_i^{\rm theory}(\bar{c}_6,\bar{c}_8)\,,\qquad 
\mu_i^{\rm theory}\equiv \frac{\int d\Phi_i (d\sigma^{\rm pheno}_{\rm NLO}/d\Phi_i)}{ \int d\Phi_i (d \sigma^{\rm SM}_{\rm LO}/d\Phi_i)   }
\,,
\eeq
where in the right equation we have understood the dependence on $\bar{c}_6$ and $\bar{c}_8$ and $\sigma^{\rm SM}_{\rm LO}=\sigma^{\rm pheno}_{\rm NLO}|_{\bar{c}_6=0,\bar{c}_8=0}$. The quantity $\Phi_i$ corresponds to the $b \bar b \gamma \gamma$ phase space such that the reconstructed $m(HH)$ value is within the bin $i$. Within all the work, unless differently specified, we take into account the selection cuts of Appendix \ref{app:cuts} in $\Phi_i$. When we say ``flat $\mu$-bin''  we precisely refer to the case where selection cuts are not taken into account for the definition of $\mu_i^{\rm theory}$. 

\clearpage

\section{Topologies of the integral expressions from non-factorisable two-loop contributions}
\label{app:topo}

\begin{figure}[h]
\centering
 \includegraphics[width=0.27\textwidth]{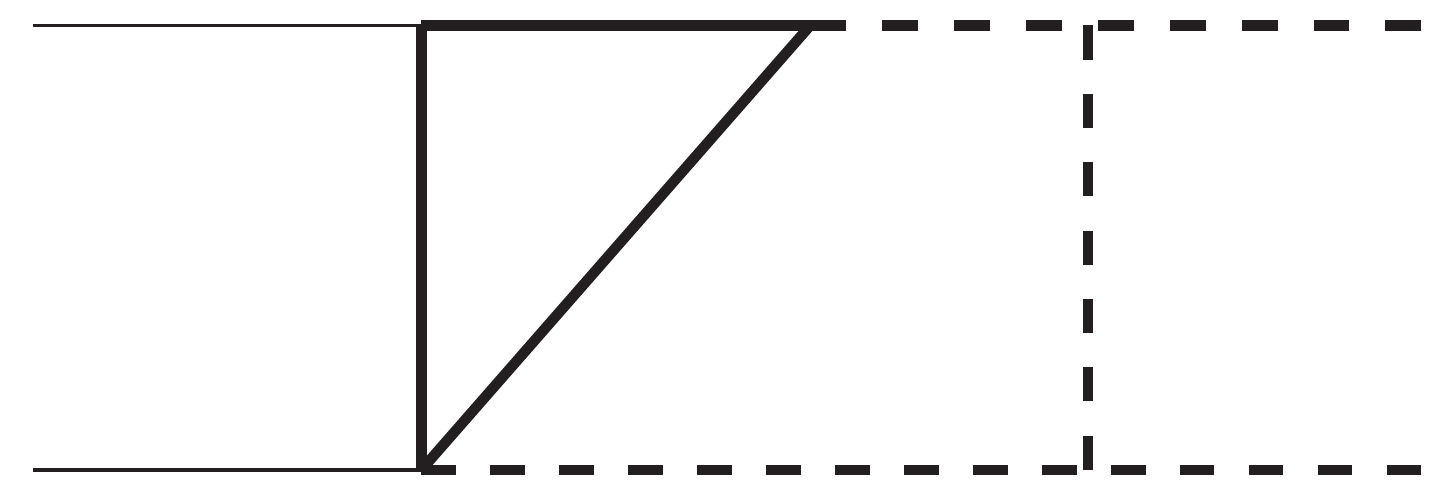}
  \includegraphics[width=0.27\textwidth]{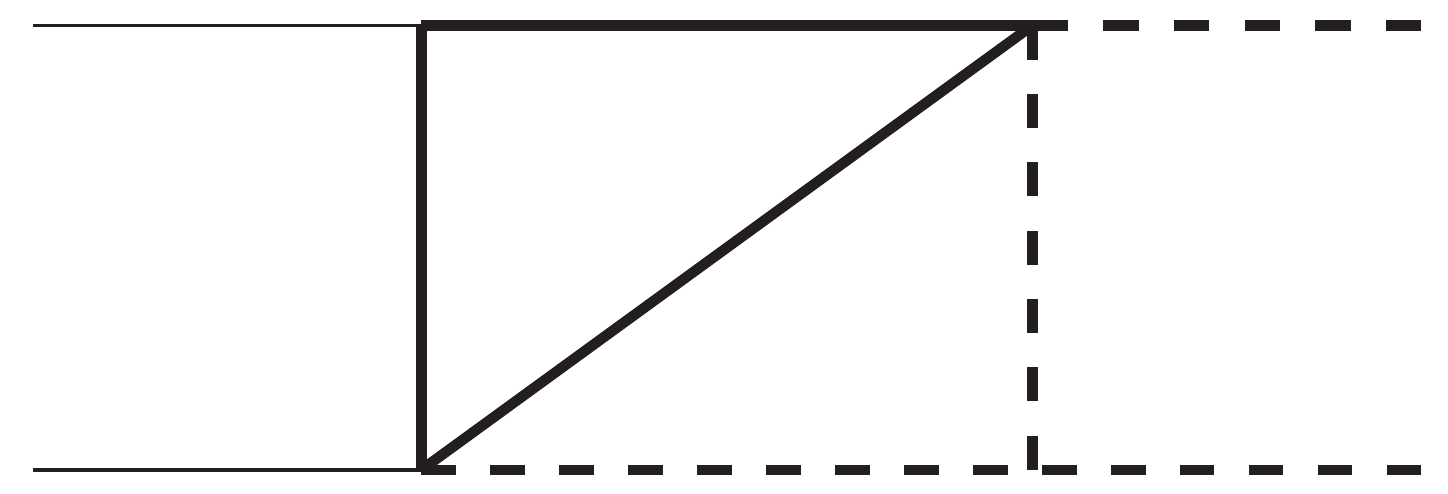}
   \includegraphics[width=0.27\textwidth]{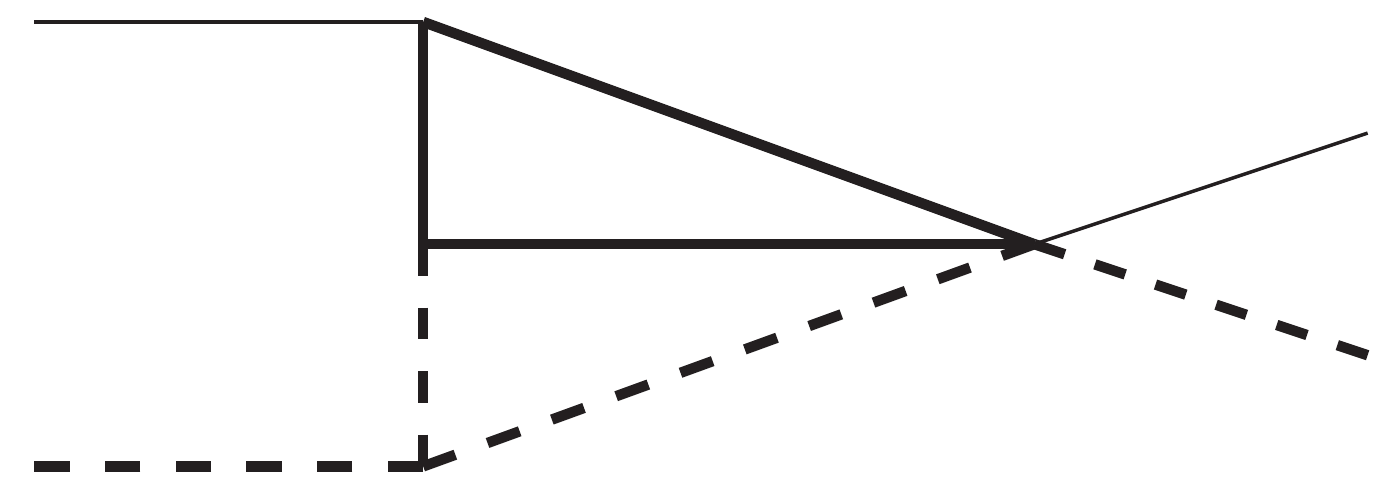}
    \includegraphics[width=0.27\textwidth]{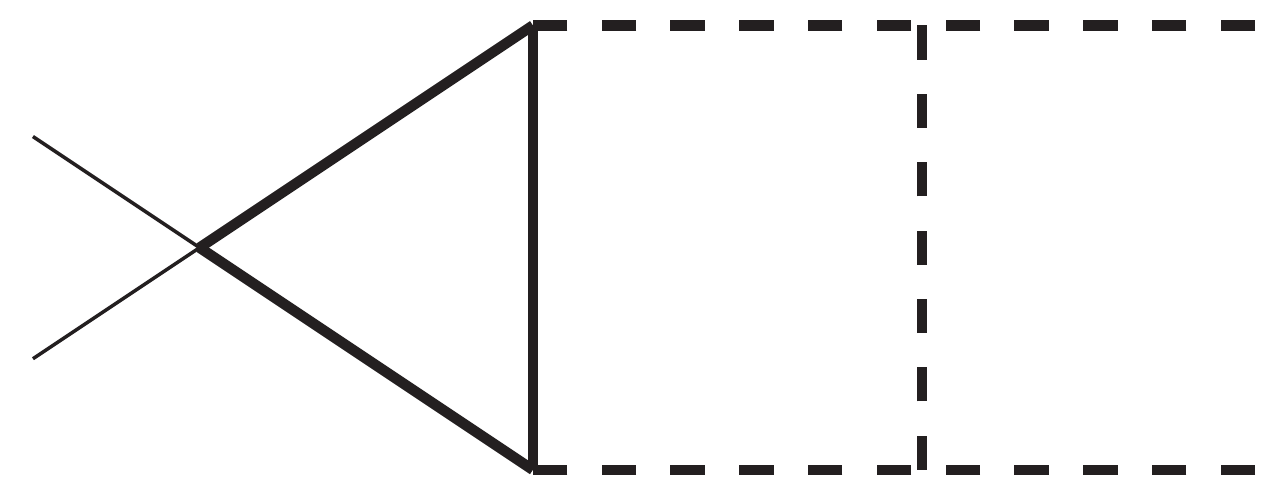}
     \includegraphics[width=0.27\textwidth]{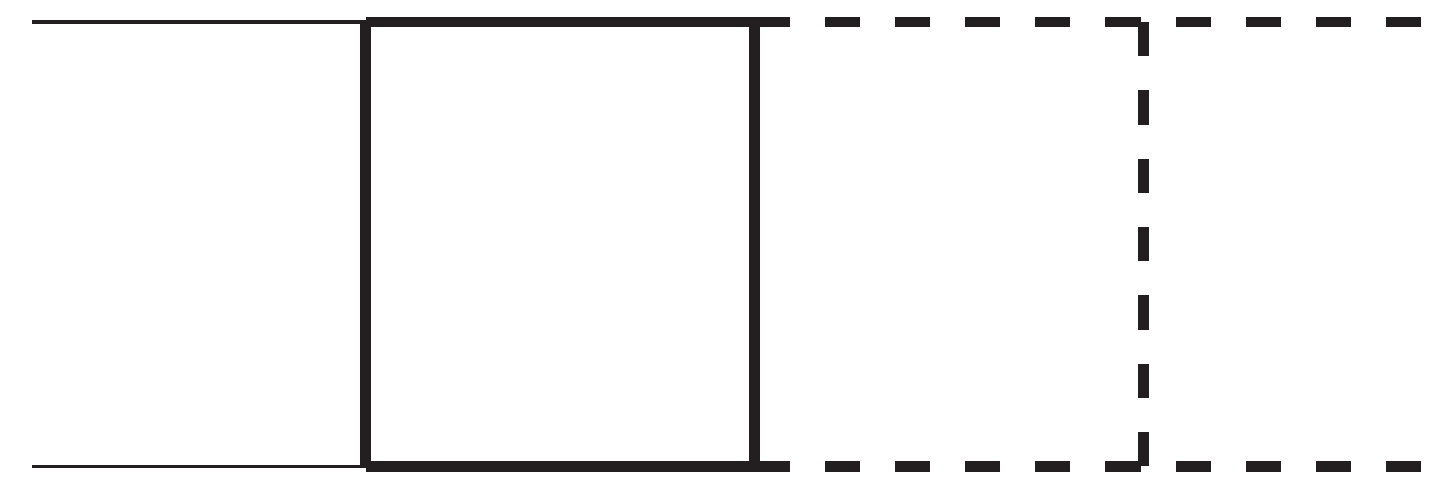}
     \includegraphics[width=0.27\textwidth]{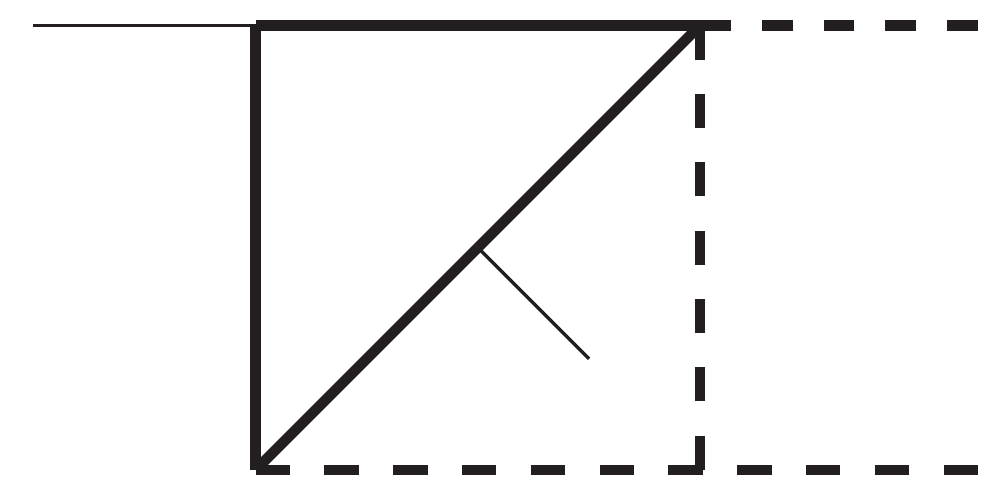}
     \includegraphics[width=0.27\textwidth]{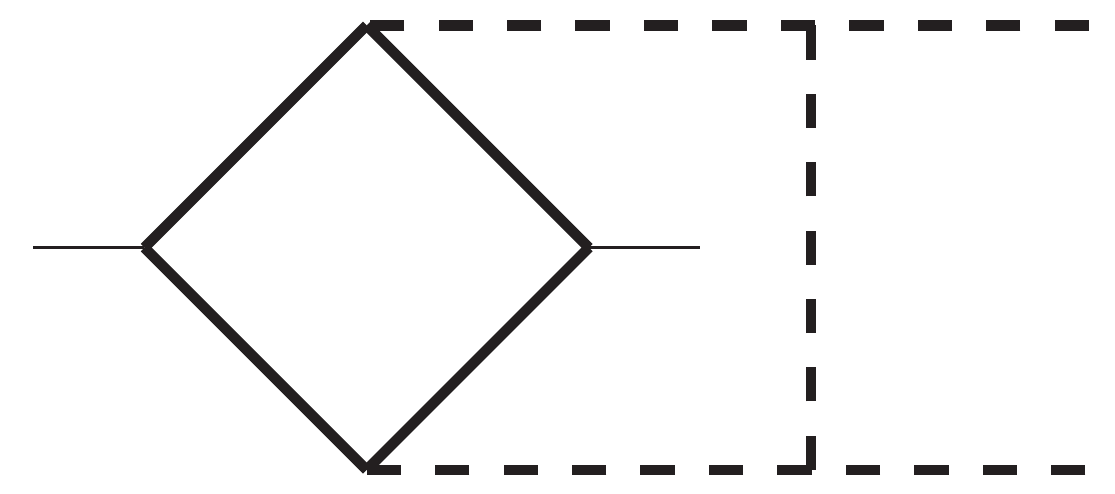}
     \includegraphics[width=0.27\textwidth]{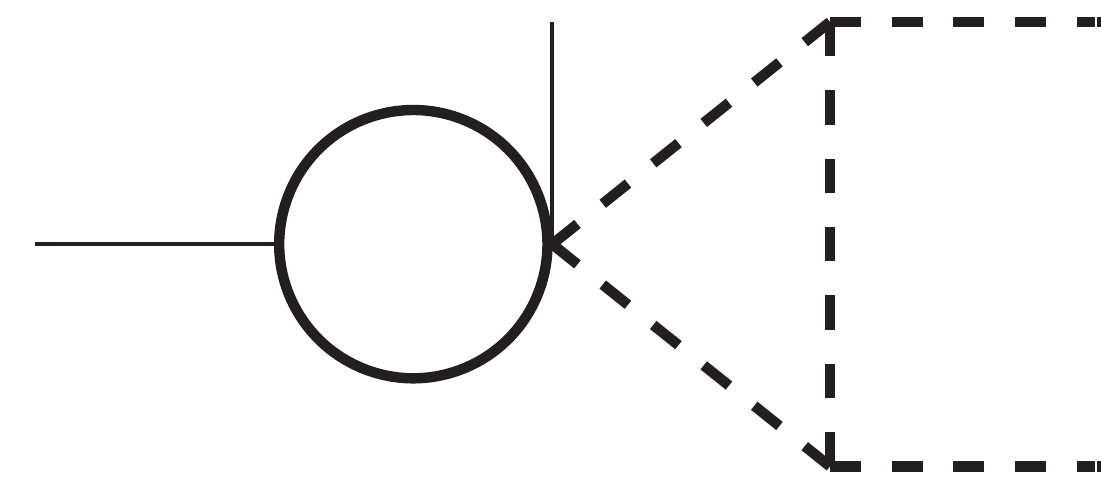}
  \includegraphics[width=0.27\textwidth]{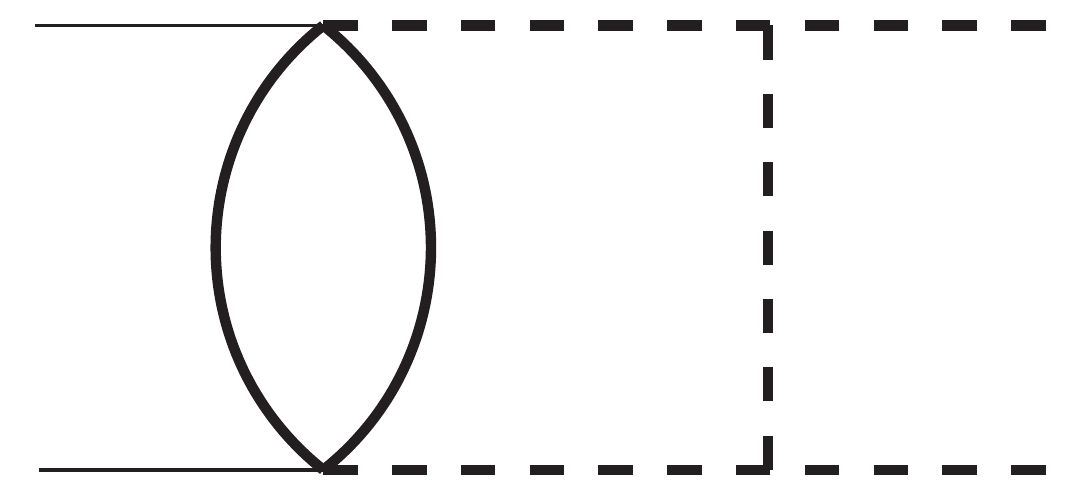}
     \includegraphics[width=0.27\textwidth]{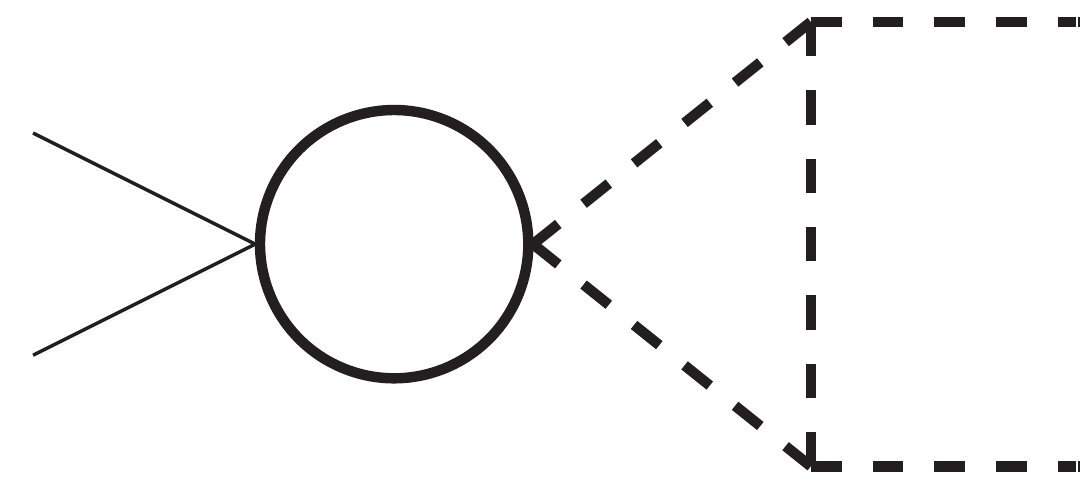}
     \includegraphics[width=0.27\textwidth]{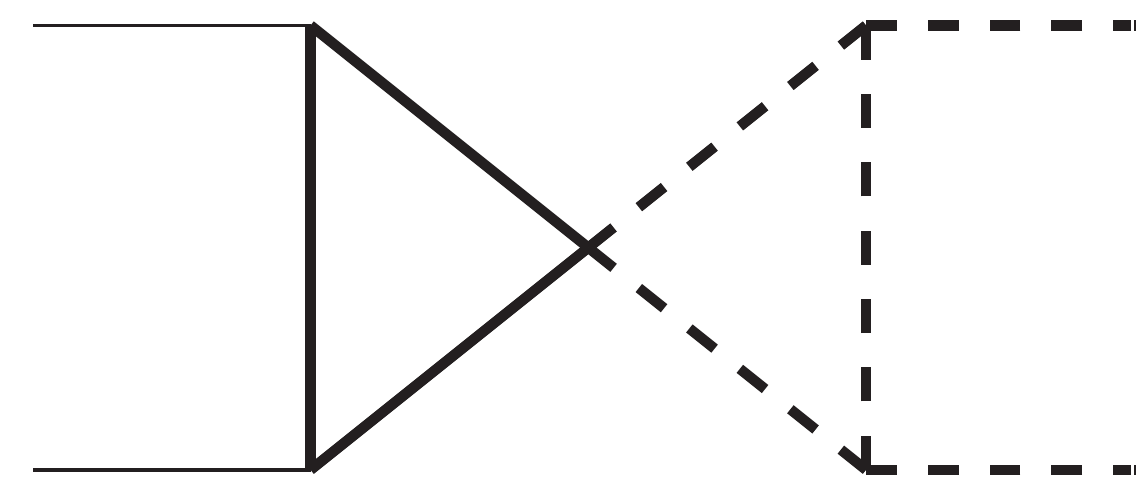}
          \includegraphics[width=0.27\textwidth]{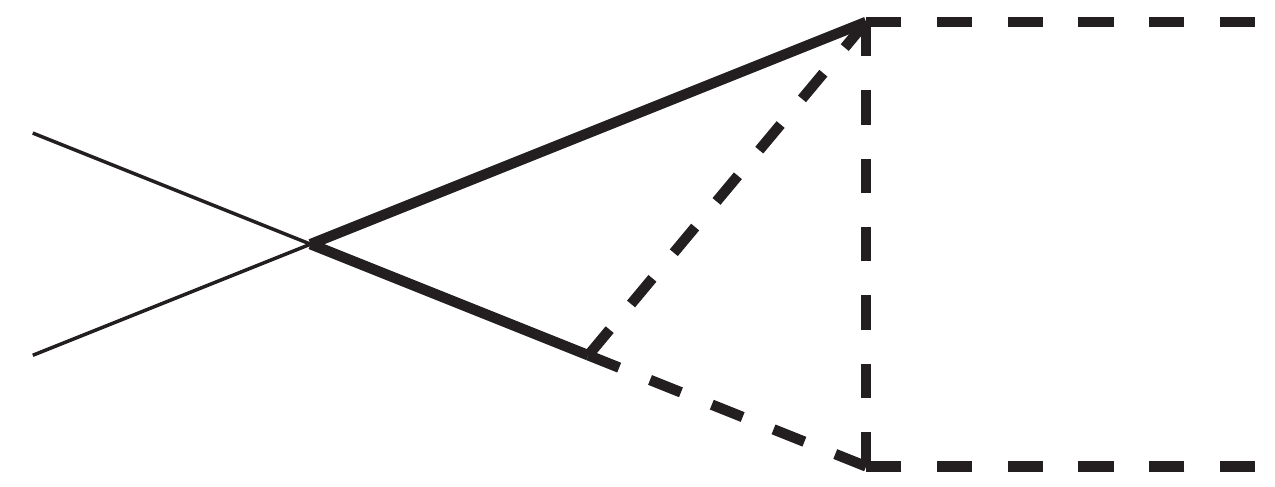}
     \includegraphics[width=0.27\textwidth]{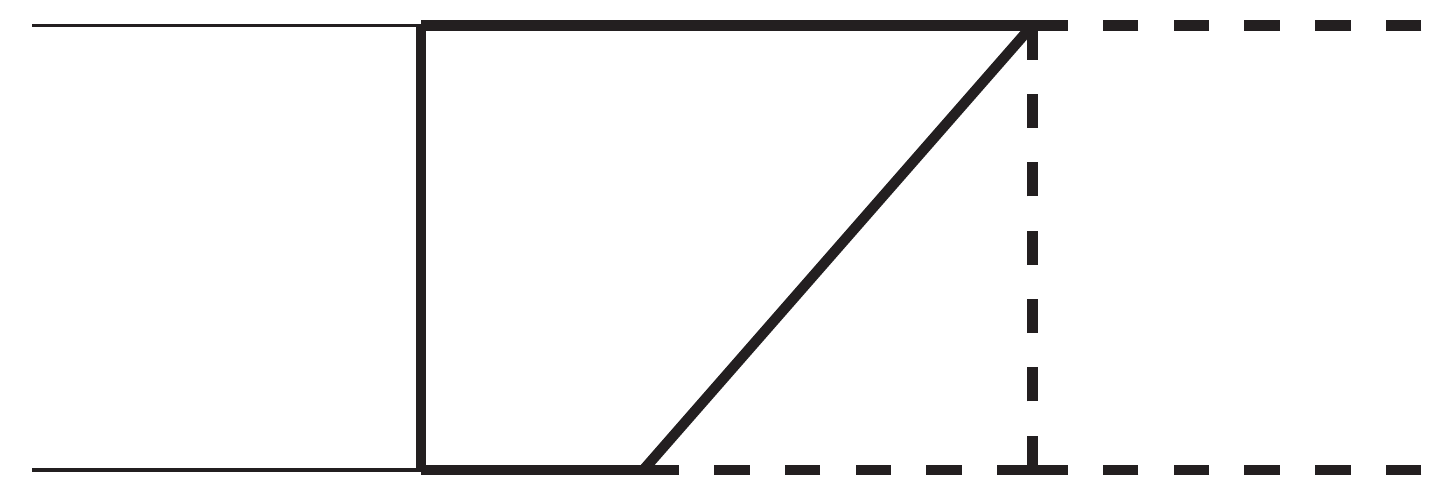}
 \caption{Topologies of the integral expressions from the form factors $\tilde F_{0,a}$ and $\tilde F_{2,a}$.}
 \label{fig:toposBa}
\end{figure}

\begin{figure}
\centering
 \includegraphics[width=0.27\textwidth]{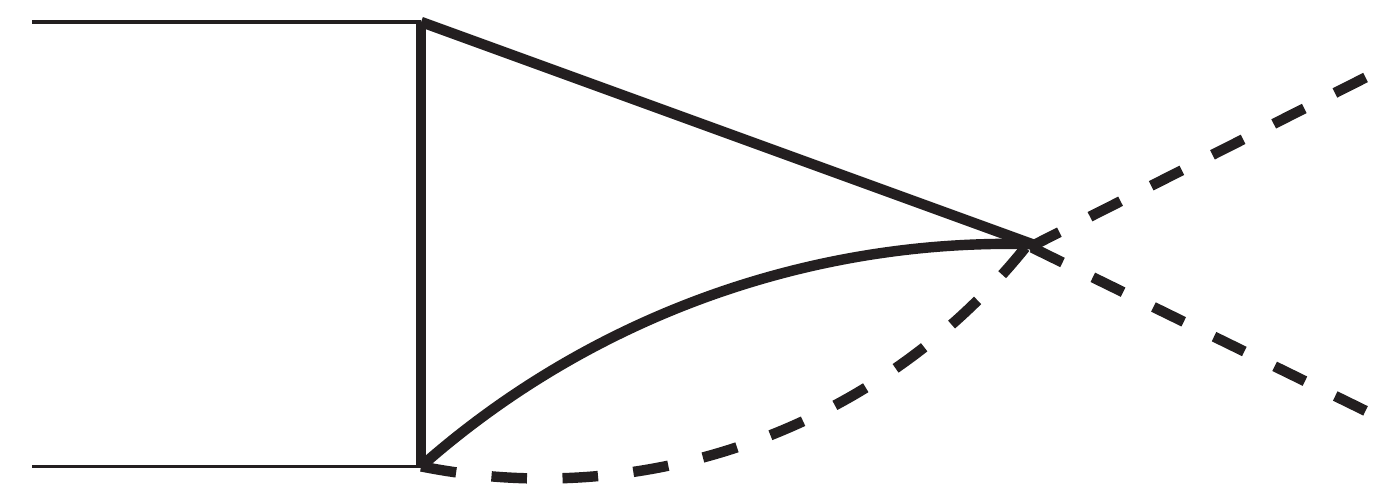}
  \includegraphics[width=0.27\textwidth]{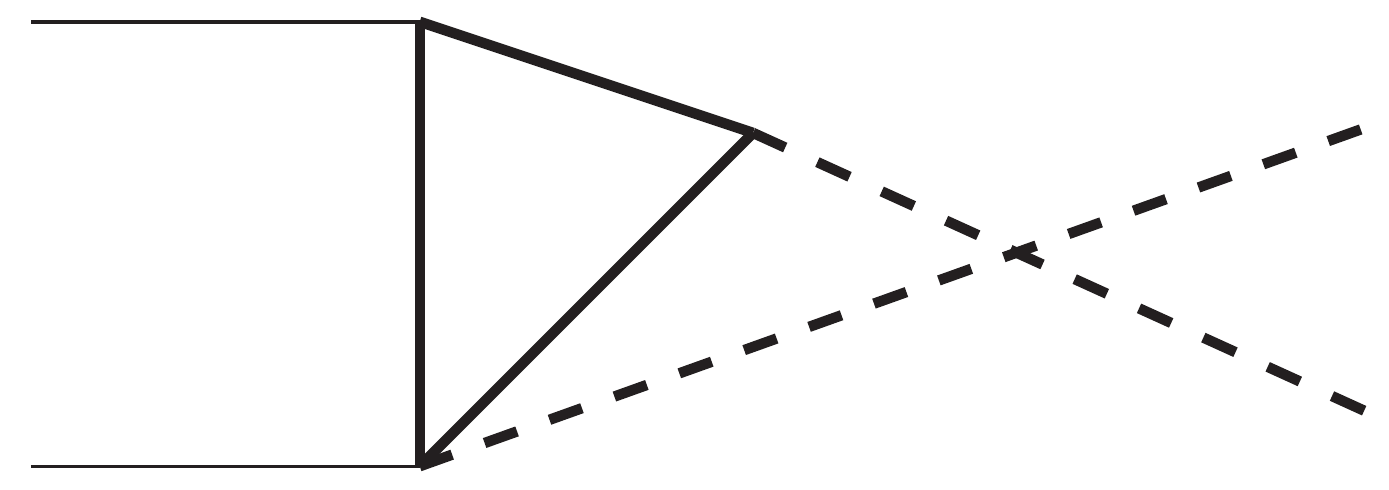}
   \includegraphics[width=0.27\textwidth]{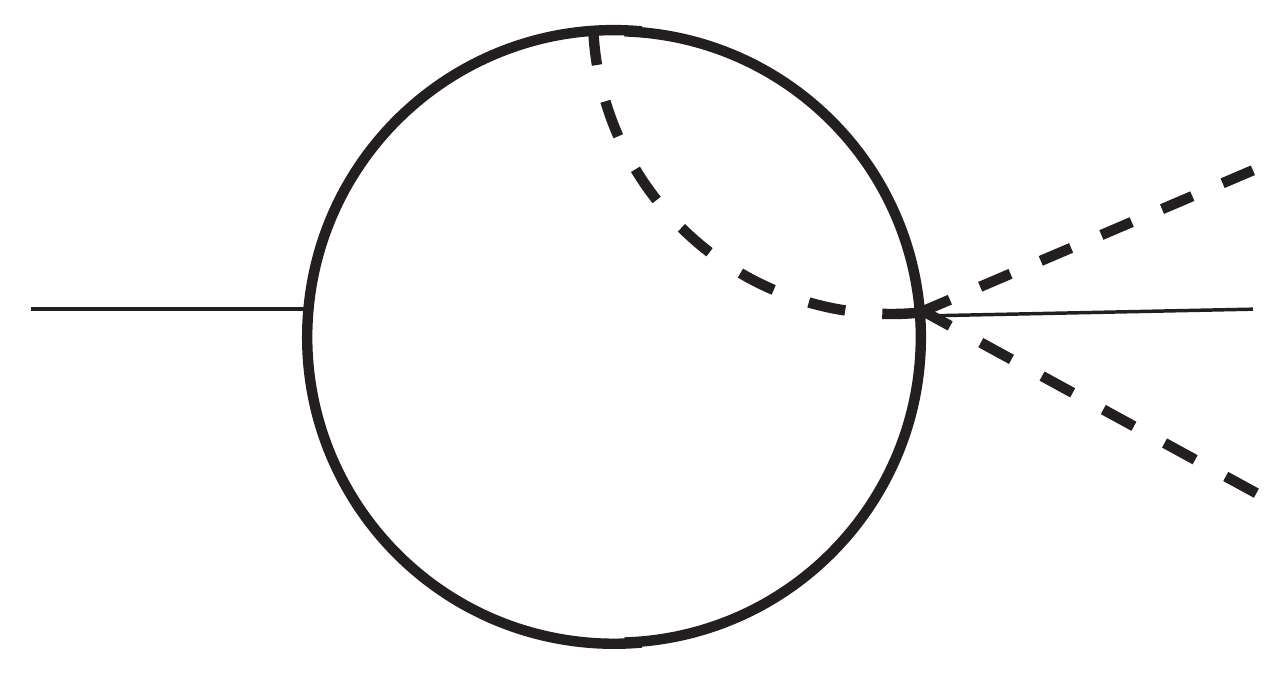}
    \includegraphics[width=0.27\textwidth]{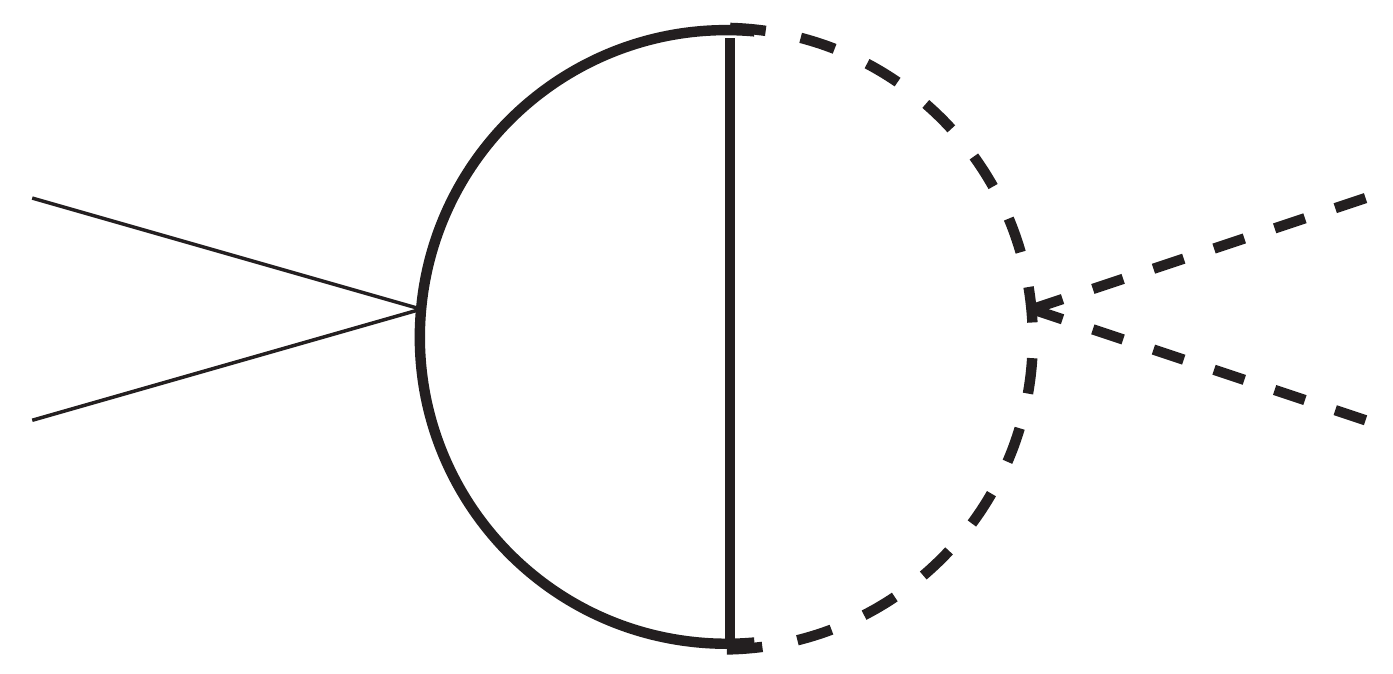}
     \includegraphics[width=0.27\textwidth]{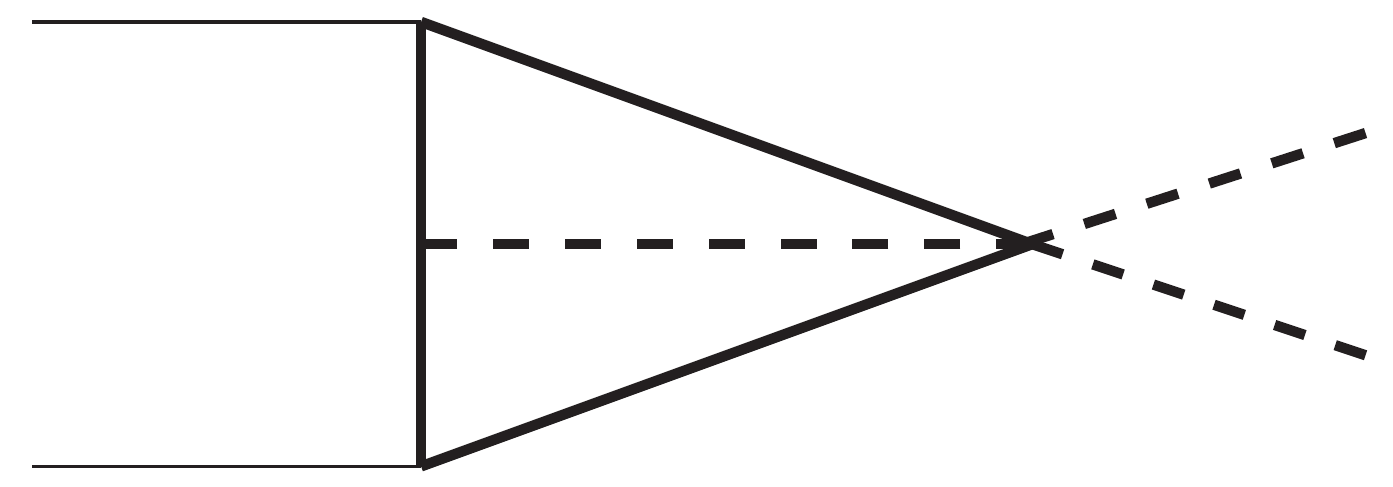}
          \includegraphics[width=0.27\textwidth]{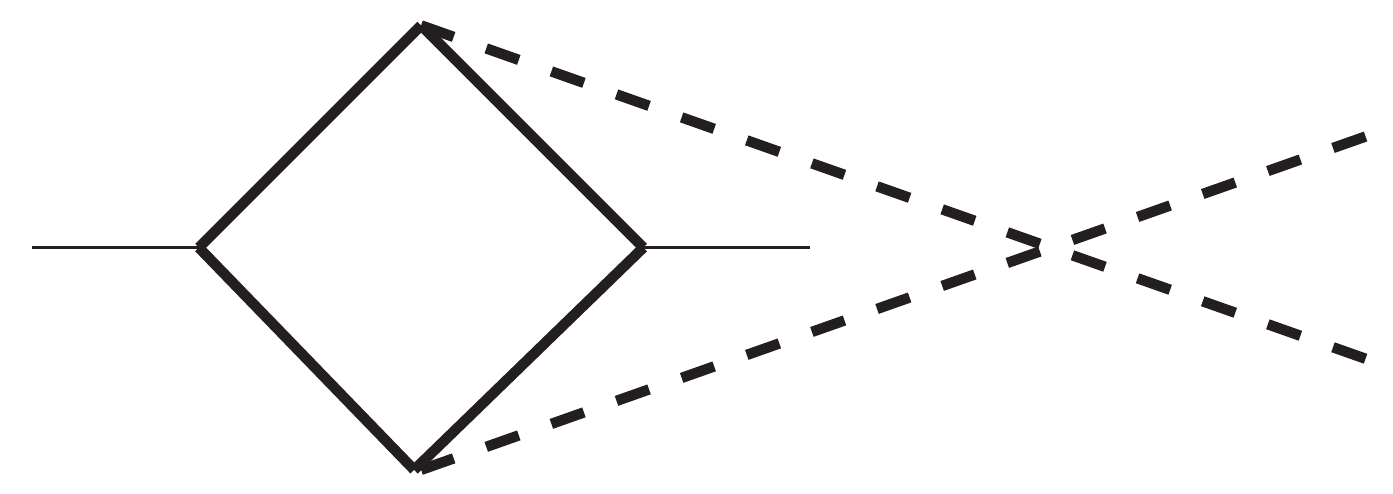}
               \includegraphics[width=0.27\textwidth]{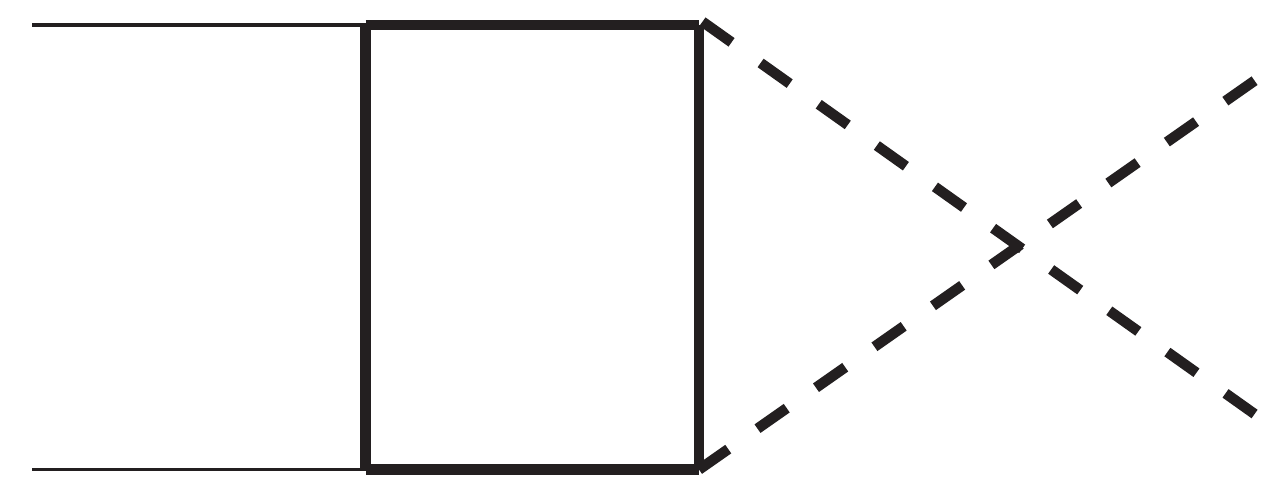}
 \caption{Topologies of the integral expressions from the form factor $\tilde F_{0,b}$.}
 \label{fig:toposBb}
\end{figure}

In this Appendix we show the topologies of the integral expressions obtained from non-factorisable two-loop contributions. Each topology can lead to more than one integral expression. In Fig.~\ref{fig:toposBa} we show those relevant for $\tilde F_{0,a}$ and $\tilde F_{2,a}$, while in Fig.~\ref{fig:toposBb} those for $\tilde F_{0,b}$, which are relevant also for $\tilde F_{0,c}$. Thick lines correspond to massive propagators, dashed with mass $\mh$ while solid with mass $\mt$. The solid-thin lines corresponds to massless propagators.

\clearpage

 \bibliographystyle{JHEP}
 \bibliography{article} 
 
\end{document}